\documentclass{natureprintstyle}
\usepackage{psfig}

\usepackage{lscape}
\usepackage{amsmath}
\usepackage{amssymb}
\usepackage{color}
\usepackage{url}
\usepackage{ulem}
\usepackage{multirow}
\usepackage{graphics,graphicx}
\usepackage[affil-it]{authblk}
\usepackage{blindtext}
\usepackage{subfig}
\usepackage{hyperref}
\usepackage{auto-pst-pdf}

\usepackage{astjnlabbrev-nature} 

\newcommand{\todo}{\ifmmode \text{\color{red}{\Huge(\bullet)}} \else {\color{red}{\Huge$\bullet$}}\fi}
\newcommand{\tido}{\ifmmode {{\color{red}\bullet}} \else \color{red}{$\bullet$}\fi}

\def\spose#1{\hbox to 0pt{#1\hss}}
\def\simlt{\mathrel{\spose{\lower 3pt\hbox{$\mathchar"218$}}
     \raise 2.0pt\hbox{$\mathchar"13C$}}}
\def\simgt{\mathrel{\spose{\lower 3pt\hbox{$\mathchar"218$}}
     \raise 2.0pt\hbox{$\mathchar"13E$}}}

\makeatletter
\renewcommand{\section}{\@startsection%
{section}{1}{0mm}{-\baselineskip}%
{0.5\baselineskip}{\normalfont\Large\bfseries}}%
\makeatother

\title{Changing-look Active Galactic Nuclei}

\author{Claudio Ricci$^{1,2,3}$ \& Benny Trakhtenbrot$^4$}

\begin{document}
\pagestyle{plain}
\pagenumbering{arabic}

\twocolumn[
  \begin{@twocolumnfalse}

\let\newpage\relax\maketitle

\begin{affiliations}
\item N\'ucleo de Astronom\'ia de la Facultad de Ingenier\'ia, Universidad Diego Portales, Av. Ej\'ercito Libertador 441, Santiago, Chile\\
\item Kavli Institute for Astronomy and Astrophysics, Peking University, Beijing 100871, China\\
\item George Mason University, Department of Physics \& Astronomy, MS 3F3, 4400 University Drive, Fairfax, VA 22030, USA \\
\item School of Physics and Astronomy, Tel Aviv University, Tel Aviv 69978, Israel
\end{affiliations}

\bigskip

    \begin{abstract}
Active Galactic Nuclei (AGN) are known to show flux variability over all observable timescales and across the entire electromagnetic spectrum. Over the past decade, a growing number of sources have been observed to show much more dramatic flux and spectral changes, both in the X-rays and in the optical/UV. Such events, commonly described as ``changing-look AGN'', can be divided into two well-defined classes. Changing-obscuration objects show strong variability of the line-of-sight column density, mostly associated with clouds or outflows eclipsing the central engine of the AGN. Changing-state AGN are instead objects in which the continuum emission and broad emission lines appear or disappear, and are typically triggered by strong changes in the accretion rate of the supermassive black hole. Here we review our current understanding of these two classes of changing-look AGN, and discuss open questions and future prospects.
      \bigskip
    \end{abstract}
  \end{@twocolumnfalse}
  ]
{
  \renewcommand{\thefootnote}%
    {\fnsymbol{footnote}}
  \footnotetext[0]{\small e-mails: claudio.ricci@mail.udp.cl, trakht@wise.tau.ac.il} 
}

\section{Background}\label{sec:intro}
Supermassive black holes (SMBHs) are ubiquitously found at the centers of most massive galaxies\cite{Magorrian:1998fr}, and grow predominantly through the accretion of material from their surroundings. The accretion process may produce highly luminous radiation across the entire electromagnetic spectrum. Such active galactic nuclei (AGN) are commonly classified based on their observed optical/UV or X-ray spectral properties. 
In the optical/UV, objects showing both broad ($\gtrsim 1000\,{\rm km\,s}^{-1}$) and narrow ($\lesssim 1000\,{\rm km\,s}^{-1}$) emission lines are classified as `Type\,1' AGN, while those displaying only narrow lines are referred to as `Type\,2' sources. 
Several studies use a finer classification scheme, based on increasingly fainter broad emission lines (i.e., Type\,1.2, 1.5, 1.8 and 1.9).
In the X-rays, objects showing line-of-sight column densities $N_{\rm H}\geq 10^{22}\rm\,cm^{-2}$ are commonly referred to as `obscured', while those with $N_{\rm H}< 10^{22}\rm\,cm^{-2}$ are `unobscured'. A very good match has been found between these two classification schemes, with the vast majority of Type\,1s (Type\,2s) being unobscured (obscured)\cite{Koss:2017gv,Ricci:2017co}.
A good, first order explanation for these basic, observationally-motivated AGN sub-classes is provided by the AGN unification model\cite{Antonucci93,Urry95,Ramos-Almeida:2017hw}, according to which in Type\,1 (unobscured) AGN the central engine is observed at low inclination angles (i.e., closer to face-on than edge-on), so that one can see both the broad and the narrow-line regions (BLR and NLR, respectively). On the other hand, Type\,2 (obscured) AGN are observed at higher inclinations (i.e., edge on), through significant columns of circumnuclear dusty gas, distributed in an anisotropic, axisymmetric structure, possibly concentrated near the plane of the central accretion disk. This so-called dusty torus would obscure the X-ray and optical/UV continuum emission emerging from the central engine, as well as the broad emission lines  emerging from the BLR (see Fig.\,1 of Ref.\,\citen{Ramos-Almeida:2017hw}). 
Over the past decades it has become clear that additional ingredients should be included in the unification model\cite{Netzer15,Ramos-Almeida:2017hw}. In particular, the covering factor of the obscuring material, which could play an important role in the probability of an AGN to be observed as Type\,1 or Type\,2 (Ref.\,\citen{Elitzur12}), shows important trends with AGN  luminosity\cite{Lawrence:1982pk} and/or Eddington ratio\cite{Ricci:2017pr} ($\lambda_{\rm Edd}$).  

Another element that could significantly affect the spectral properties of AGN, and their classification into obscured/Type\,2 or unobscured/Type\,1, is variability. AGN variability has been observed across the entire electromagnetic spectrum\cite{Vanden-Berk:2004bw,Uttley:2005tn}, and on timescales ranging from hours to years\cite{Ulrich:1997ep}. 
While usually this variability is stochastic and does not exceed 10s of per-cent, over the past twenty years an increasing number of accreting SMBHs have been observed as they undergo much more dramatic and coherent changes in their spectral properties, over a range of (exceedingly short) timescales. 
These changes defy the expectations from our basic understanding of AGN structure, and specifically the classical unification model. 
These events, commonly called {\it Changing-look} AGN to denote their varying appearance, have been discovered both in the X-rays and in the optical/UV regimes, and are caused by very different physical processes. In the X-ray band these transformations can be mostly ascribed to changes in the line-of-sight column density ($N_{\rm H}$), which are typically produced by gas located near the SMBH ({\it Changing-obscuration} AGN\cite{Mereghetti:2021aa}, CO-AGN). On the other hand, in the optical/UV, these events are usually associated to rapid changes in the accretion-driven radiation field itself, which in turn lead to the suppression, enhancement, or indeed (dis-)appearance of the blue continuum and broad optical/UV lines typical of Type 1 AGN ({\it Changing-state}\cite{Graham:2020nz} AGN, CS-AGN).
\begin{figure*}
\centering
\textbf{\huge Changing-look AGN -- Basics}\par\bigskip
{\it {\LARGE Spectral classes identified in:}}\par\smallskip
\begin{minipage}{.49\textwidth}
\centering
\textbf{\Large \hspace{0.75cm}X-rays}\par\smallskip
\end{minipage}
\begin{minipage}{.49\textwidth}
\centering
\textbf{\Large \hspace{0.75cm} UV/optical}\par\smallskip
\end{minipage}
\includegraphics[width=0.48\textwidth]{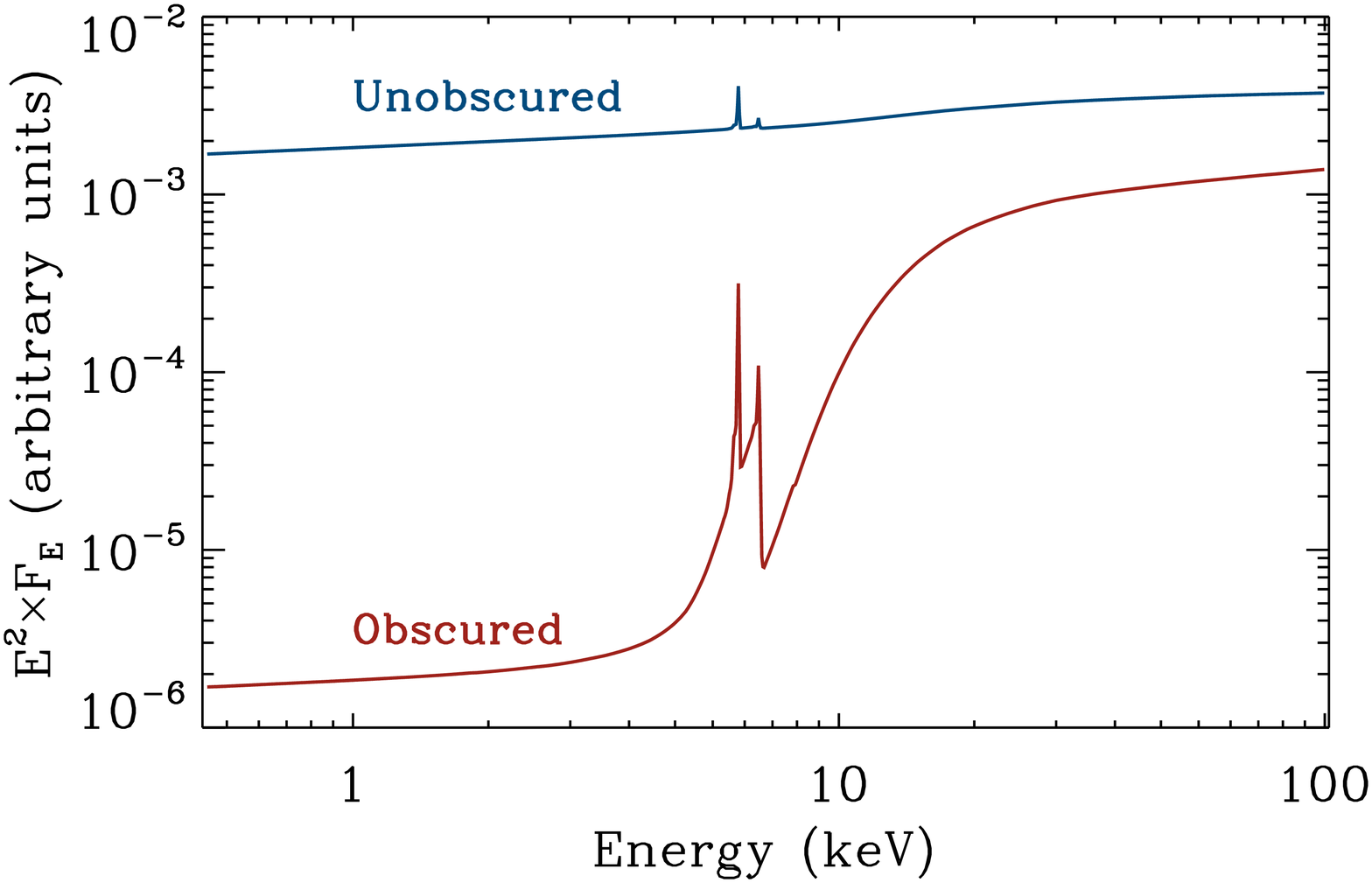}\hfill
\includegraphics[width=0.48\textwidth]{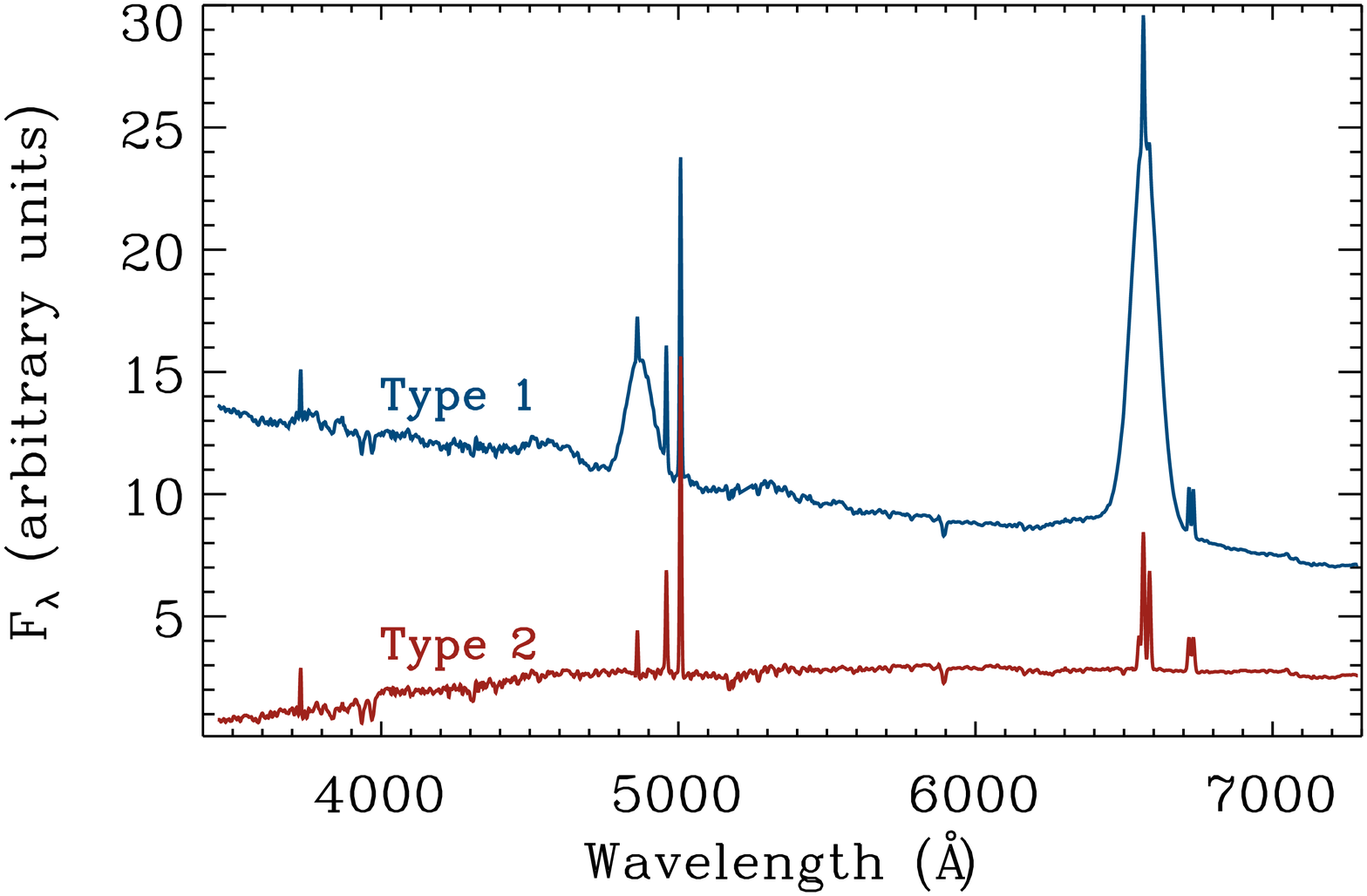}
\put(-187,130){\bf \Large{b)}\hspace{-9cm} \Large{\bf a)}}\\
\smallskip
{\it {\LARGE Spectral transitions usually driven by:}}\par\medskip
\begin{minipage}{.49\textwidth}
\centering
\textbf{\Large \hspace{1cm}Changing obscuration  (CO-AGN)}\par\smallskip
\end{minipage}
\begin{minipage}{.49\textwidth}
\centering
\textbf{\Large \hspace{0.75cm} Changing accretion state (CS-AGN)}\par\smallskip
\end{minipage}
\caption{{\bf Spectral transitions in different types of Changing-look AGN.} a) Changes in the X-ray emission from an unobscured state (blue line) to a heavily obscured, Compton-thick state (red line). b) Spectral transition from a type\,1 (blue line) to a type\,2 (red line) AGN in the optical band. The term CL-AGN can be used to refer to all AGNs that show drastic spectral transitions, while CO-AGN and CS-AGN can be used whenever robust evidence for the corresponding physical mechanism is available.}
\label{fig:schematic_spec}
\end{figure*}
Figure~\ref{fig:schematic_spec} illustrates the basic spectral states of AGN, and thus the potential transitions between them, as viewed in the X-ray and in the optical regimes.
With the advent of new facilities and surveys focused on the transient universe, including the Vera Rubin Observatory\cite{Ivezic:2019go}, the {\it Einstein Probe}\cite{Yuan:2015zd}, {\it ULTRASAT}\cite{Ben-Ami:2022kh}, and many others, a full understanding of these peculiar sources will be one of the main novel efforts within the study of AGN. 
In this Review we summarize our current understanding of changing-look AGN, discussing first changing-obscuration AGN (\S\ref{sect:COagn}) and then changing-state AGN (\S\ref{sect:CSagn}).
We stress that throughout this Review, we use the term CL-AGN to refer to all AGNs that show drastic spectral transitions, regardless of the physical mechanisms driving these changes, and refer to CO-AGN and CS-AGN only whenever robust evidence for the corresponding physical mechanism is in hand.

\begin{figure*}
\centering
\includegraphics[height=0.4\textwidth,angle=-90]{Miniutti.ps}
\includegraphics[width=0.45\textwidth]{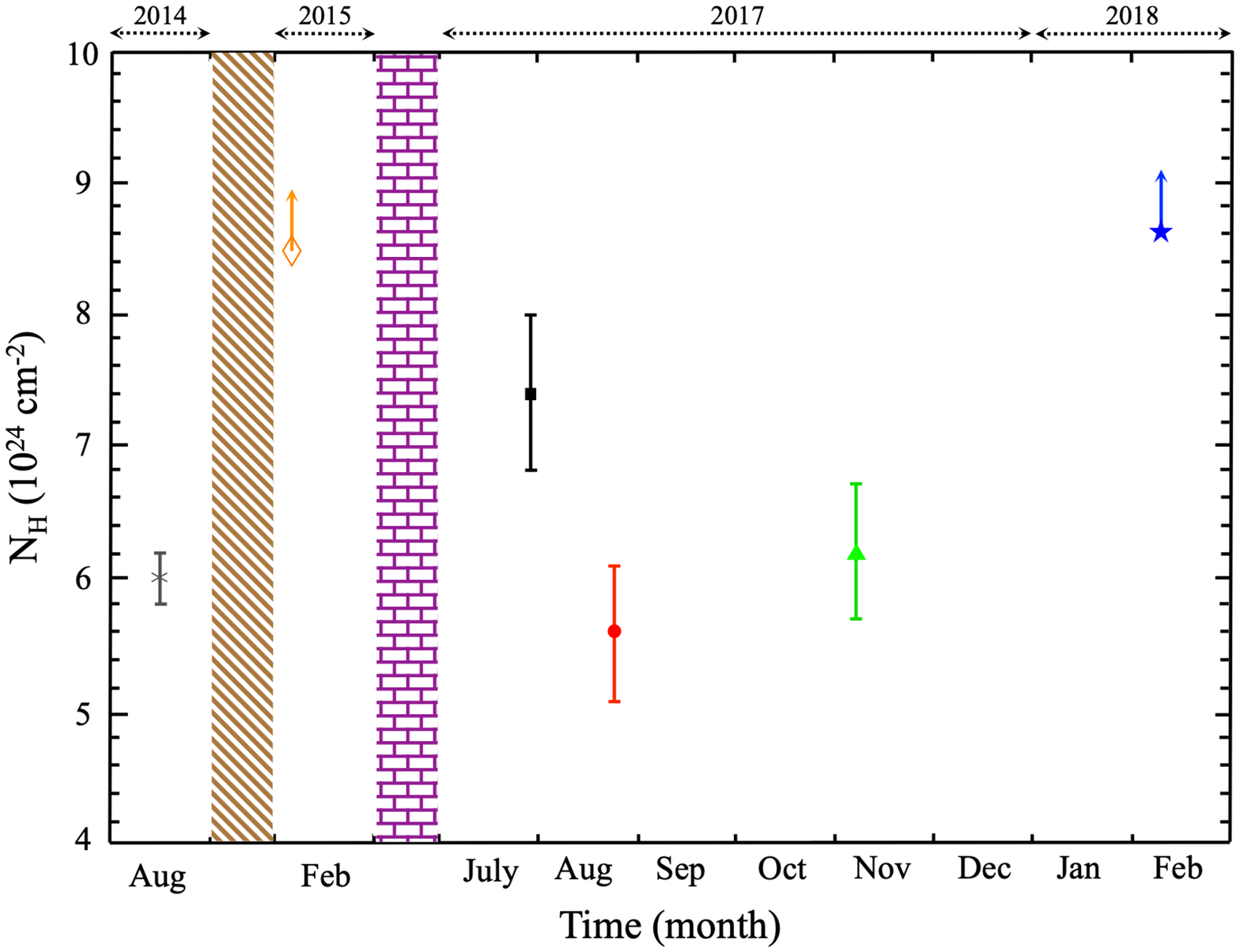}
\put(-180,165){\bf \Large{b)}\hspace{-9cm} \Large{\bf a)}}
\caption{{\bf X-ray spectra and $N_{\rm H}$ variations of changing-obscuration events.} a) Evolution of the X-ray spectrum of the Type\,1 AGN ESO\,$323-$G77\cite{Miniutti:2014yt} during a CO event. The spectra show the progression of the event, going from the least obscured ({\it Chandra}; green) to the most obscured ({\it Suzaku}, blue) phase. As the column density increases the X-ray flux decreases, and the Fe\,K$\alpha$ line at 6.4\,keV becomes more prominent. b) Variations of the column density as a function of time for NGC\,1068, obtained from a dedicated {\it XMM-Newton}/{\it NuSTAR} campaign\cite{Marinucci:2016eu,Zaino:2020cz}, which allowed to probe changes of $N_{\rm H}\gtrsim3\times10^{24}\rm\,cm^{-2}$.}
\label{fig:COAGN1}
\end{figure*}

\subsection{The typical scales of key AGN physical elements}\label{sect:sizes}

We first briefly review the scales of the different AGN regions that are relevant for the changing-look AGN phenomena. Most of the X-ray emission in AGN is produced within $\sim 10\rm r_{\rm g}=10\,GM_{\rm BH}/c^2$ from the SMBH\cite{Risaliti:2007aa,Risaliti:2009mi} by seed (optical/UV) photons, created in the accretion flow, and then Comptonized in a corona of hot electrons\cite{Haardt:1993ln}. The size of the accretion disk has been inferred by micro-lensing\cite{Morgan:2010na} and reverberation\cite{Jha:2022yv} studies to be of order $\sim 10-100\rm\,r_{\rm g}$.
Both the BLR and the dusty obscuring torus have sizes that are known to scale with luminosity. The typical emission-weighted size of the BLR ($R_{\rm BLR}$), and its dependence of the AGN luminosity, can be inferred through intensive BLR reverberation mapping studies. A useful example is the following relation, involving the X-ray luminosity\cite{Kaspi:2005uh}:
\begin{equation}\label{eq:Rblr}
R_{\rm BLR}=7.2\times10^{-3}\left(\frac{L_{2-10}}{10^{43}\rm\,erg\,s^{-1}} \right)^{0.532}\rm\,pc,
\end{equation}
where $L_{2-10}$ is the AGN luminosity in the 2--10\,keV band. 
The scale typically associated with the inner edge of the torus is dictated by the sublimation of dust at sufficiently high temperatures ($T_{\rm sub}$). 
This can be estimated theoretically, for certain dust types\cite{Mor:2009kw}:
\begin{equation}\label{eq:Rblr}
R_{\rm sub}=a\left(\frac{L_{\rm Bol}}{10^{46}\rm\,erg\,s^{-1}} \right)^{0.5}\left(\frac{T^{0}_{\rm sub}}{T_{\rm sub}}\right)^{2.6}\rm\,pc,
\end{equation}
where $L_{\rm Bol}$ is the bolometric luminosity of the central engine, $a$ is 0.5 (1.3) for graphite (silicate) grains, while $T^{0}_{\rm sub}$ corresponds to the temperature at which grains sublimate, i.e., $1800$K and $1500$K for graphite and silicate grains, respectively. 
Observational approaches to quantify the location of the dusty clouds in AGN, for a given luminosity, focus either on the hottest dust emission in the inner torus, observed through $K$-band reverberation mapping\cite{Suganuma06} and direct imaging\cite{Gravity-Collaboration:2020by}, or on the cooler dust observed through 12$\mu$m interferometry\cite{Tristram:2011zx}.
Note that the torus is generally thought to reside just outside of the BLR, and indeed many studies consider the clouds in the latter to be the dust-free extension of the former.

\section{Changing-obscuration AGN (CO-AGN)}\label{sect:COagn}

Obscuration by neutral intervening gas can leave a clear imprint on the X-ray emission observed in AGN: as $N_{\rm H}$ increases the photoelectric cutoff is shifted towards higher energies. At $N_{\rm H}\gtrsim 10^{24}\rm cm^{-2}$ the effect of Compton scattering becomes important, and the primary X-ray continuum becomes increasingly suppressed and more difficult to detect. Most of the AGN in the nearby Universe are obscured (i.e., $\sim 70\%$ with $N_{\rm H}\geq 10^{22}\rm\,cm^{-2}$; Ref.\,\citen{Ricci:2015tg}), and $\sim 20-30\%$ of all AGN are Compton-thick (CT, i.e. with $N_{\rm H}\geq 10^{24}\rm\,cm^{-2}$; Ref.\,\citen{Ricci:2015tg}). As the column density increases, reprocessed X-ray radiation produced by the circumnuclear material (i.e., a Fe K$\alpha$ line and a Compton hump at $\sim 20-30$\,keV), emerges above the suppressed primary X-ray continuum. Some of the most obscured CT AGN, with $N_{\rm H}\gtrsim 10^{24.5}\rm\,cm^{-2}$, can show a reflection-dominated spectrum, in which most of the emission can be attributed to reprocessed X-ray radiation (left panel of Fig.\,\ref{fig:COAGN1}).
Thus, the broad-range X-ray spectra of AGN offer clear and robust measures of line-of-sight obscuration.

Some AGN have been discovered to show temporal variations in the line-of-sight column density\cite{Risaliti:2002hj}. These changing-obscuration AGN (CO-AGN) can, in the most extreme cases, transition from/to being obscured by Compton-thin ($N_{\rm H}\leq 10^{24}\rm\,cm^{-2}$) material to/from CT [$\Delta \log (N_{\rm H}/\rm cm^{-2})\sim 1-2$ dex]. Such transitions are typically observed in the X-rays: given its small size, the X-ray corona is considerably easier to obscure than e.g. the significantly more extended BLR. Variability in $N_{\rm H}$ has provided clear indication that the material around the SMBH is clumpy, and offers unique insights into the its dynamics. 
This dynamic obscuring material could be located either inside the dust sublimation radius of the AGN, and then be associated with the BLR; or further out, beyond the sublimation radius, and then be associated with the putative torus. Interestingly, at least in some cases, these changes in the line-of-sight obscuration could explain\cite{Burtscher:2016ys} the large deviation from the Galactic $N_{\rm H}/A_{\rm V}$ value observed in numerous AGN\cite{Maiolino:2001ex}.

\subsection{Early works}
One of the first known cases of clear column density variability was discovered in the nearby broad-line AGN NGC\,4151. X-ray observations of this source carried out in December 1976\cite{Barr:1977rr} found a $N_{\rm H}$ that is ${\sim} 4$ times higher than that determined from data taken in January of the same year\cite{Ives:1976ty}. Such a spectral change was confirmed by a subsequent study\cite{Yaqoob:1989rk}, which argued that the X-ray absorbing medium could be partially ionized by the UV/X-ray continuum. It was later proposed\cite{Holt:1980kc} that the X-ray absorption in this source is related to clouds in the BLR, with the variability being caused by some of these clouds moving in or out of the line-of-sight. The timescale expected from clouds moving at $10,000\rm\,km\,s^{-1}$ ($\sim 12$\,days) agreed rather well\cite{Malizia:1997vy} with the most stringent constraints obtained for NGC\,4151 ($\sim 1$\,week\cite{Yaqoob:1989rk}). A similar variation was detected in the AGN ESO\,103$-$G35, which was found to show a change in $N_{\rm H}$ of a factor ${\sim} 2$ over a period of 90\,days\cite{Warwick:1988si}. 
The discovery of yet more extreme $N_{\rm H}$ variations came a few years later, from observations of NGC\,1365. This AGN was observed by {\it ASCA} in 1994 and found to be in a reflection-dominated state, with $N_{\rm H}\gtrsim 1.5\times10^{24}\rm\,cm^{-2}$ (Ref.\,\citen{Iyomoto:1997af}). Three years later, {\it BeppoSAX} observations found the source to be in a Compton-thin state\cite{Risaliti:2000rg}, implying a change in column density of $\Delta N_{\rm H} \sim 10^{24}\rm\,cm^{-2}$. Since then this source has been observed repeatedly by all major X-ray telescopes, which confirmed further strong $N_{\rm H}$ variability on a wide range of timescales and column densities\cite{Risaliti:2005cd,Risaliti:2007aa,Risaliti:2009mi}.

In the years following these first discoveries, several other nearby AGN were observed to undergo large-amplitude variations in $N_{\rm H}$. From the analysis of {\it XMM-Newton} and {\it ASCA} observations, the nearby AGN UGC\,4203 was found to transition from a reflection dominated state, to a Compton-thin state in less than six years\cite{Guainazzi:2002mz}. The presence of $N_{\rm H}$ variability was later confirmed by {\it Suzaku} observations\cite{Matt:2009uk}. Another nearby AGN, NGC\,6300 was found to be CT by {\it RXTE}\cite{Leighly:1999td} in 1997, while observations carried out 2.5 years later by {\it BeppoSAX} found the source in a Compton-thin state\cite{Guainazzi:2002oq}. More recent observations confirmed the presence of obscuration variability, finding a variation in $N_{\rm H}$ of a factor 2 between 2007 and 2016\cite{Jana:2020yk}.
We note that CO events were detected on very short timescales, including days and even hours, when higher cadence observations were available (see NGC\,4151 above and \S\ref{sec:NHvar} below).

The first large study of a CO-AGN {\it sample} was carried out in the late 1990s: by studying $\sim 50$\,AGN with multiple X-ray observations, equally split between obscured and unobscured, it was found that $\sim 70\%$ of the objects show variable $N_{\rm H}$ on timescales of months to years\cite{Malizia:1997vy}. A dedicated study of 25 narrow line (Type\,2) AGN\cite{Risaliti:2002hj}, collecting X-ray observations spanning timescales from months to several years, found that 22 sources (i.e. nearly 90\%) showed large amplitude ($\sim20\%-100\%$) variations in their $N_{\rm H}$. For a sub-sample of 11 objects with at least five X-ray observations, the typical variation timescales were found to be shorter than several months, with the obscuring material varying by $\Delta N_{\rm H}\sim 10^{22}-10^{23}\rm cm^{-2}$. These studies confirmed the importance and high occurrence of CO events in nearby AGN.

\subsection{The origin of CO events}

While the most widely accepted explanation for CO events is related to intrinsic column-density variability, due to clouds moving in and out the line-of-sight (\S\ref{sec:NHvar}), several additional explanations have been proposed over the years. One such scenario is that the observed $N_{\rm H}$ changes are driven by changes in the ionization state of the material (\S\ref{sec:ionizstate}), associated with an increase or decrease of the intrinsic AGN luminosity. In some other cases, CO events have been instead attributed to powerful AGN outflows (\S\ref{sec:outflows}). The most extreme CO events, where the AGN goes from/to a transmission-dominated to/from a reflection-dominated state in the X-rays (i.e., $\Delta N_{\rm H}\gtrsim 10^{24}\rm\,cm^{-2}$), could be associated to a switch off/on of the AGN (see \S\ref{sec:switchoff}). 
We briefly discuss each of these possible scenarios.
\subsubsection{Eclipses}\label{sec:NHvar}
Eclipses of the X-ray source are likely one of the most common causes for CO events. This was initially suggested based on the detection of rapid (i.e., a few hours timescale\cite{Elvis:2004jg}) variations of column density in nearby AGN, and then confirmed by the observation of covering and uncovering events. A {\it Chandra} monitoring campaign of NGC\,1365, with six observations carried out over 10\,days, found that the source was first in a Compton-thin state, then showing a reflection-dominated X-ray spectrum two days later\cite{Risaliti:2007aa}. Subsequent observations found that the source was back to its initial, Compton-thin, state. A continuous 5-day {\it XMM-Newton} monitoring campaign of NGC\,1365 recorded the uncovering of the central X-ray source\cite{Risaliti:2009mi}. The source was in a reflection-dominated state for the first 1.5\,days of the campaign, then showing a $\sim 10$\,hours long increase in flux, followed by a rather symmetric decrease. The X-ray spectral analysis showed that the increase in flux was due to a decrease of the column density, with the nuclear source being less obscured for several hours. Another, albeit incomplete, transit was detected in a previous, shorter {\it XMM-Newton} observation\cite{Risaliti:2009yq}. The frequency of these events strongly argues in favour of the clumpy nature of the obscurer in NGC\,1365. Long X-ray observations of other objects (e.g., SWIFT\,J2127.4+5654 by Ref.~\citen{Sanfrutos:2013xt}, Mrk\,766 by Ref.~\citen{Risaliti:2011jl} and NGC\,6814 by Ref.~\citen{Gallo:2021bw}) allowed to detect the transit of clouds with $N_{\rm H}\sim 10^{22}-10^{23}\rm\,cm^{-2}$ on time-scales of a few hours.

The aforementioned X-ray facilities, operating at $E<10$\,keV, cannot observe and identify $N_{\rm H}$ transitions of $\sim 3\times10^{24}-10^{25}\rm\,cm^{-2}$. The advent of {\it NuSTAR}, with its novel focusing capabilities in the hard X-ray band ($>10$\,keV), has opened a new window to study obscured AGN, allowing to detect and characterize the absorption properties of a growing number of AGN, and to discover new CO-AGN\cite{Ricci:2016lq}. A joint {\it XMM-Newton} and {\it NuSTAR} campaign of the reflection-dominated AGN NGC\,1068\cite{Marinucci:2016eu} carried out in 2014 and 2015 found a transient excess in the hard X-ray band. This event was associated to temporary decrease of the obscuring material, from $N_{\rm H}\simeq 10^{25}$ to ${\simeq} 6.7\times10^{24}\rm\,cm^{-2}$. This allowed, for the first time, to observe the primary X-ray continuum from the AGN in this object. Mid-IR observations carried out before and after the X-ray event showed that the intrinsic, accretion-driven radiation from the source did not change during the transient  event\cite{Lopez-Gonzaga:2017kh}, thus confirming that it was due to a change in the line-of-sight column density. A follow-up {\it NuSTAR} campaign identified two additional CO events\cite{Zaino:2020cz}: one unveiling [$\Delta N_{\rm H}\sim 1.8\times10^{24}\rm\,cm^{-2}$] and one eclipsing [$\Delta N_{\rm H}\gtrsim 2.4\times10^{24}\rm\,cm^{-2}$] the X-ray source, thus confirming that the obscuring medium in NGC\,1068 is rather dynamic (right panel of Fig.\,\ref{fig:COAGN1}). 

Eclipses of the X-ray source can be used to put constraints on the location of the clouds producing the variable obscuration (see \S\ref{sec:location}), and to infer the size of the X-ray corona. This can be done assuming that the clouds are moving in a Keplerian orbit around the SMBH, and that the size of the obscuring cloud is larger than (or consistent with) that of the X-ray source. In the case of NGC\,1365, it was found that the size of the X-ray source is $<10^{13}\rm\,cm$, corresponding to a few gravitational radii\cite{Risaliti:2007aa,Risaliti:2009mi}. Similar scales were also derived for Mrk\,766 (Ref.~\citen{Risaliti:2011jl}) and SWIFT\,J2127.4+5654 (Ref.~\citen{Sanfrutos:2013xt}). Interestingly, the X-ray coronae sizes obtained through this approach are consistent with the of micro-lensing studies\cite{Chartas:2009fx}.

\subsubsection{Changes in the ionization state of the obscuring gas}\label{sec:ionizstate}
An increase of the AGN luminosity could lead to an enhancement in the ionization state of the obscuring material, making it more transparent to X-ray radiation, which would in turn lead to an apparent decrease of the line-of-sight column density. This was one of the first explanations proposed for the $N_{\rm H}$ variations observed in NGC\,4151 (Ref.~\citen{Yaqoob:1989rk}). For NGC\,1365, it was argued that a change of two orders of magnitude in intrinsic AGN luminosity was required to explain the $N_{\rm H}$ variations between some of the observations\cite{Risaliti:2005cd}. However, such a strong luminosity variation was not observed. In general, this explanation does not appear to be viable for most CO-AGN events, as they usually do not display large variations of their bolometric luminosity associated with the CO events.

\begin{figure*}
\centering
\includegraphics[height=0.8\textwidth,angle=-90]{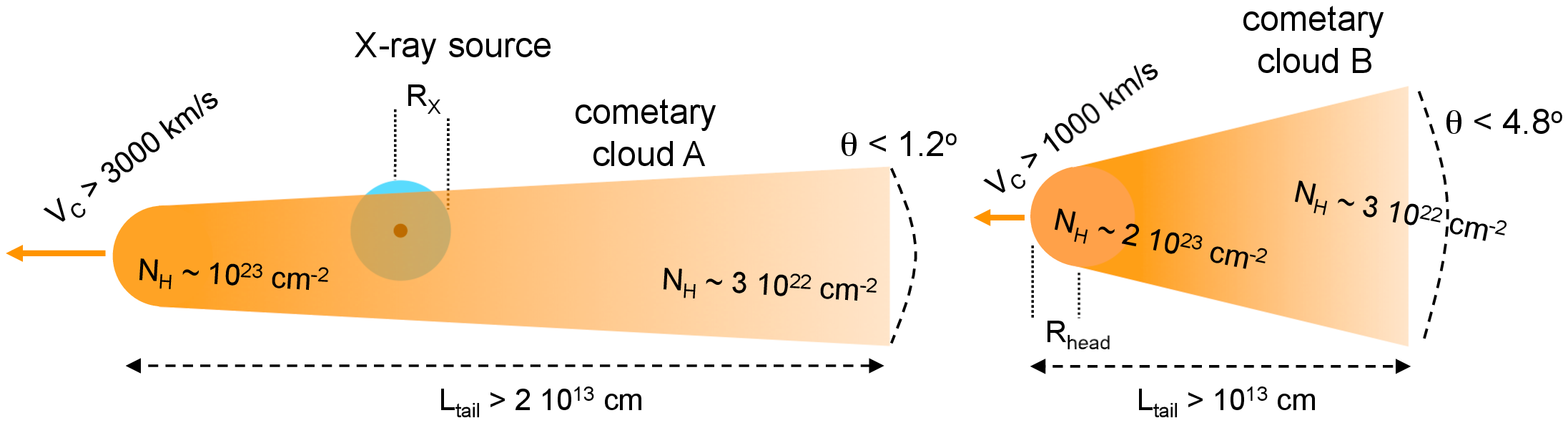}
\caption{{\bf The cometary-shaped absorbers in NGC\,1365.} Geometry and physical properties of the two clouds obscuring the X-ray source in NGC\,1365, as inferred by a long {\it Suzaku} observation\cite{Maiolino:2010fu}. The clouds are thought to have a cometary shape, with a dense core and an elongated, lower-density, tail.}
\label{fig:COAGN2}
\end{figure*}

\subsubsection{Outflows}\label{sec:outflows} 

While most objects discussed so far showed variations of neutral absorption, in some cases, and particularly in unobscured/Type\,1 AGN, it has been found that the X-ray spectrum is strongly affected by variable {\it ionized} absorbing gas along the line of sight. Such absorbers are associated with outflows, which are common in AGN, and have been routinely detected though absorption troughs both in the X-ray and optical/UV regimes\cite{King:2015oc}. An intensive multi-wavelength monitoring campaign of the Type\,1 AGN NGC\,5548 carried out in 2013 showed a clear CO event. A detailed study of the X-ray spectra revealed two obscuring components\cite{Kaastra:2014wa}: one with a low ionization level and a column density of $\sim 10^{22}\rm\,cm^{-2}$ covering $86\%$ of the X-ray source, and another one almost neutral, with a higher column density ($\sim 10^{23}\rm\,cm^{-2}$) and smaller covering factor ($\sim 30\%$). The simultaneous UV spectroscopy allowed to observe broad absorption lines due to the low-ionization component, which showed that the material was outflowing with velocities of up to $\sim 5000\rm\,km\,s^{-1}$. This event, which lasted several years, was ascribed to a wind from the accretion disk extending beyond the BLR. A similar obscuration event was later found in NGC\,3783 (Ref.~\citen{Mehdipour:2017tv}). This event lasted about one month, with material partially obscuring the central source with a column density of $\sim 10^{23}\rm\,cm^{-2}$, and outflowing with a velocity of a few $1000\rm\,km\,s^{-1}$. An analysis of previous X-ray observations of this source identified several additional CO events\cite{Kaastra:2018vu}. Dedicated campaigns have recently found similar CO events in other nearby AGN, such as NGC\,3227 (Ref.~\citen{Beuchert:2015fg}) and Mrk\,817 (Ref.~\citen{Kara:2021bk}). Some of these campaigns have also been able to observe the outflows obscuring the accretion disk\cite{Kara:2021bk}, and might be able in the future to shed light on the interplay between the accretion disk and the outflows. It should be noted that low signal-to-noise ratio and spectral resolution X-ray observations could sometimes associate spectral changes due to outflowing ionized absorbers to eclipses from neutral material.

\subsubsection{Switch-off/switch-on of the central engine}\label{sec:switchoff}
In addition to the scenarios discussed above, at least in some cases the dramatic changes observed in the X-ray spectral appearance of AGN could be directly related to changes in the (intrinsic) X-ray luminosity\cite{Matt:2003tc}. If the accretion power of an AGN decreases rapidly, the primary X-ray continuum (i.e. the power-law component) would also decrease. On the other hand, the reprocessed X-ray radiation, produced by material on scales of $\gtrsim 0.1-1$\,pc, would react more slowly to changes in the irradiating X-ray flux, and could dominate the X-ray emission when the primary flux is at a low level. Therefore, for a certain time after a significant drop in the intrinsic X-ray radiation, the source could be observed as a reflection-dominated AGN, and could be confused for a CT AGN. A re-increase of the accretion-powered radiation would then lead the source to again appear as a continuum-dominated X-ray AGN. This switch-off/switch-on process could (mistakenly) lead to identifying the source as a CO-AGN.

Two cases in which this mechanism could explain the spectral transitions observed in AGN are those of NGC\,2992 (Ref.~\citen{Gilli:2000bw}) and the highly variable AGN NGC\,4051 (Ref.~\citen{Guainazzi:1998yy}). Both sources are switch-off AGN that appeared reflection-dominated in several epochs of observation. Multiple X-ray observations of NGC\,2992, starting from 1978, suggested that the X-ray emission faded over $\sim 15-20$\,years, and then went back to its originally observed flux level in 1998\cite{Gilli:2000bw}. The X-ray spectrum of the Type\,1 AGN NGC\,4051 was found to be reflection-dominated in a 1998 observation\cite{Guainazzi:1998yy}, which was associated to a prolonged low-flux state\cite{Guainazzi:1998yy,Uttley:1999dy}. On the other hand, UGC\,4203  was initially classified as a possible switch-off AGN\cite{Guainazzi:2002mz}, while subsequent studies showed that the appearance of this AGN is likely affected by  line-of-sight eclipses\cite{Matt:2009uk}. In order to clearly differentiate between switch-on/switch-off sources and sources with varying $N_{\rm H}$, it is fundamental to carefully analyze the long-term X-ray light curves and have a simultaneous broad spectral coverage, to best decipher the rich spectral information available in the X-rays.

\subsection{Recent studies}\label{sect:COrecent}

Over the past two decades, repeated observations of nearby AGN carried out by {\it XMM-Newton}, {\it NuSTAR} {\it Chandra}, {\it RXTE}, {\it Swift} and {\it Suzaku}, found a growing number of CO transitions, showing that this phenomenon is rather common in AGN. These variations have been found either by repeated or very long observations. The best studied CO-AGN, NGC\,1365, has been the subject of several dedicated monitoring campaigns. These studies found variability of $N_{\rm H}$ on timescales down to $\sim 10$\,hours, with a variable column density going from $\sim 10^{23}$ to $\sim 10^{24}\rm\,cm^{-2}$. Similarly, rapid variations of $N_{\rm H}\sim 10^{22}-10^{23}\rm\,cm^{-2}$ down to timescales of $\sim 4$\,hours and $\sim 10-30$\,hours were observed in NGC\,4388 (Ref.~\citen{Elvis:2004jg}) and in  NGC\,4151 (Ref.~\citen{Puccetti:2007zr}). While typically these events are detected in AGN showing obscuration signatures in the optical regime (Type\,1.8$-$2 sources), in a few instances they have also been observed in narrow-line Seyfert\,1 (e.g., Mrk\,766, Ref.~\citen{Risaliti:2011jl} and SWIFT\,J2127.4+5654, Ref.~\citen{Sanfrutos:2013xt}) and Type\,1 (e.g., ESO323$-$G77, Ref.~\citen{Miniutti:2014yt}) sources, although their occurrence rate appear to be lower. 
Polarimetry studies\cite{Schmid:2003nk} of ESO323$-$G77 suggested it is observed at an intermediate inclination angle ($\sim 45^{\circ}$) with respect to the normal to the disk. This might imply that the detection of CO events in unobscured AGN could be associated with lines of sight grazing the edge of the torus.

Several other studies have focused on searching for CO events in large samples of AGN, to better understand their occurrence rates and links to other AGN and SMBH properties. A careful analysis of {\it RXTE} observations of $\sim 40$ nearby Type\,1 and Compton-thin Type\,2 AGN uncovered 12 eclipse events in eight of the sources\cite{Markowitz:2014oq}, with typical column densities of $\sim 10^{22}-10^{23}\rm\,cm^{-2}$. Analyzing hardness ratio light-curves of 40 nearby AGN with long {\it XMM-Newton} and {\it Suzaku} observations, it was found that spectral variability that could be ascribed to eclipses by BLR-like clouds was detected in all the cases where the length of the observations was sufficiently long\cite{Torricelli-Ciamponi:2014do}. More recently, $N_{\rm H}$ variability was detected in seven out of 20 objects observed by various X-ray satellites\cite{Laha:2020sd}, with variations in the range $\sim 10^{21}-10^{23}\rm\,cm^{-2}$.
The CO-AGN phenomena, as probed by X-ray observations, is thus rather common.

\subsection{The location of the dynamic absorber}\label{sec:location} 
The distance of an eclipsing cloud from the central X-ray source ($R_{\rm C}$) can be estimated by considering that the clouds are moving in Keplerian orbits around the SMBH, and by making some further simple assumptions\cite{Risaliti:2007aa,Marinucci:2013iz}, specifically that the size of the cloud ($D_{\rm c}$) is similar to that of the X-ray source ($D_{\rm s}$), and that the transverse velocity ($v_{\rm k}$) is given by the ratio between the size of the X-ray source and that of the crossing time ($v_{\rm k}=D_{\rm C}/\Delta t_{\rm cr}$). The X-ray source is known to be compact (\S\ref{sect:sizes}) and, setting the radius of the cloud to be $D_{\rm C}\simeq D_{\rm s} = 10\,r_{\rm g}$, one would then obtain:
\begin{equation}\label{eq:COagnR}
R_{\rm C}=\frac{GM_{\rm BH}}{v^2_{\rm k}}=\frac{GM_{\rm BH}\Delta t^{2}_{\rm cr}}{D^2_{\rm s}}=6.3\,M_{8}\,R_{10}^{-2}\,\Delta t_{10}^2\rm\,pc \, ,
\end{equation}
where $\Delta t_{10}$ is the crossing time in units of 10 days, $M_{8}$ is the black hole mass in units of $10^{8}\,M_{\odot}$, and $R_{10}= D_{\rm s}/10\,r_{\rm g}$. 

If the obscuring material is ionized, its distance from the SMBH can be simply inferred, under the assumption that the cloud is in photoionization equilibrium with the radiation field, by the definition of the ionization parameter $\xi$:
\begin{equation}\label{eq:distanceoutflow}
R_{\rm C}=\left(\frac{L_{\rm ion}}{\xi n_{\rm H}}\right)^{0.5},
\end{equation}
where $L_{\rm ion}$ is the ionizing radiation, commonly integrated over 1--1000\,Ryd, and $n_{\rm H}$ is the density of the obscurer. Since constraining $n_{\rm H}$ is difficult, Equation\,\ref{eq:distanceoutflow} can be rewritten, considering that the depth of the absorber $l_{\rm C}$ is smaller than its distance from the central source ($l_{\rm C}  \simeq N_{\rm H}/n_{\rm H}\lesssim R_{\rm C}$):
\begin{equation}\label{eq:distanceoutflow2}
R_{\rm C}\lesssim \frac{L_{\rm ion}}{\xi N_{\rm H}}.
\end{equation}

While for many objects only very loose constraints on the timescales of the CO event (i.e., $\lesssim$months-years) could be found, for a few sources very stringent constraints were obtained.
NGC\,1365 showed CO transitions on time-scales of weeks\cite{Risaliti:2005cd}, days\cite{Risaliti:2007aa}, and down to $\sim 10$\,hours\cite{Risaliti:2009mi}. The distance of the varying CT absorber in this object was constrained to be $3,000-10,000\rm\,r_{\rm g}$, i.e. consistent with what is expected for the BLR. The CO-AGN NGC\,7582 (Ref.~\citen{Piconcelli:2007bh}) showed column density variations on timescales shorter than a day\cite{Bianchi:2009au} ($\lesssim 20$\,hours), and associated with clouds in the BLR, or just outside it. Similarly, the varying Compton-thin and CT absorbers were associated to the BLR in several other objects, such as NGC\,4388 (Ref.~\citen{Elvis:2004jg}), SWIFT\,J2127.4+5654 (Ref.~\citen{Sanfrutos:2013xt}),  Mrk\,766 (Ref.~\citen{Risaliti:2011jl}) and NGC\,4151 (Ref.~\citen{Puccetti:2007zr}).
In the case of ESO323$-$G77, two components were distinguished\cite{Miniutti:2014yt}. The less variable one, with $N_{\rm H}\sim 2-6\times10^{22}\rm\,cm^{-2}$, a density of $\leq 1.7\times 10^{8}\rm\,cm^{-2}$, varying on timescales longer than a few months and shorter than 3.6\,years, was associated to clouds in the clumpy torus. The larger amplitude variations ($N_{\rm H}\sim 0.3-3\times10^{24}\rm\,cm^{-2}$) detected on shorter timescales were instead linked to clouds in the BLR. Assuming a size of the obscuring cloud comparable to that of the X-ray source, for NGC\,3783 a distance of $\sim 10$\,light days from the SMBH was inferred\cite{Mehdipour:2017tv} for the outflowing gas, which is also consistent with that of the BLR.

A large study of dozens of nearby AGN found that the transient obscuring clouds are located at $R_{\rm C}\sim 0.3-140\times 10^{4}\rm\,r_{\rm g}$ from the SMBH, which is comparable to the dust sublimation radius\cite{Markowitz:2014oq}, with the mean of the individual values indeed indicating $R_{\rm C}\simeq R_{\rm sub}$. The object in which clouds are located farthest away from the sublimation radius is Cen\,A, with $R_{\rm C}/R_{\rm sub}\sim 2-7$. For the three objects in the sample for which IR interferometry or optical-to-near-IR reverberation mapping was performed, $R_{\rm C}$ was found to be consistent with the radial distance of the IR-emitting material. It should be noted however that this study might preferentially detect eclipses lasting tens of days\cite{Markowitz:2014oq} and, similarly, long continuous observations might be biased towards detecting CO events on timescales of hours to days. Interestingly, these longer-timescales eclipses have typically lower column densities ($\simeq 10^{22}-10^{23}\rm\,cm^{-2}$) than those observed on shorter timescales, which could imply that a significant fraction of the obscuration is caused by material within the sublimation radius.

\subsection{The structure of the BLR from X-ray eclipses}\label{sec:BLRcoagn}

Studies of some of the most extreme CO events have shown that the density of the obscuring clouds tends to be high. For example, in the case of NGC\,1365 (Ref.~\citen{Risaliti:2009mi}) and ESO\,323$-$G77 (Ref.~\citen{Miniutti:2014yt}) the clouds were found to have a density of $\sim 10^{11}\rm\,cm^{-3}$ and  $0.1-8\times10^{9}\rm\,cm^{-3}$, respectively. Similarly high densities were found for objects in which only a lower limit could be obtained, i.e. $>3\times10^{8}\rm\,cm^{-3}$ and $\geq 1.5\times10^{9}\rm\,cm^{-3}$  for NGC\,7582 (Ref.~\citen{Bianchi:2009au}) and SWIFT\,J2127.4+5654 (Ref.~\citen{Sanfrutos:2013xt}), respectively. These values are consistent with the typical density of BLR gas ($\gtrsim 10^{9.5}\rm\,cm^{-3}$, Ref:\citen{Netzer:2013ix}), and higher than what is expected for the dusty torus. Combined with the aforementioned BLR-like distances of the transient obscurers, this strongly supports the idea that these CO-causing clouds are associated with those producing broad optical/UV lines.

Since a significant fraction of CO events could be caused by BLR clouds, X-ray spectroscopy can provide an important tool to shed light on this region. A crucial step forward in our understanding of the relation between BLR clouds and CO events came from a 300\,ks {\it Suzaku} observation of NGC\,1365\cite{Maiolino:2010fu}, which found that the rapidly-moving clouds in that system appear to have a cometary shape, with a high-density ($\sim 10^{11}\rm\,cm^{-3}$) core and an elongated ($\sim 10^{13}\rm\,cm$), lower density, structure (Fig.\,\ref{fig:COAGN2}). The opening angle of the tail was estimated to be small, i.e., only a few degrees. 
Similarly, in Mrk\,766, besides dense ($\sim 10^{10}-10^{11}\rm\,cm^{-3}$) neutral clouds, a tail of lower density, highly ionized gas, causing blueshifted iron lines (Fe XXV and Fe XXVI with $\Delta v\simeq 3,000-15,000 \rm\,km\,s^{-1}$), was also detected\cite{Risaliti:2011jl}. 
As the mass loss rate of the cloud core material into the tail would imply their destruction within a few months, it was speculated that the clouds should be continuously replenished, possibly through outflows from the accretion disk\cite{Maiolino:2010fu}. This would be in agreement with some recent models of the formation of BLR clouds\cite{Czerny:2011ek}.

\bigskip


\section{Changing-state AGN (CS-AGN)}\label{sect:CSagn}

\begin{figure*}
\centering
\includegraphics[scale=.237]{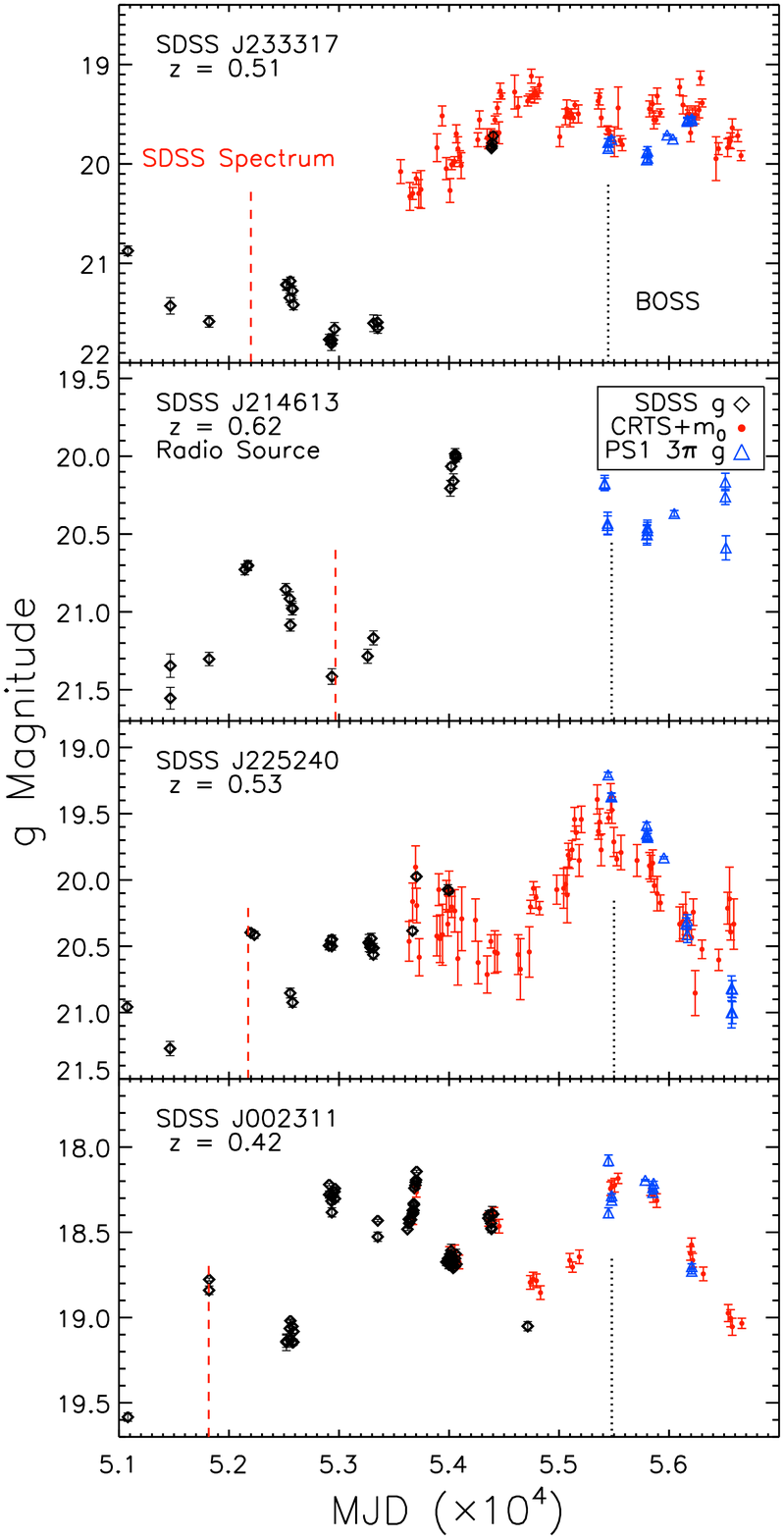}
\includegraphics[scale=.237]{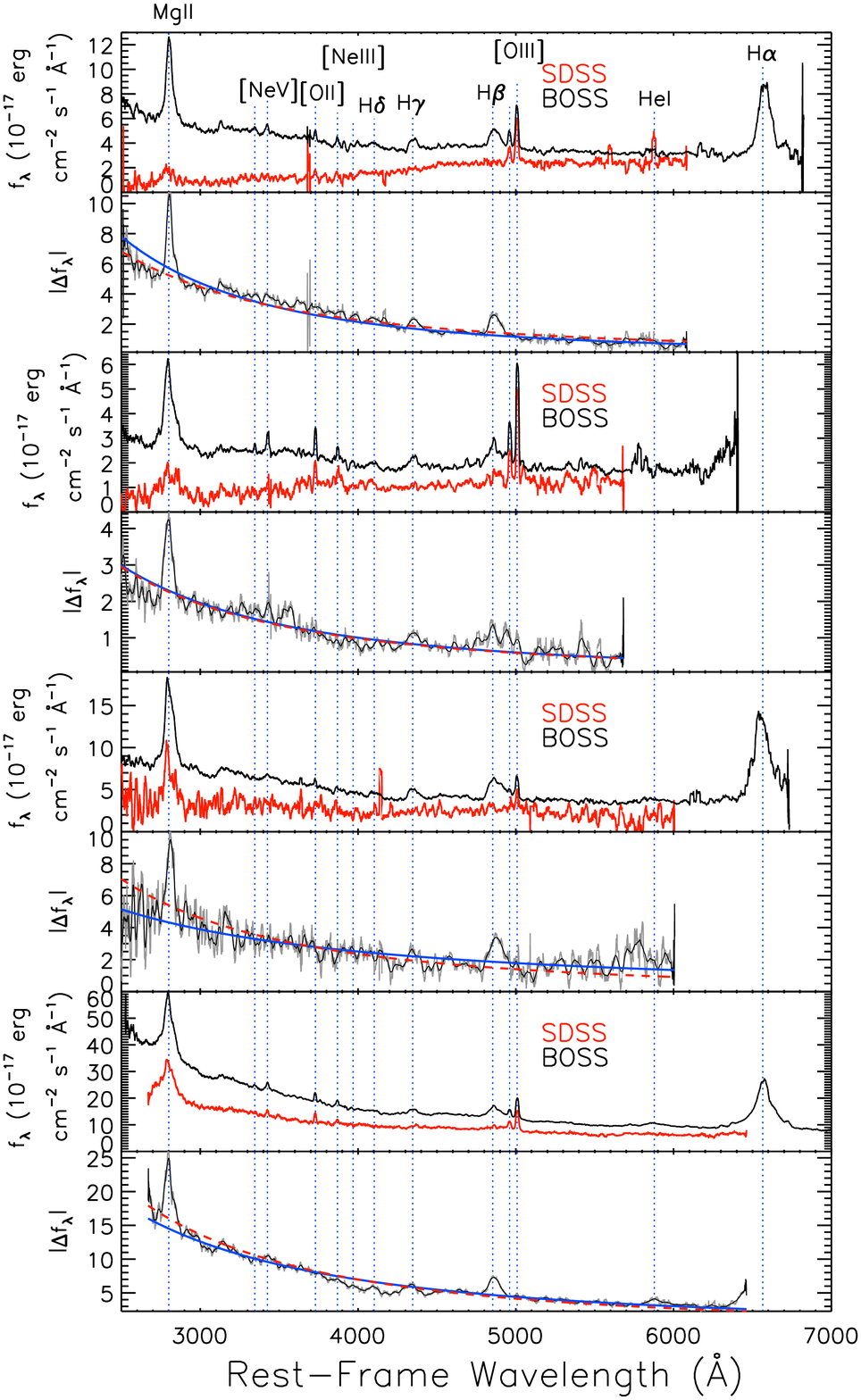}
\includegraphics[scale=.3]{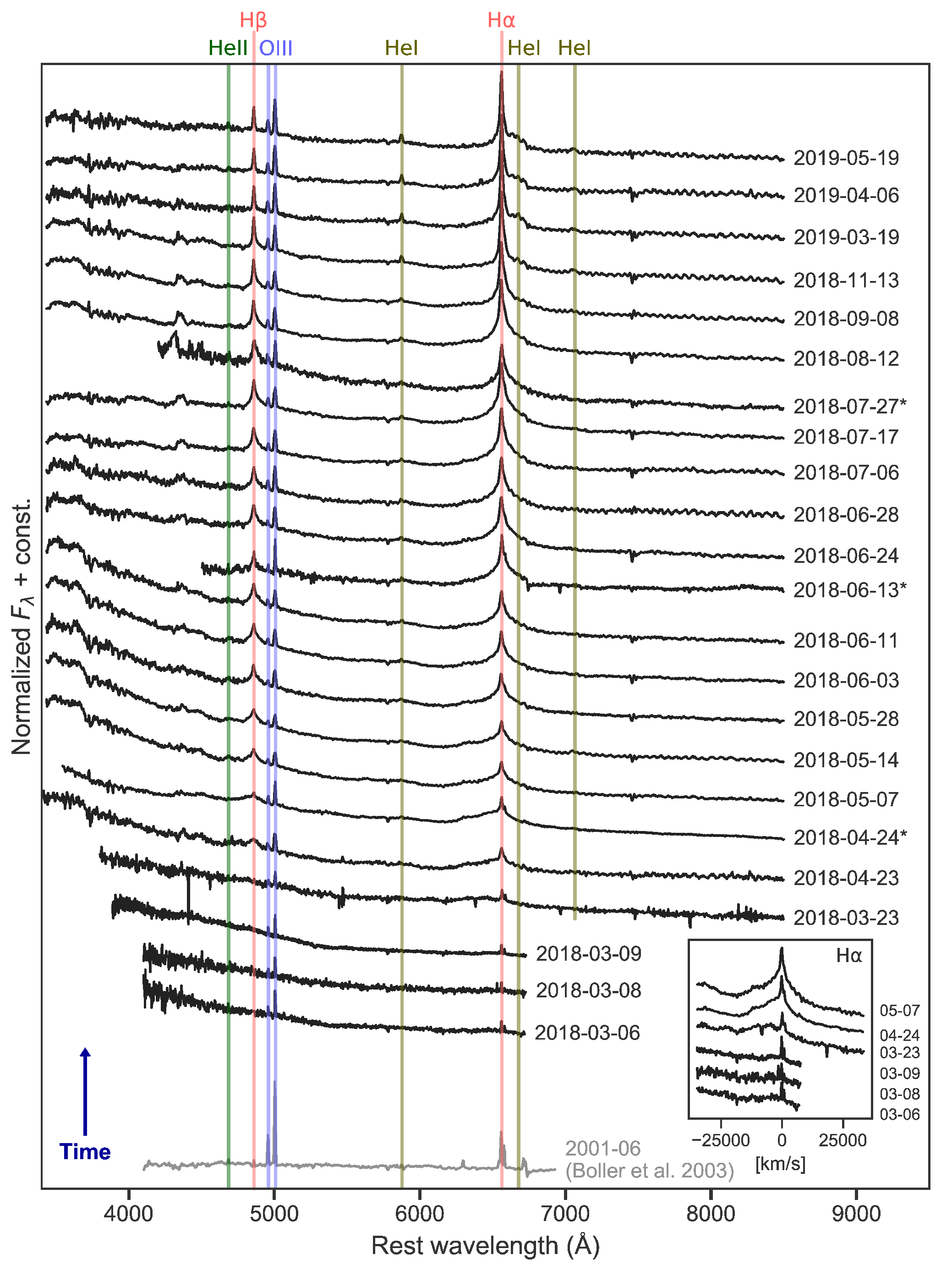}
\put(-170,235){\bf \Large{c)}\hspace{-5.3cm} \Large{\bf b)}\hspace{-5cm} \Large{\bf a)}}
\caption{{\bf Light curves and spectra of CS-AGN showing an emergent BLR.} a) SDSS (black diamonds), Pan-STARRS  (blue triangles), and CRTS (red dots) light curves in the g band\cite{MacLeod:2016pu}. b) 
SDSS (red) and BOSS (black) spectra, corresponding to the epochs indicated by the vertical dotted lines in the left panel\cite{MacLeod:2016pu}. The lower sub-panels show the flux difference ($|\Delta f_{\lambda}|$) between the two spectra, together with the best-fit power-law $f_{\nu} \propto \nu^{\beta}$, with the red-dashed curve showing a power-law with $\beta = 1/3$, as expected for a standard thin disk\cite{Shakura:1973ex}.
c) Transition from a Type-2 AGN (bottom) to a Type-1 AGN in 1ES\,1927+654, with an intermediate stage of a blue-continuum-dominated emission\cite{Trakhtenbrot:2019ay}.}
\label{fig:CSAGN1}
\end{figure*}

The accretion phase onto SMBH is expected to last $\sim10^{5}-10^{9}$\,years\cite{Shen:2007dh,Schawinski:2015cs}, and during this period AGN are well known to show variability, with the typical amplitude of optical continuum variations being tens of percents\cite{Vanden-Berk:2004bw}. A growing population of AGN has been recently found to show much stronger flux variability, over relatively short timescales, in both the optical\cite{Lawrence:2016bi,Rumbaugh:2018iv,Trakhtenbrot:2019wj,Shen:2021se} and X-ray\cite{Timlin:2020bv} regimes. In some cases, these strong flux variations have been associated to CS transitions, with the (dis-)appearance of broad optical/UV emission lines being closely linked to the (dis-)appearance of the accretion-powered continuum emission (Fig.\,\ref{fig:CSAGN1}).

\medskip

\subsection{Early works}\label{sect:earlyworks}

One of the first observed CS transition was found in 1975 from repeated observations of NGC\,7603, which showed that the broad H$\beta$ line weakened by a factor $\sim 3$ within one year\cite{Tohline:1976gd}. The interacting galaxy NGC\,2992, one of the first candidate CO-AGN (see also \S\ref{sec:switchoff}), was also found to undergo a CS event. Early optical spectra published in the 1980s\cite{Ward:1980ed} showed the presence of a weak broad H$\alpha$ component but no detectable broad H$\beta$, marking the source a Type\,1.9 AGN. In 1994 the source appeared to also lose its broad H$\alpha$ line\cite{Allen:1999yl}, transitioning to a Type\,2 class. The broad H$\alpha$ line then reappeared 5 years later\cite{Gilli:2000bw}. Interestingly, the variations were showed to be associated with the X-ray flux: in the 1980s the source was in a high X-ray flux state\cite{Mushotzky:1982nt}, which declined over time, reaching its minimum in 1994, before going back to the originally observed flux level in 1999\cite{Gilli:2000bw}. Other systems found to show disappearing broad line emission include 3C\,390.3\cite{Penston:1984tl}, Mrk\,372\cite{Gregory:1991dq} and Mrk\,993\cite{Tran:1992hb}. In several other nearby sources (e.g., NGC\,7582\cite{Aretxaga:1999ro} and NGC\,3065 \cite{Eracleous:2001pu}), the opposite process was observed, with the broad line emission {\it appearing}.

A small fraction of CS-AGN have been found to undergo a full cycle, transitioning twice. Mrk\,1018 went from Type\,1.9 to Type\,1 over a time span of $<5$\,years\cite{Cohen:1986sa}, and then transitioned back to Type 1.9 after 30 years\cite{McElroy:2016hs}. NGC\,1566, which was recently found to have transitioned from Type\,1.9-1.8 to Type\,1.2\cite{Oknyansky:2019le}, shows recurrent activity\cite{Alloin:1986iu}: between 1970 and 1985, four separate periods of activity of $\sim 1300$\,days each were identified\cite{Alloin:1986iu}, with the flux of both the continuum and the broad lines increasing and then decreasing, leading to CS transitions. 
NGC\,4151, which was originally found to be a Type\,1.5 AGN\cite{Osterbrock:1977pd}, first lost its broad lines (with the exception of weak and possibly asymmetric wings) during the 1980s\cite{Penston:1984tl}, and then regained them\cite{Shapovalova:2010an}. Similar transitions were also observed for Mrk\,590\cite{Denney:2014do} and NGC\,3516\cite{Ilic:2020jj}.
Importantly, optical \cite{LaMassa:2015rt} and X-ray\cite{Ricci:2020fp} continuum emission studies showed that these transitions are usually not associated with CO events (see \S\ref{sect:accretionVSobs}).

\medskip

\subsection{The advent of large surveys and statistical studies}\label{sect:largesurveys}

Over the past few years, with the advent of large imaging and spectroscopic surveys, there has been a sharp increase in the number of CS-AGN identifications, greatly expanding the scope of early studies in terms of redshift and key SMBH properties. 
One prominent example is the first changing-look quasar\cite{LaMassa:2015rt} (at $z=0.31$), SDSS\,J0159+0033, which was observed in two different programs within the Sloan Digital Sky Survey (SDSS\cite{York:2000kn}). 
The original SDSS spectrum, taken in 2000 (and reported in the first SDSS data release\cite{Abazajian:2003iw}) showed broad emission lines, which however disappeared by 2010, when another spectrum was taken as part of SDSS-III. This transition was confirmed by further spectroscopy obtained during 2014 (Ref.~\citen{LaMassa:2015rt}). 
Several additional studies pursued a similar approach, systematically searching the (repeated) spectroscopy of SDSS programs and/or other spectroscopic surveys, and revealing a growing number of CS-AGN\cite{Runnoe:2016hb,MacLeod:2016pu,Yang:2018mn,MacLeod:2019rk,Green:2022cv}.

In most such studies, the identification of CS events is based on drastic changes in the H$\alpha$ and H$\beta$ lines in spectra that are taken several years apart. 
At higher redshifts ($z\gtrsim1.5$) CS-AGN were also identified based on the Mg\,\textsc{ii}\,$\lambda2798$,  C\,\textsc{iv}\,$\lambda1549$, and other (rest-frame) UV broad emission lines\cite{Guo2020_hiz_CLQs,Ross:2020gn}.
These emission lines probe somewhat different regimes of the variable ionizing AGN continuum and/or the BLR (see e.g., Table~1 in Ref.~\citen{Ross:2020gn}), and may thus provide invaluable, complementary insights regarding the nature of extreme variability, and indeed CS transitions, in AGN\cite{Shen:2021se}.
For example, several works studied the somewhat peculiar variability properties of the broad Mg\,\textsc{ii}\,$\lambda2798$ emission line\cite{Sun2015}, including its persistence in optical spectra that show weak (or no) other AGN indicators\cite{Roig2014}. 
Such systems can be interpreted as recently turned-off AGN\cite{Guo:2020ll}, where the Mg\,\textsc{ii} line is not as responsive to the CS event as the Balmer lines\cite{KoristaGoad2004,Yang2020_MgII}, in what seems like an extreme version of some reverberation mapping results\cite{Clavel1991,Cackett2015}. 
Another examples of this behaviour is 1ES\,1927+654 (Ref.~\citen{Trakhtenbrot:2019ay}), for which the appearance of a blue continuum was followed by the appearance of broad Balmer emission lines, while neither Mg\,\textsc{ii} nor C\,\textsc{iv} were detected.

The samples obtained with such spectroscopically-focused searches for CS-AGN can already provide some insights regarding their occurrence rates and the typical key  properties for AGN that experience CS transitions.
Re-observing 102 local ($0.02 \leq z \leq 0.1$) broad-line AGN with previous SDSS spectroscopy, it was found\cite{Runco:2016po} that $\sim 38\%$ of the sources had changed spectral (sub-)classification, and that in $\sim3\%$ of the sources the broad H$\beta$ disappeared completely, within $\sim 3-9$\,years. 
A spectroscopic effort to re-observe SDSS quasars that exhibited extreme variability in multi-year optical photometry [$\Delta(g) > 1$ mag] identified 17 robust CS-AGN among 130 carefully-selected candidates\cite{MacLeod:2019rk}.
Another search, within SDSS-IV repeated spectroscopy (and no photometric pre-selection)\cite{Green:2022cv} resulted in a total of 19 (15 newly identified) CS-AGN, out of a parent sample of $\sim$64,000 broad-line quasars. This translates to a significantly lower occurrence rate of $\sim$0.03\%.
The discrepancy in occurrence rate may be attributed not only to the different search methods (see also below), but also to the well-known anti-correlation between AGN luminosity and variability amplitude\cite{Vanden-Berk:2004bw,Caplar2017}.
Some of these relatively large, spectroscopy-led samples of CS-AGN, for which some form of comparison samples can be constructed, suggest that CS-AGN have preferentially relatively low Eddington ratios\cite{MacLeod:2019rk,Green:2022cv}. 
This, however, should be treated with caution, as such studies may be subject to various observational biases (see below).

Since CS events are typically associated with drastic changes in both the optical and X-ray fluxes (\S\ref{sect:X-ray}), time-domain imaging surveys can be a powerful tool to identify candidate highly variable AGN which can then be confirmed as CS-AGN through spectral observations. 
In particular, several wide-field optical imaging surveys, including
the Catalina Real-time Transient Survey (CRTS\cite{Drake:2009ts}), 
the Palomar Transient Factory (PTF\cite{Law:2009it}), 
Panoramic Survey Telescope and Rapid Response System (Pan-STARRS\cite{Kaiser:2010wj}),
the All-Sky Automated Survey for SuperNovae (ASAS-SN\cite{Shappee:2014yp}), 
the Zwicky Transient Facility (ZTF\cite{Bellm:2019il}), have greatly increased the number of known CS-AGN.
One of the first CS-AGN events to be identified by these facilities was NGC\,2617, for which ASAS-SN identified a significant flux increase in 2013. Spectroscopic follow up showed that the source, which was originally classified as a Type\,1.8 AGN in 2003, had transitioned to Type\,1\cite{Shappee:2014yp}. Some of these surveys have discovered extremely peculiar objects, such as 1ES\,1927+654\cite{Trakhtenbrot:2019ay}, the first source in which the CS transition was clearly temporarily resolved. This source, a known nearby AGN which historically showed no broad emission lines and no X-ray obscuration\cite{Gallo:2013vw}, underwent an outburst at the end of 2017, which was detected by ASAS-SN. Intensive spectroscopic follow up showed the source first developing a blue, quasar-like continuum, then showing the appearance of broad Balmer lines\cite{Trakhtenbrot:2019ay} (right panel of Fig.\,\ref{fig:CSAGN1}). Photometric surveys have also found cases in which the AGN rapidly turned on, transitioning from LINER to broad-lined AGN\cite{Gezari:2017md,Frederick:2019vf}.  

The study of infrared variability has proven to be an effective approach to detect rapidly varying AGN, particularly thanks to the repeated all-sky imaging of the {\it WISE} mission. 
Among the $>$4 million AGN identified with {\it WISE}\cite{Assef:2018vj}, about 700 (0.015\%) showed strong variability, and some of those were indeed associated with CS transitions\cite{Assef:2018vj,Stern:2018su,Ross:2018ah}. 
In one exemplary {\it WISE}-identified source, the broad H$\alpha$ emission has disappeared within $\lesssim 10$\,years\cite{Assef:2018vj}. 
By using the various {\it WISE}-based datasets, another CS-AGN, in which the broad H$\beta$ disappeared on a timescale of $\lesssim 11$\,years, was identified\cite{Stern:2018su}. The flux of this source has faded by a factor of $\gtrsim 2$ (at $3.4$ and $4.6\mu$m) between 2010 and 2014. Follow-up  observations showed that the X-ray flux of the source had decreased by a factor of $\approx$10 since 1995.
Recently, by combining CRTS and {\it WISE} photometry with spectroscopic observations, 73 additional CS-AGN were identified\cite{Graham:2020nz}, 36 (37) of which with fading (increasing) H$\beta$, greatly increasing the number of known CS-AGN. These systems were found to be typically more luminous and at a slightly higher redshift than previously known CS-AGN.

\medskip

The growing samples of CS-AGN, from all observational approaches, would allow to probe population properties and thus be very insightful for understanding the mechanisms behind such extreme events.
Perhaps the most basic property to consider is the occurrence rate of CS events in AGN, with some preliminary findings mentioned above.
All efforts to understand the (statistical) population properties of CS events are naturally limited by the diversity of identification methods, as well as the associated observational challenges related to (irregular) cadence and/or availability of follow-up data.
Moreover, there are some biases that are intrinsic to the search methodology of choice. 
For example, (spectroscopic) monitoring of previously-known luminous AGN (quasars) is expected to be biased towards turn-off events (and against turn-on events), given the historically high accretion rates.
In contrast, some imaging time-domain surveys are geared to alert on rapid (and/or extreme) brightening sources, thus potentially providing a biased view regarding turn-on events, and neglecting turn-off events.
These important limitations require further analysis\cite{Shen2021_bias} if observations of samples are to be used to constrain models, and specifically those that have distinct predictions in terms of occurrence rates, turn-on or turn-off behavior, and/or relevant SMBH properties (as discussed in \S\ref{sect:CSAGN_origin}).

\bigskip

\begin{figure*}[t!]
\centering
\includegraphics[width=0.96\textwidth]{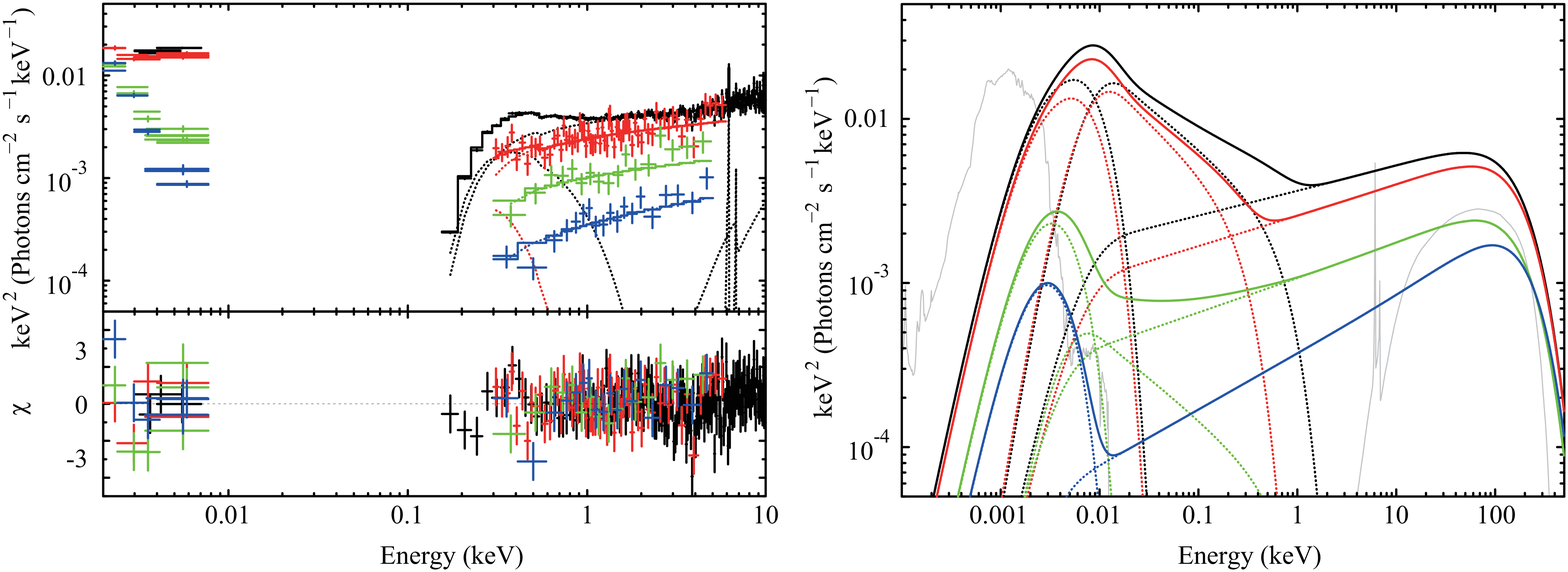}
\put(-185,153){\bf \Large{b)}\hspace{-8.6cm} \Large{\bf a)}}
\par\bigskip
\includegraphics[width=0.48\textwidth]{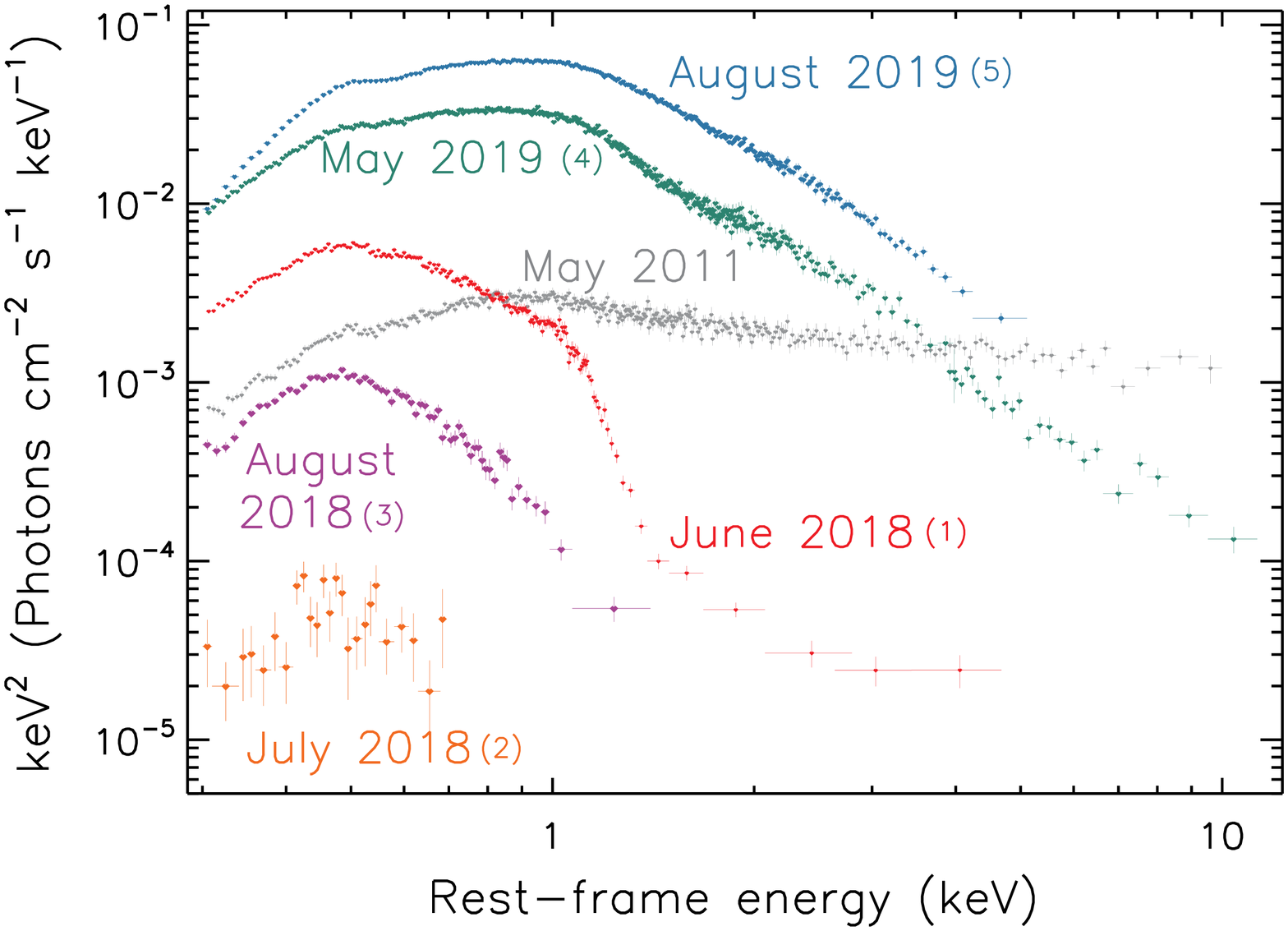}
\includegraphics[width=0.48\textwidth]{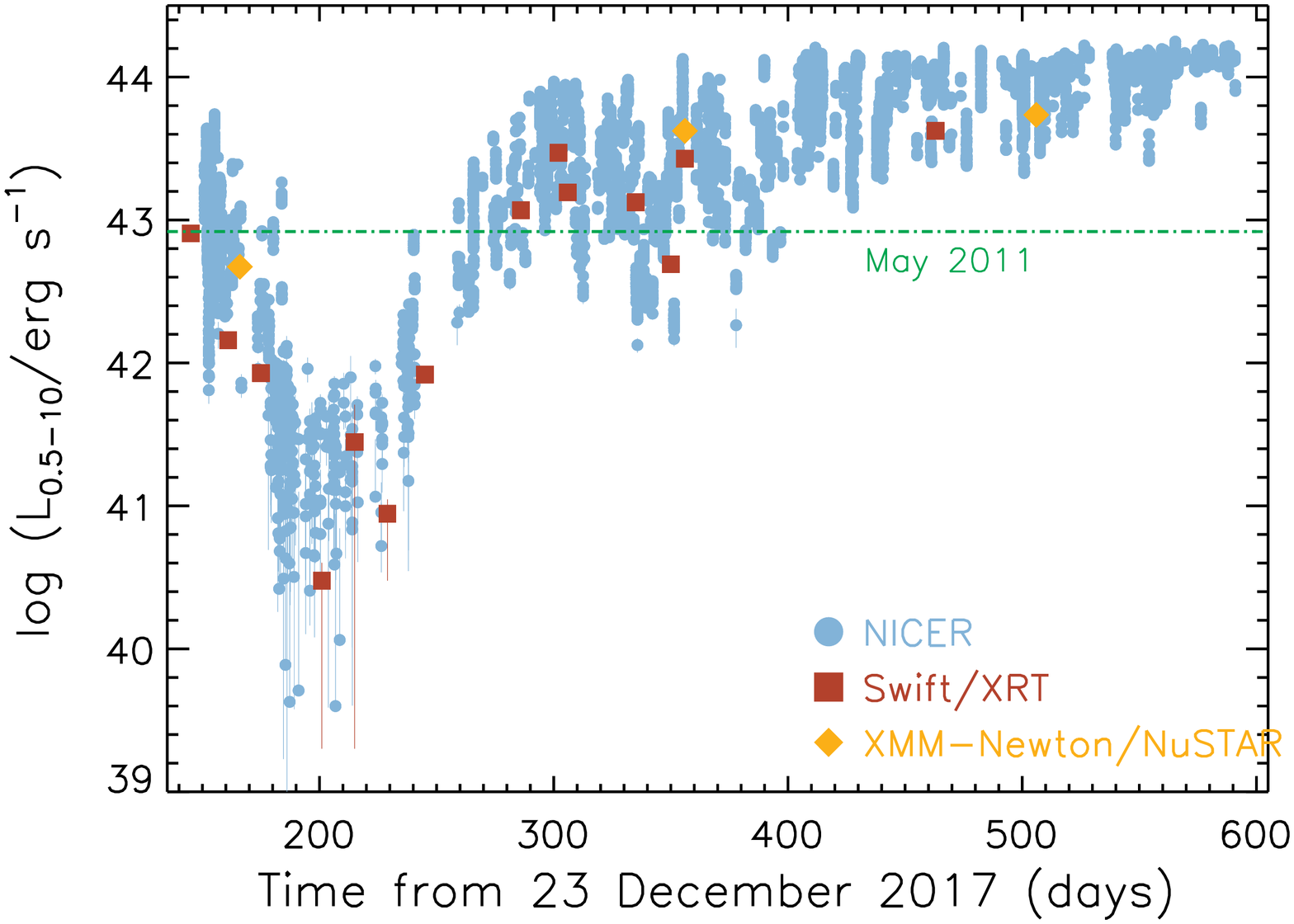}
\put(-185,146){\bf \Large{d)}\hspace{-8.6cm} \Large{\bf c)}}
\caption{{\bf X-ray spectra and light curve of CS-AGN} a) Optical, UV, and X-ray spectra of Mrk\,1018 in 2008 August (black), 2008 June (red), 2013 (green), and 2016 (blue)\cite{Noda:2018jz}. b) Best-fit model of the different optical/UV and X-ray observations, showing an increasing soft X-ray component for increasing luminosities\cite{Noda:2018jz}. The grey points show the spectrum of an S0-type host galaxy.
c) Spectral evolution of 1ES\,1927+654\cite{Ricci:2020fp,Ricci:2021ro} showing the disappearance and reappearance of the power-law component after the CS event (2018/2019), compared with pre-outburst observations\cite{Gallo:2013vw}. d) 0.5--10\,keV emission of 1ES\,1927+654, compared to the historical value (green horizontal dot-dashed line)\cite{Ricci:2020fp,Ricci:2021ro}, showing extreme X-ray variability (a factor $\sim10,000$ and $\sim100$ on timescales of $\sim 100$\,days and a few hours, respectively).}
\label{fig:CSxray}
\end{figure*}

\subsection{Multi-wavelength observations of CS-AGN}

\subsubsection{X-rays}\label{sect:X-ray}

When observed in the X-rays, most CS-AGN show a change in flux correlated with that observed in the optical regime. This has been observed for several nearby objects such as Mrk\,1018\cite{Krumpe:2017ss} and NGC\,1566\cite{Parker:2019sp,Jana:2021wk}, with the X-ray flux varying up to a factor ${\approx}30$. At higher redshifts, SDSS\,J0159+0033 showed a dimming by a factor of $\sim10$ in the X-rays between 2000 and 2005, similar to that observed in the optical\cite{LaMassa:2015rt}. 
In a few cases the change in X-ray flux was found to be associated to significant X-ray spectral  variability\cite{Noda:2018jz}, with a strong soft excess appearing during the brightest intervals, when the broad optical lines are observed, and disappearing in the lowest flux periods, when the lines fade (top panels of Fig.\ref{fig:CSxray}). While no significant changes in the line-of-sight (neutral) $N_{\rm H}$ were observed during or after CS events, in NGC\,1566, {\it XMM-Newton}/RGS shows the presence of absorption lines from an outflow moving at ${\approx} 500\rm\,km\,s^{-1}$, which could be associated with the drastic increase in flux and in radiation pressure traced in this source\cite{Parker:2019sp}.

One very clear exception to the generally ``synchronized'' behaviour discussed was observed in the system 1ES\,19267+654\cite{Ricci:2020fp,Ricci:2021ro}. After the optical/UV outburst that led to the CS event, the X-ray spectral shape of the source changed drastically, with the power-law component almost completely disappearing (i.e. dropping by a factor of at least 100), and the emission becoming dominated by a soft blackbody component, with a temperature $kT\simeq 80-120$\,eV (bottom left panel of Fig.\,\ref{fig:CSxray}). This was the first time that the power-law emission component, produced in the hot X-ray corona and ubiquitously observed in AGN, was observed to vanish. 
The source was monitored for over 450\,days, and showed further drastic variability on a broad range of timescales, varying by up to $\sim 4$\,dex on timescales of $\sim 100$\,days, and by $\sim 2$\,dex  within as little as $\sim 8$\,hours (bottom right panel of  Fig.\,\ref{fig:CSxray}). 
Following the strong dip in the X-ray flux that followed the UV/optical CS event, the X-ray emission increased rapidly, with the power-law component recovering its dominance over the X-ray regime $\sim 300$\,days after the identification of the CS  event\cite{Ricci:2020fp,Ricci:2021ro}. Another unexpected feature of this AGN is the clear harder-when-brighter behaviour, with both the temperature of the blackbody component and the relative strength of the power-law component increasing with source luminosity\cite{Ricci:2021ro,Masterson:2022wy}.

\subsubsection{Infrared}\label{sect:IR}

Infrared observations have been extremely useful to pinpoint the origin of CS-AGN, and to exclude a significant contribution of reddening events associated to transient obscuration\cite{Yang:2018mn}.
In general, the properties of CS-AGN in the infrared appear to be similar to those of the general AGN population, specifically showing evidence for (1) the mid-infrared emission echoing the variability observed in the optical band, with time lags that are consistent with what would be expected for reprocessing by torus dust\cite{Sheng:2017of}; and (2) a redder-when-brighter behaviour, likely associated to hotter dust for increasing luminosities\cite{Yang:2018mn}. 
Infrared observations of CS-AGN can also be used, exploiting their rapid and large-amplitude variability, to infer the size of the dusty torus through reverberation mapping measurements\cite{Kokubo:2020kr}. A study of the dust around Mrk\,509 was also able to infer that the replenishment time-scale of the dust in the innermost regions of the system is very rapid\cite{Kokubo:2020kr} ($\sim 4$\,years).

\subsubsection{Radio}\label{sect:radio}
Due to the current scarcity of radio studies of CS events in (unbeamed) AGN, it is still unclear whether the radio emission varies at all following a CS event and, if so, whether it is tightly correlated to the optical and X-ray flux in such events, or what are the typical lags. 
In principle, understanding the radio behaviour could provide unprecedented insights regarding disc/corona-related radio (or sub-mm) emission mechanisms\cite{Panessa:2019vp}, jet launching and/or quenching.
For well-studied nearby sources, it was found that NGC\,2617 showed no significant radio core emission variability after the 2013 CS outburst\cite{Yang:2021jr}, while Mrk\,590 showed correlated radio and optical/UV/X-ray variability\cite{Koay:2016jy}, besides a weak radio jet that could be the byproduct of the recurring AGN activity\cite{Yang:2021oe}. In Mrk\,1018 the 5 GHz radio flux decreased by $\sim 20\%$ during the Type 1.9 dimming phase (2016$-$2017)\cite{Lyu:2021ag}. 
Radio observations of the UV/optical-flaring AGN 1ES\,1927+654 have instead revealed a dimming, by a factor $\sim 4$, of the core radio component during the CS event, followed by a re-brightening\cite{Laha:2022yu}. In this case the radio-to-X-ray flux ratio continuously decreased during the CS event. 
A study of a sample of  CS-AGN at $z<0.83$ found that these objects unlikely to be more active in the radio than normal AGN\cite{Yang:2021jr}.

\medskip

\subsection{CS events: generally accretion, not obscuration}\label{sect:accretionVSobs}

The idea that (most) CS events can be attributed to changes in the level of obscuration of the BLR has been generally discarded over the past few years by numerous lines of evidence, which have thus promoted the view that most CS events are driven by changes in the intrinsic AGN luminosity. 
The main arguments relate to the observed transition timescales, unobscured spectral appearances, BLR lags, and other multi-wavelength properties of CS-AGN, as we detail next.

Specifically, these challenges include
i) The long timescales ($\gtrsim 10$\,years) that would be needed for a dusty cloud to efficiently cover the BLR\cite{LaMassa:2015rt}, assuming reasonable cloud velocities. 
ii) The absence of clear signatures of reddening in the optical spectra after the broad lines have weakened\cite{LaMassa:2015rt,MacLeod:2016pu}. 
iii) The observation of the optical broad lines appearing in 1ES\,1927+654 $\sim 2-3$ months after the UV/optical continuum flare, and following the development of a blue continuum\cite{Trakhtenbrot:2019ay}. 
iv) The lack of obscuration in the X-rays for sources in which the broad lines are not clearly observed\cite{Husemann:2016iu}. 1ES\,1927+654 showed a similar level of X-ray obscuration before and after the CS event\cite{Gallo:2013vw,Ricci:2020fp,Ricci:2021ro}, while Mrk\,590 did not show strong X-ray obscuration after the nuclear luminosity dropped by two orders of magnitude and the broad optical lines disappeared\cite{Denney:2014do}. 
v) The lack of a clear difference in $E(B-V)/N_{\rm H}$ between CS-AGN and ``normal'' AGN\cite{Jaffarian:2020qt}.  
vi) The detection of rapid infrared variability (\S\ref{sect:IR}), specifically the identification of IR flux drops in several sources that exhibited  H$\alpha$ or H$\beta$ disappearance\cite{Sheng:2017fu,Assef:2018vj,Stern:2018su}. The infrared-emitting region is too large to be obscured on timescales of $\lesssim 10$\,years (similarly to the BLR mentioned above), and---importantly---the radiation at these mid-IR wavelengths is much less affected by dust obscuration (compared with the optical/UV regime). 
vii) The correlation between radio and optical/UV (and/or X-ray) variability observed in several objects\cite{Koay:2016jy} (\S\ref{sect:X-ray},\S\ref{sect:radio}). 
viii) The low levels of optical polarization found in CS-AGN\cite{Hutsemekers:2019zw}. 
ix) The relation between optical/IR colours and flux in CS-AGN being similar to that of normal AGN\cite{Yang:2018mn}. 
x) The broad line widths varying in a similar fashion as normal AGN\cite{Runnoe:2016hb}, with the line being broader (narrower) when the source is dimmer (brighter). 
This is consistent with a picture where, as the continuum flux increases (decreases), the line is produced from parts within the BLR that are further (closer) to the SMBH. 
Quantitatively, for a virialized BLR the expected trend follows  $FWHM \sim L_{\rm UV/opt}^{-1/4}$, which is indeed supported by several reverberation mapping studies\cite{Bentz:2013cn}.

While some sources, such as  NGC\,7582 and NGC\,4151, have been found to show both CO and CS transitions, these phenomena are not correlated, and appear on very different timescales, suggesting that the two processes are generally independent. 
One sort of exception are CO-AGN in which the {\it apparent} change in $N_{\rm H}$ is caused by a switch-off/switch-on of the central engine (see \S\ref{sec:switchoff}), such as seen in NGC\,2992. 
Another possible exception is if a transient (dusty) cloud obscures a part of the BLR (but not the whole of it). This scenario may be more likely to occur in systems that are observed with lines-of-sight that ``graze'' the torus, and was invoked to explain the spectral transitions related to some Type 1.8/1.9 AGN\cite{Goodrich90}.
A recent study of a CL-AGN identified in the new SDSS-V survey provides another example of how a transient obscuration may explain spectral transitions in the optical regime\cite{Zeltyn:2022gr}.
It is possible that future high cadence spectroscopic studies will reveal more events of this type. Otherwise, it appears that the vast majority of CS events detected so far are driven by changes in the accretion flow itself. 

Depending on the origin of the BLR, a change in the accretion flow (or rate) could either dramatically transform the BLR or change its ionization properties. 
First, the line emission from a pre-existing BLR is expected to respond to any (strong) ionizing continuum variability, as unambiguously seen in reverberation mapping studies. CS events could thus be regarded as the most extreme cases of such BLR response, where it essentially turns on/off (with the associated light-travel lags).
Second, it has been proposed that the BLR, and indeed even the torus, could be created by outflows produced in the accretion disk\cite{Emmering:1992ut}. In this framework, the disappearance of broad optical/UV lines in CS-AGN would arise straightforwardly from a strong decrease in the accretion rate\cite{Elitzur:2006ec}, which would lead to a decrease in the mass outflow rate of the disk wind. In particular, the disappearance of the BLR would be expected to happen at bolometric luminosities of $L_{\rm bol}\lesssim2.3\times10^{40}\rm\,erg\,s^{-1} M_{8}^{2/3}\rm\,erg\,s^{-1}$ (Ref. \citen{Elitzur:2009hh}). Therefore, as $L_{\rm bol}$ increases, the optical classification of AGN would transition following the sequence Type\,2.0 $\rightarrow$ 1.9/1.8 $\rightarrow$ 1.5/1.2 $\rightarrow$ 1.0  (Ref. \citen{Elitzur:2014kc}).

\medskip

\subsection{Accretion disks and variability timescales}\label{sect:ADtimescales}

To contextualize the accretion-driven (dramatic) variability seen in CS-AGN, we rely on the  geometrically-thin, optically-thin accretion disk (aka ``$\alpha$-disk) model\cite{Shakura:1973ex}, which is commonly applied to AGN, and specifically invoked in many studies that (try to) interpret CS-AGN.
The thin nature of the disk implies an aspect ratio (at a given radius $R$) of $H/R\ll1$, where the height of the disk is given by $H=c_{\rm s}/\Omega$, with $c_{\rm s}$ being the sound speed and $\Omega$ the Keplerian orbital angular frequency, $\Omega_{\rm K}=\sqrt{GM_{\rm BH}/R^{3}}$. In this model the angular momentum transport is driven by (pseudo-)viscosity, which is likely associated to magneto-rotational instability\cite{Balbus:1991qv}. Despite its many successes, observational evidence gathered over the past 30\,years has shown that this model fails to explain several important aspects of the AGN emission, such as their UV luminosity and optical variability, suggesting that accretion disks are not in a quasi-steady state equilibrium\cite{Alloin:1985it,Lawrence:2012sd}.

The observed timescales of CS events, typically $\sim 1-10$\,years or even shorter\cite{Noda:2018jz,Gezari:2017md,Trakhtenbrot:2019ay}, are generally much shorter than the timescales over which the accretion flow in an AGN $\alpha$-disk is expected to vary drastically\cite{Lawrence:2018gc}. Depending on the physical process responsible for the changes in the AGN accretion-driven radiation, different timescales would be at play. In the following we review the most important timescales associated to thin disks in accreting SMBHs, in line with many CS-AGN studies\cite{Netzer:2013ix,LaMassa:2015rt,Stern:2018su}.

When dealing with casually-connected varying zones, for a region of size $R$ the shortest timescale to consider is the {\it light crossing timescale}:
\begin{equation}\label{eq:lc}
t_{\rm lc} \equiv \frac{R}{c}\simeq 0.875\,\left(\frac{R}{150\,r_{\rm g}}\right)\,M_{8}\rm\,\,days \, ,
\end{equation}
where $c$ is the speed of light, and $M_{8}$ is (again) the black hole mass in units of $10^{8}\rm\,M_{\rm \odot}$. This timescale is typically associated with radiation reprocessing, and is significantly shorter than that of CS events.

Gas motion within the disk can occur over the {\it dynamical timescale} ($t_{\rm dyn}$) or the closely related {\it orbital timescale} ($t_{\rm orb} = 2 \pi t_{\rm dyn}$).
Standard relations for Keplerian (circular) motion yield: 
\begin{equation}\label{eq:tdyn}
t_{\rm dyn} \equiv \frac{R}{v} \simeq \left(\frac{GM_{\rm BH}}{R^{3}} \right)^{-0.5}\simeq 10\,\left(\frac{R}{150\,r_{\rm g}}\right)^{3/2}\,M_{8}\rm\,\,days
\end{equation}
where $G$ is the Gravitational constant. 
The timescale at which sound waves would propagate radially through the disk would be longer than $t_{\rm dyn}$ by a factor ($R/H$): $t_{\rm snd}\simeq t_{\rm dyn} (H/R)^{-1}$, which for a postulated $(H/R)<0.1$ yields timescales of order several months.
Both these timescales are thus typically too short to explain most CS events.

The timescale associated with (significant parts of) the disk heating up or cooling down, i.e. the {\it thermal timescale}, can be estimated as the ratio between $t_{\rm dyn}$ and the viscosity parameter\cite{Shakura:1973ex} ($\alpha$):
\begin{equation}\label{eq:tth}
t_{\rm th} \simeq \frac{t_{\rm dyn}}{\alpha}\simeq \left(\frac{\alpha}{0.03}\right)^{-1}
\left(\frac{R}{150\,r_{\rm g}}\right)^{3/2}\,M_{8}\rm\,\,yrs.
\end{equation}
This timescale is suggested as the origin of stochastic optical variability observed on $\sim$months timescales in AGN\cite{Kelly:2009bo}, and it has been argued that this timescale could agree with some CS-AGN\cite{Stern:2018su}.

Cooling and heating fronts can propagate radially through the accretion disk at $v_{\rm front} \simeq c_{\rm s} \alpha$, implying timescales of 
$t_{\rm front}\simeq t_{\rm th} (H/R)^{-1}$, or:
\begin{equation}\label{eq:theating}
t_{\rm front}\simeq 20 \left(\frac{H/R}{0.05}\right)^{-1} \left(\frac{\alpha}{0.03}\right)^{-1} \left(\frac{R}{150\,r_{\rm g}}\right)^{3/2} M_{8} \rm\,\, yrs
\end{equation}

Finally, changes in the global accretion rate through the disk are associated with the {\it radial inflow timescale}, or so-called  {\it viscous timescale}:
\begin{equation}\label{eq:tvis}
t_{\rm vis} \simeq 400 \left(\frac{H/R}{0.05}\right)^{-2} \left(\frac{\alpha}{0.03}\right)^{-1} \left(\frac{R}{150\,r_{\rm g}}\right)^{3/2}\,M_{8}\rm\,\,yrs.
\end{equation}
We note that since all these timescales depend on the location within the disk (namely, $R/r_{\rm g}$), and since in such disk the continuum emission at any given wavelength is preferentially emitted from certain disk regions, one may want to consider a more detailed derivation of wavelength-dependent timescales, which in principle further depend on $M_{\rm BH}$, $\lambda_{\rm Edd}$ and even BH spin.
The $R \simeq 150\,r_{\rm g}$ scaling used above is generally appropriate for the near-UV-to-optical continuum.

The large-amplitude and coherent nature of the radiation changes seen in CS-AGN are generally thought to be associated with larger-scale processes, which would act on timescales of either $t_{\rm front}$ or $t_{\rm vis}$.
However, both these timescales are longer than what is typically observed (see \S\ref{sect:earlyworks} and \S\ref{sect:largesurveys}). 
For example, in the nearby system Mrk\,1018 a drop in the luminosity of a factor of $\sim 13-15$ was observed over a timescale of $\sim 8$ years\cite{Noda:2018jz}. Assuming the nominal values of Eqs.\ref{eq:theating}--\ref{eq:tvis}, for this object the cooling/heating front and viscous timescales would be expected to be $\sim 15$ and $\sim 280$ years, respectively.
Likewise, the CS-AGN transitions in many other systems are constrained to occur within $\lesssim10$ years\cite{Tohline:1976gd,Gilli:2000bw,Gezari:2017md}.
Perhaps the most extreme case reported to date is that of 1ES\,1927+654, where the UV/optical luminosity increased by more than a factor of $\sim 50$ within a few months\cite{Trakhtenbrot:2019ay}.

Several possible mechanisms have been proposed to overcome this inconsistency, typically by considering thicker disks, since increasing the aspect ratio ($H/R$) would lead to shorter timescales. 
Among the mechanisms explored in the literature, thickening the disk was suggested to be possible through: 
i) an enhanced total opacity in the disk, due to line opacity from heavy elements, which could alter the thermal stability of the disk and increase its scale height\cite{Jiang:2016ql};
ii) magnetic torques in the innermost regions of the disk, which would cause the inner disk to heat up and inflate\cite{Agol:2000pz}; 
iii) magnetic pressure dominating in the upper atmosphere of the disk, so that the magnetic field could support the disk vertically, increasing its height\cite{Dexter:2019nr}. 
Magnetically-driven disk-winds \cite{Feng:2021iy}, carrying away most of the angular momentum of the gas, would also considerably shorten the viscous timescales of the accretion disk.

Many recent studies present more elaborate calculations and/or simulations of globally stable accretion flows, which however present large-amplitude variability on exceedingly short timescales, reminiscent of what is seen in CS-AGN\cite{Noda:2018jz,Jiang2019,Sniegowska:2020an}
We discuss some of these models in more detail in what follows.

\medskip

\subsection{The origin of CS transitions}\label{sect:CSAGN_origin}

Several different scenarios have been proposed to explain CS events, which can be roughly divided into two main categories: instabilities in the (globally stable) accretion disk (\S\ref{sect:instabilities}), which could lead to changes in the emission properties of the accretion flow; 
and major accretion disk perturbations (\S\ref{sect:perturbations}), such as stellar tidal disruption events (TDEs). 
As discussed below, it is possible that these two different types of mechanisms could be at work in different CS events.

\medskip

\subsubsection{Disk instabilities}\label{sect:instabilities}

Disk instabilities can be triggered by various effects, and can thus occur over a range of timescales. One possibly relevant mechanism for CS-AGN is the thermal instability\cite{Lin:1986yt}, associated with changes in the surface density of the disk, driven by changes in opacity, which could lead to significant, though localized, variations in temperature. The magnetorotational instability\cite{Balbus:1991qv} could also lead to changes in the local heating on thermal timescales. It has been shown\cite{Noda:2018jz} that, in the case of Mrk\,1018, the disappearance/appearance of the broad optical lines is directly related to the absence/presence of a strong soft X-ray excess at $\lesssim 1-2$\,keV (top panels of Fig.\,\ref{fig:CSxray}). Such a component, possibly associated to a warm Comptonizing material, could extend to the UV, and be responsible for most of the ionizing photons that create the broad emission lines. In Mrk\,1018 the Eddington ratio goes from $\lambda_{\rm Edd}\approx 8\times10^{-2}$ to $\approx 6\times10^{-3}$ during the event that leads to the disappearance of the broad lines and, while the $\gtrsim 2$\,keV emission drops only of a factor $\sim 7$, the soft X-ray excess decreases of a factor of $\gtrsim 60$. 
Qualitatively similar results were obtained for other CS-AGN\cite{Ai:2020wv}. This spectral hardening could be similar to the soft-to-hard state transitions observed in black hole binaries, associated to the hydrogen ionisation disk instability, where the inner disk puffs up and transitions into an advection dominated accretion flow (ADAF) at $\lambda_{\rm Edd}\lesssim 0.02$.
These state transitions would therefore produce large variations of the amount of ionizing radiation emitted by the AGN. 
A consistent behaviour was also found\cite{Ruan:2019pa} by studying the relation between the optical-to-X-ray spectral index ($\alpha_{\rm OX}$) and $\lambda_{\rm Edd}$, which showed similarities with what is observed in black hole binaries, including an inversion of the $\alpha_{\rm OX}-\lambda_{\rm Edd}$ correlation at $\lambda_{\rm Edd}\simeq 10^{-2}$.  This scenario would imply that CS events typically happen at an Eddington ratio of a few percent, which appears to be in agreement with some recent studies\cite{MacLeod:2019rk,Graham:2020nz}. 
The timescale of these transitions would be the viscous timescale (Eq.\,\ref{eq:tvis}), which is considerably longer than the timescale of CS events. However, since radiation pressure in AGN is more important than in binaries\cite{Noda:2018jz}, the sound speed in the disk would be higher, leading to a decrease in $t_{\rm vis}$. Including magnetic pressure would further increase the sound speed and decrease the timescales. A similar effect would be produced by thicker disks, as outlined in \S\ref{sect:ADtimescales}.

Another viable mechanism for CS events is the propagation of a cooling front in an inflated accretion disk, associated to a sudden change in the magnetic torques in the innermost regions of the accretion flow, close to the innermost stable circular orbit\cite{Ross:2018ah,Stern:2018su}. Magneto-hydrodynamical instability could cause the torque to decrease, leading to an  outwards-propagating cold front, which would decrease the optical/UV luminosity of the AGN. At some point the cooling front would stop, and a heating front is expected to propagate inward, leading to a re-brightening of the AGN\cite{Ross:2018ah}. The inflated inner disk would have a larger aspect ratio ($H/R\simeq 0.2$) than typically assumed for standard accretion disk\cite{Stern:2018su}, which would lead to shorter timescales (see Eq.~\ref{eq:theating}), consistent with those observed in some CS events. This model was shown to explain well the fading and re-brightening of SDSS\,J1100$-$0053, which was associated to the disappearance and reappearance of the UV continuum and broad hydrogen emission lines\cite{Ross:2018ah}.
There are other suggestions for how advection and radiation pressure regions in the inner disk may drive significant instabilities, relevant for driving CS transitions, even in sources accreting at Eddington rates of a few percent. Specifically, instabilities are expected to develop in a narrow region that links the inner ADAF-like part of the accretion flow and the standard, outer disk, and could explain the recurrent CS events observed in objects such as, e.g. NGC\,1566 and NGC\,4151 (Ref.~\citen{Sniegowska:2020an}).
A sudden change of polarity of the magnetic flux advected onto the SMBH could also lead to a CS event, triggering an increase of UV/optical flux while at the same time destroying the X-ray corona\cite{Scepi:2021yc}, as observed in 1ES\,1927+654. 
For a magnetically supported and highly viscous ($\alpha \simeq 1$) disk this would lead to changes on short timescales ($<$1 year), consistent with those observed in 1ES\,1927+654.

\begin{table*}[h!]
\caption[]{Probing different regions of accreting SMBHs with CL-AGN.}\vspace{-0.6cm}
\label{tab:CLAGN}
\begin{center}
\begin{tabular}{lcc}
\hline
\hline
\noalign{\smallskip}
  & {\bf Changing-obscuration AGN} (\S\ref{sect:COagn})	&		{\bf Changing-state AGN} (\S\ref{sect:CSAGN_origin}) \\ 
\noalign{\smallskip}
\hline
\noalign{\smallskip}
\noalign{\smallskip}
\noalign{\smallskip}
{\bf X-ray corona}   &	 	Size\cite{Risaliti:2007aa} (\S\ref{sec:NHvar})			&	Creation\cite{Ricci:2020fp}, evolution with changes  	\\	%
   				&												 				&	 in accretion\cite{Noda:2018jz} (\S\ref{sect:X-ray})	\\	%
\noalign{\smallskip}
{\bf Accretion disk}   &	 		Interplay disk/outflows (\S\ref{sec:outflows}) 		&	Variability timescales (\S\ref{sect:ADtimescales}), mechanisms	\\	%
   &	 		 		&	 triggering instabilities and perturbations (\S\ref{sect:CSAGN_origin})	\\	%
\noalign{\smallskip}
{\bf Broad-line region}   &	 	Physical and kinematical properties 			&	Creation and evolution\cite{Elitzur:2006ec} (\S\ref{sect:accretionVSobs})	\\	%
   &	 	of the clouds\cite{Maiolino:2010fu,Risaliti:2011jl} (\S\ref{sec:BLRcoagn})		&		\\	%
\noalign{\smallskip}
{\bf Torus}   &	 		Physical and kinematical properties 	&	Size and dust replenishment\cite{Kokubo:2020kr} (\S\ref{sect:IR})	\\	%
 &	 		of the clouds\cite{Markowitz:2014oq} (\S\ref{sect:COrecent})		&	  	\\	%
\noalign{\smallskip}
\hline
\noalign{\smallskip}
\end{tabular}
\end{center}
\end{table*}

\subsubsection{Major disk perturbations}\label{sect:perturbations}

Some CS-AGN could be powered by the tidal disruption of a star passing near a SMBH. This explanation was first proposed for the changing look quasar SDSS\,J0159+0033\cite{Merloni:2015ew} which, after the outburst that caused the CS event, showed a decline in flux consistent with the $t^{-5/3}$ trend expected for TDEs (and supported by some observations; Ref.~\citen{vanVelzen2020}). 
The estimated virial  black hole mass of this object ($\sim 1.7\times10^{8}\,M_{\odot}$, Ref:\citen{LaMassa:2015rt}) was however found to be considerably higher than what is observed in, and expected for, typical TDEs. This could pose a problem, since the tidal radius ($r_{\rm tid}$), i.e. the radius at which a star of mass $M_*$ and radius $R_*$ would likely be disrupted is inversely proportional to $M_{\rm BH}$:
\begin{equation}\label{eq:tidalradius}
r_{\rm tid}\simeq 2.2 r_{\rm g} \left(\frac{M_*}{M_{\odot}}\right)^{-1/3} \left(\frac{R_{*}}{R_{\odot}}\right) M_{8}^{-2/3}.
\end{equation}
Therefore, if the star is compact and/or the black hole is massive, then the tidal disruption would only take place inside the event horizon, and it would not be detected. For SDSS\,J0159+0033 the tidal radius would be expected to be at only a few $r_{\rm g}$. It was however argued that the virial mass could be significantly overestimated in this objects, possibly due to the contribution of the stellar debris to the H$\alpha$ line\cite{Zhang:2021fc}, since it would lie significantly above the $M_{\rm BH}-\sigma$ relation.
Another explanation may involve a BH with a non-zero spin, which would lead to a smaller ergosphere and allow for (observable) TDEs with smaller $r_{\rm tid}$ (Ref.~\citen{Kesden2012}).

TDEs that occur in an already-active SMBH, and that interact with a pre-existing accretion disk, were also argued to explain the disappearance and reappearance of the X-ray corona\cite{Ricci:2020fp,Ricci:2021ro}, the strong blackbody component\cite{Ricci:2020fp,Ricci:2021ro}, and the peculiar behaviour of the broad optical lines\cite{Li:2022pp} seen in 1ES\,1927+654. 
Some models suggest that the occurrence rate of TDEs would be higher in (persistently accreting) AGN\cite{Karas2007}, and that TDEs-in-AGN could be associated with the presence of stars within the accretion flow itself\cite{McKernan:2022yd}. Recent simulations\cite{Chan:2019aa} demonstrated how the debris stream of the tidally disrupted star would hit an accretion disk, disrupting the gas in the disk and further depleting a large fraction of the angular momentum of the disk gas (including through shocks). This, in turn, would rapidly increase the accretion rate of material through the inner regions of the disk, significantly affecting the disk UV/optical emission\cite{Chan2020}, and may lead to the disappearance of the X-ray corona, which is linked to the innermost regions of the accretion flow. 
As the inner disk is replenished the corona would be ``reignited''. If the gas pressure falls sharply from the outer disk to the inner disk, then magnetic loops could pull the gas, allowing it to spiral inward in tens of disk orbits (Eq.\,\ref{eq:tdyn}), roughly in agreement with the observed timescales.
We note that detailed numerical studies of such processes are still scarce\cite{Chan:2019aa} and large parts of the parameter space are yet to be explored.

Other mechanisms proposed to explain CS events include the tidal interaction between disks in {\it binary} SMBHs\cite{Wang:2020si}, and the interaction between the disk and a recoiling SMBH\cite{Kim:2018sa}. The latter explanation has been proposed for Mrk\,1018, with the recoiling SMBH having a very eccentric orbit and perturbing the accretion flow with a period of 29 years. Finally, stellar-mass point perturbers in the disk (i.e., stars or stellar mass BHs) could cause extreme mass ratio inspiral events or changes in the accretion flow\cite{Stern:2018su}, triggering (potentially recurring) CS events.

\medskip
\medskip

\section{Open questions and future prospects}\label{sec:openquestionsandprospects}

While CO and CS events are very different in nature, both types of transitions allow us to probe the close vicinity of active SMBHs (Table\,\ref{tab:CLAGN}). In the following we discuss some of the key open questions that could be addressed to better understand changing-look AGN in the coming years, and to improve our overall understanding of the AGN phenomenon.

\smallskip
\subsection{CO-AGN}

\begin{itemize}
\item {\it Is there a relation between the host galaxy properties and the fraction of AGN showing CO events?}\newline
AGN in the final stages of a major galaxy merger are typically heavily obscured\cite{Ricci:2017aa}, likely due to the efficient depletion of the angular momentum of galactic gas caused by tidal forces. This could make the nuclear environment of AGN in mergers particularly dynamic, and could enhance the fraction of CO events\cite{Yamada:2021to}. While several\cite{Yamada:2021to} CO-AGN have been found in mergers, but it is not yet clear whether the rate of CO events changes with the host galaxy properties.

\item {\it What can we learn about properties of the circumnuclear material from CO events?}\newline
High-resolution X-ray spectroscopy can shed light on some fundamental properties (e.g., ionization state, velocity, structure) of the BLR and torus clouds. This will be possible in the coming years with the new generation of X-ray calorimeters, such as the one on-board {\it XRISM}\cite{XRISM-Science-Team:2020gd}.

\item {\it Do low-$\lambda_{\rm Edd}$ AGN show eclipses?}\newline
Several theoretical studies have predicted that at $\lambda_{\rm Edd}\lesssim 10^{-3}$ the AGN would be too weak to produce outflows that refill both the BLR and the torus\cite{Elitzur:2006ec}. Therefore one would expect that in these objects eclipses are absent, or extremely rare. Large systematic studies of nearby AGN would allow us to better understand the relation between the likelihood of observing CO events and the physical properties of accreting SMBHs.

\end{itemize}

One fundamental issue with constraining the properties of the obscuring clouds in CO-AGN has been the non-continuous monitoring in the X-ray band. Thanks to the advent of surveys such as those carried out by {\it eROSITA}\cite{Merloni:2020ec} and in the future by the {\it Einstein Probe}\cite{Yuan:2015zd}, it will be possible to significantly increase the sample of CO-AGN, and to better understand their occurrence rates and typical timescales.

\smallskip
\subsection{CS-AGN}

\begin{itemize}
\item {\it How common are CS transitions in AGN, and what are their typical timescales?}\newline
Large samples of CS-AGN, together with high-cadence monitoring, will allow us to understand their occurrence rates and typical timescales. The on-going SDSS-V survey\cite{Kollmeier:2017dx} is expected to identify a large number of CS-AGN in the next few years, over timescales ranging days-years. The Vera Rubin Observatory\cite{Ivezic:2019go}, which will provide deep multi-band optical photometry with several-day cadence over the entire southern sky, will largely increase the number of known extreme AGN variability events. The identification of CS-AGN will be possible thanks to machine-learning techniques\cite{Sanchez-Saez:2021bq} and spectroscopic follow-ups carried out with the next generation of highly multiplexed spectrographs, such as 4MOST\cite{de-Jong:2019fp}.

\item {\it Are CS transitions related to certain physical properties of the AGN?}\newline
It is still unclear whether there is a particular set of physical parameters of the accreting SMBH (e.g., $M_{\rm BH}$, $\lambda_{\rm Edd}$) that increase the likelihood of a CS transition, as it would be expected by some models\cite{Noda:2018jz}. Future large and well-understood samples of CS-AGN will allow us to identify and/or test such links, which would be a fundamental step towards understanding the origin of CS transitions.

\item {\it What drives extreme AGN variability? Is there a unique physical mechanism driving CS transitions?}\newline
The large samples that will be available in the near future will allow to refine our understanding of the various physical mechanisms that drive extreme AGN variability. The various types of mechanisms, associated with e.g. disk instabilities (\S\ref{sect:instabilities}) or perturbations (\S\ref{sect:perturbations}), can produce CS events with markedly different properties. For example, the CS event in 1ES\,1927+654\cite{Trakhtenbrot:2019ay} is clearly very different from other nearby CS-AGNs, showing very unexpected properties in the X-ray band\cite{Ricci:2020fp,Ricci:2021ro,Masterson:2022wy}. Other events seem to necessitate a mechanism that can produce {\it recurring} CS transitions\cite{Alloin:1986iu,Shapovalova:2010an,McElroy:2016hs,Ilic:2020jj}.
The combination of optical and X-ray surveys will therefore be extremely important to fully characterise CS events.

\end{itemize}

With the advent of large photometric (Vera Rubin Observatory\cite{Ivezic:2019go}) and spectroscopic (SDSS-V\cite{Kollmeier:2017dx}, 4MOST\cite{de-Jong:2019fp}) optical surveys, as well as with that of wide-field surveys in the X-rays (i.e., with {\it eROSITA}\cite{Merloni:2020ec} and the {\it Einstein Probe}\cite{Yuan:2015zd}) and the UV (e.g., with {\it ULTRASAT}\cite{Ben-Ami:2022kh}), in the next few years a very large number of CS-AGN will be identified and characterised, allowing us to discover new extreme variability and transient phenomena in AGN, and to understand what physical mechanisms are at play in these fascinating objects.

\section*{Acknowledgements}
C.R. acknowledges support from the Fondecyt Iniciacion (grant 11190831) and ANID BASAL (project FB210003). 
B.T. acknowledge support from the European Research Council (ERC) under the European Union's Horizon 2020 research and innovation program (grant agreement 950533) and from the Israel Science Foundation, (grant 1849/19). 
We thank Grisha Zeltyn for his help with Figure\,1.\\

\noindent{Correspondence should be addressed to Claudio Ricci.}

\section*{Author contributions}
CR and BT devised, together, the concept and structure of the Review.


\begin{thebibliography}{100}
\expandafter\ifx\csname url\endcsname\relax
  \def\url#1{\texttt{#1}}\fi
\expandafter\ifx\csname urlprefix\endcsname\relax\def\urlprefix{URL }\fi
\providecommand{\bibinfo}[2]{#2}
\providecommand{\eprint}[2][]{\url{#2}}

\bibitem{Magorrian:1998fr}
\bibinfo{author}{{Magorrian}, J.} \emph{et~al.}
\newblock {The Demography of Massive Dark Objects in Galaxy Centers}.
\newblock \emph{\bibinfo{journal}{\aj}} \textbf{\bibinfo{volume}{115}},
  \bibinfo{pages}{2285--2305} (\bibinfo{year}{1998}).
\newblock \eprint{astro-ph/9708072}.

\bibitem{Koss:2017gv}
\bibinfo{author}{{Koss}, M.} \emph{et~al.}
\newblock {BAT AGN Spectroscopic Survey. I. Spectral Measurements, Derived
  Quantities, and AGN Demographics}.
\newblock \emph{\bibinfo{journal}{\apj}} \textbf{\bibinfo{volume}{850}},
  \bibinfo{pages}{74} (\bibinfo{year}{2017}).
\newblock \eprint{1707.08123}.

\bibitem{Ricci:2017co}
\bibinfo{author}{{Ricci}, C.} \emph{et~al.}
\newblock {BAT AGN Spectroscopic Survey. V. X-Ray Properties of the Swift/BAT
  70-month AGN Catalog}.
\newblock \emph{\bibinfo{journal}{\apjs}} \textbf{\bibinfo{volume}{233}},
  \bibinfo{pages}{17} (\bibinfo{year}{2017}).
\newblock \eprint{1709.03989}.

\bibitem{Antonucci93}
\bibinfo{author}{{Antonucci}, R.}
\newblock {Unified models for active galactic nuclei and quasars}.
\newblock \emph{\bibinfo{journal}{\araa}} \textbf{\bibinfo{volume}{31}},
  \bibinfo{pages}{473--521} (\bibinfo{year}{1993}).

\bibitem{Urry95}
\bibinfo{author}{{Urry}, C.~M.} \& \bibinfo{author}{{Padovani}, P.}
\newblock {Unified Schemes for Radio-Loud Active Galactic Nuclei}.
\newblock \emph{\bibinfo{journal}{\pasp}} \textbf{\bibinfo{volume}{107}},
  \bibinfo{pages}{803} (\bibinfo{year}{1995}).
\newblock \eprint{astro-ph/9506063}.

\bibitem{Ramos-Almeida:2017hw}
\bibinfo{author}{{Ramos Almeida}, C.} \& \bibinfo{author}{{Ricci}, C.}
\newblock {Nuclear obscuration in active galactic nuclei}.
\newblock \emph{\bibinfo{journal}{Nature Astronomy}}
  \textbf{\bibinfo{volume}{1}}, \bibinfo{pages}{679--689}
  (\bibinfo{year}{2017}).
\newblock \eprint{1709.00019}.

\bibitem{Netzer15}
\bibinfo{author}{{Netzer}, H.}
\newblock {Revisiting the Unified Model of Active Galactic Nuclei}.
\newblock \emph{\bibinfo{journal}{\araa}} \textbf{\bibinfo{volume}{53}},
  \bibinfo{pages}{365--408} (\bibinfo{year}{2015}).
\newblock \eprint{1505.00811}.

\bibitem{Elitzur12}
\bibinfo{author}{{Elitzur}, M.}
\newblock {On the Unification of Active Galactic Nuclei}.
\newblock \emph{\bibinfo{journal}{\apjl}} \textbf{\bibinfo{volume}{747}},
  \bibinfo{pages}{L33} (\bibinfo{year}{2012}).
\newblock \eprint{1202.1776}.

\bibitem{Lawrence:1982pk}
\bibinfo{author}{{Lawrence}, A.} \& \bibinfo{author}{{Elvis}, M.}
\newblock {Obscuration and the various kinds of Seyfert galaxies.}
\newblock \emph{\bibinfo{journal}{\apj}} \textbf{\bibinfo{volume}{256}},
  \bibinfo{pages}{410--426} (\bibinfo{year}{1982}).

\bibitem{Ricci:2017pr}
\bibinfo{author}{{Ricci}, C.} \emph{et~al.}
\newblock {The close environments of accreting massive black holes are shaped
  by radiative feedback}.
\newblock \emph{\bibinfo{journal}{\nat}} \textbf{\bibinfo{volume}{549}},
  \bibinfo{pages}{488--491} (\bibinfo{year}{2017}).
\newblock \eprint{1709.09651}.

\bibitem{Vanden-Berk:2004bw}
\bibinfo{author}{{Vanden Berk}, D.~E.} \emph{et~al.}
\newblock {The Ensemble Photometric Variability of
  \raisebox{-0.5ex}\textasciitilde25,000 Quasars in the Sloan Digital Sky
  Survey}.
\newblock \emph{\bibinfo{journal}{\apj}} \textbf{\bibinfo{volume}{601}},
  \bibinfo{pages}{692--714} (\bibinfo{year}{2004}).
\newblock \eprint{astro-ph/0310336}.

\bibitem{Uttley:2005tn}
\bibinfo{author}{{Uttley}, P.}, \bibinfo{author}{{McHardy}, I.~M.} \&
  \bibinfo{author}{{Vaughan}, S.}
\newblock {Non-linear X-ray variability in X-ray binaries and active galaxies}.
\newblock \emph{\bibinfo{journal}{\mnras}} \textbf{\bibinfo{volume}{359}},
  \bibinfo{pages}{345--362} (\bibinfo{year}{2005}).
\newblock \eprint{astro-ph/0502112}.

\bibitem{Ulrich:1997ep}
\bibinfo{author}{{Ulrich}, M.-H.}, \bibinfo{author}{{Maraschi}, L.} \&
  \bibinfo{author}{{Urry}, C.~M.}
\newblock {Variability of Active Galactic Nuclei}.
\newblock \emph{\bibinfo{journal}{\araa}} \textbf{\bibinfo{volume}{35}},
  \bibinfo{pages}{445--502} (\bibinfo{year}{1997}).

\bibitem{Mereghetti:2021aa}
\bibinfo{author}{{Mereghetti}, S.} \emph{et~al.}
\newblock {Time domain astronomy with the THESEUS satellite}.
\newblock \emph{\bibinfo{journal}{Experimental Astronomy}}
  (\bibinfo{year}{2021}).
\newblock \eprint{2104.09533}.

\bibitem{Graham:2020nz}
\bibinfo{author}{{Graham}, M.~J.} \emph{et~al.}
\newblock {Understanding extreme quasar optical variability with CRTS - II.
  Changing-state quasars}.
\newblock \emph{\bibinfo{journal}{\mnras}} \textbf{\bibinfo{volume}{491}},
  \bibinfo{pages}{4925--4948} (\bibinfo{year}{2020}).
\newblock \eprint{1905.02262}.

\bibitem{Ivezic:2019go}
\bibinfo{author}{{Ivezi{\'c}}, {\v{Z}}.} \emph{et~al.}
\newblock {LSST: From Science Drivers to Reference Design and Anticipated Data
  Products}.
\newblock \emph{\bibinfo{journal}{\apj}} \textbf{\bibinfo{volume}{873}},
  \bibinfo{pages}{111} (\bibinfo{year}{2019}).
\newblock \eprint{0805.2366}.

\bibitem{Yuan:2015zd}
\bibinfo{author}{{Yuan}, W.} \emph{et~al.}
\newblock {Einstein Probe - a small mission to monitor and explore the dynamic
  X-ray Universe}.
\newblock \emph{\bibinfo{journal}{arXiv e-prints}}
  \bibinfo{pages}{arXiv:1506.07735} (\bibinfo{year}{2015}).
\newblock \eprint{1506.07735}.

\bibitem{Ben-Ami:2022kh}
\bibinfo{author}{{Ben-Ami}, S.} \emph{et~al.}
\newblock {The scientific payload of the Ultraviolet Transient Astronomy
  Satellite (ULTRASAT)}.
\newblock \emph{\bibinfo{journal}{arXiv e-prints}}
  \bibinfo{pages}{arXiv:2208.00159} (\bibinfo{year}{2022}).
\newblock \eprint{2208.00159}.

\bibitem{Miniutti:2014yt}
\bibinfo{author}{{Miniutti}, G.} \emph{et~al.}
\newblock {The properties of the clumpy torus and BLR in the polar-scattered
  Seyfert 1 galaxy ESO 323-G77 through X-ray absorption variability}.
\newblock \emph{\bibinfo{journal}{\mnras}} \textbf{\bibinfo{volume}{437}},
  \bibinfo{pages}{1776--1790} (\bibinfo{year}{2014}).
\newblock \eprint{1310.7701}.

\bibitem{Marinucci:2016eu}
\bibinfo{author}{{Marinucci}, A.} \emph{et~al.}
\newblock {NuSTAR catches the unveiling nucleus of NGC 1068}.
\newblock \emph{\bibinfo{journal}{\mnras}} \textbf{\bibinfo{volume}{456}},
  \bibinfo{pages}{L94--L98} (\bibinfo{year}{2016}).
\newblock \eprint{1511.03503}.

\bibitem{Zaino:2020cz}
\bibinfo{author}{{Zaino}, A.} \emph{et~al.}
\newblock {Probing the circumnuclear absorbing medium of the buried AGN in NGC
  1068 through NuSTAR observations}.
\newblock \emph{\bibinfo{journal}{\mnras}} \textbf{\bibinfo{volume}{492}},
  \bibinfo{pages}{3872--3884} (\bibinfo{year}{2020}).
\newblock \eprint{2001.05499}.

\bibitem{Risaliti:2007aa}
\bibinfo{author}{{Risaliti}, G.} \emph{et~al.}
\newblock {Occultation Measurement of the Size of the X-Ray-emitting Region in
  the Active Galactic Nucleus of NGC 1365}.
\newblock \emph{\bibinfo{journal}{\apjl}} \textbf{\bibinfo{volume}{659}},
  \bibinfo{pages}{L111--L114} (\bibinfo{year}{2007}).
\newblock \eprint{astro-ph/0703173}.

\bibitem{Risaliti:2009mi}
\bibinfo{author}{{Risaliti}, G.} \emph{et~al.}
\newblock {The XMM-Newton long look of NGC 1365: uncovering of the obscured
  X-ray source}.
\newblock \emph{\bibinfo{journal}{\mnras}} \textbf{\bibinfo{volume}{393}},
  \bibinfo{pages}{L1--L5} (\bibinfo{year}{2009}).
\newblock \eprint{0811.1594}.

\bibitem{Haardt:1993ln}
\bibinfo{author}{{Haardt}, F.} \& \bibinfo{author}{{Maraschi}, L.}
\newblock {X-Ray Spectra from Two-Phase Accretion Disks}.
\newblock \emph{\bibinfo{journal}{\apj}} \textbf{\bibinfo{volume}{413}},
  \bibinfo{pages}{507} (\bibinfo{year}{1993}).

\bibitem{Morgan:2010na}
\bibinfo{author}{{Morgan}, C.~W.}, \bibinfo{author}{{Kochanek}, C.~S.},
  \bibinfo{author}{{Morgan}, N.~D.} \& \bibinfo{author}{{Falco}, E.~E.}
\newblock {The Quasar Accretion Disk Size-Black Hole Mass Relation}.
\newblock \emph{\bibinfo{journal}{\apj}} \textbf{\bibinfo{volume}{712}},
  \bibinfo{pages}{1129--1136} (\bibinfo{year}{2010}).
\newblock \eprint{1002.4160}.

\bibitem{Jha:2022yv}
\bibinfo{author}{{Jha}, V.~K.} \emph{et~al.}
\newblock {Accretion disc sizes from continuum reverberation mapping of AGN
  selected from the ZTF survey}.
\newblock \emph{\bibinfo{journal}{\mnras}} \textbf{\bibinfo{volume}{511}},
  \bibinfo{pages}{3005--3016} (\bibinfo{year}{2022}).
\newblock \eprint{2109.05036}.

\bibitem{Kaspi:2005uh}
\bibinfo{author}{{Kaspi}, S.} \emph{et~al.}
\newblock {The Relationship between Luminosity and Broad-Line Region Size in
  Active Galactic Nuclei}.
\newblock \emph{\bibinfo{journal}{\apj}} \textbf{\bibinfo{volume}{629}},
  \bibinfo{pages}{61--71} (\bibinfo{year}{2005}).
\newblock \eprint{astro-ph/0504484}.

\bibitem{Mor:2009kw}
\bibinfo{author}{{Mor}, R.}, \bibinfo{author}{{Netzer}, H.} \&
  \bibinfo{author}{{Elitzur}, M.}
\newblock {Dusty Structure Around Type-I Active Galactic Nuclei: Clumpy Torus
  Narrow-line Region and Near-nucleus Hot Dust}.
\newblock \emph{\bibinfo{journal}{\apj}} \textbf{\bibinfo{volume}{705}},
  \bibinfo{pages}{298--313} (\bibinfo{year}{2009}).
\newblock \eprint{0907.1654}.

\bibitem{Suganuma06}
\bibinfo{author}{{Suganuma}, M.} \emph{et~al.}
\newblock {Reverberation Measurements of the Inner Radius of the Dust Torus in
  Nearby Seyfert 1 Galaxies}.
\newblock \emph{\bibinfo{journal}{\apj}} \textbf{\bibinfo{volume}{639}},
  \bibinfo{pages}{46--63} (\bibinfo{year}{2006}).
\newblock \eprint{astro-ph/0511697}.

\bibitem{Gravity-Collaboration:2020by}
\bibinfo{author}{{Gravity Collaboration}} \emph{et~al.}
\newblock {The resolved size and structure of hot dust in the immediate
  vicinity of AGN}.
\newblock \emph{\bibinfo{journal}{\aap}} \textbf{\bibinfo{volume}{635}},
  \bibinfo{pages}{A92} (\bibinfo{year}{2020}).
\newblock \eprint{1910.00593}.

\bibitem{Tristram:2011zx}
\bibinfo{author}{{Tristram}, K.~R.~W.} \& \bibinfo{author}{{Schartmann}, M.}
\newblock {On the size-luminosity relation of AGN dust tori in the
  mid-infrared}.
\newblock \emph{\bibinfo{journal}{\aap}} \textbf{\bibinfo{volume}{531}},
  \bibinfo{pages}{A99} (\bibinfo{year}{2011}).
\newblock \eprint{1105.4875}.

\bibitem{Ricci:2015tg}
\bibinfo{author}{{Ricci}, C.} \emph{et~al.}
\newblock {Compton-thick Accretion in the Local Universe}.
\newblock \emph{\bibinfo{journal}{\apjl}} \textbf{\bibinfo{volume}{815}},
  \bibinfo{pages}{L13} (\bibinfo{year}{2015}).

\bibitem{Risaliti:2002hj}
\bibinfo{author}{{Risaliti}, G.}, \bibinfo{author}{{Elvis}, M.} \&
  \bibinfo{author}{{Nicastro}, F.}
\newblock {Ubiquitous Variability of X-Ray-absorbing Column Densities in
  Seyfert 2 Galaxies}.
\newblock \emph{\bibinfo{journal}{\apj}} \textbf{\bibinfo{volume}{571}},
  \bibinfo{pages}{234--246} (\bibinfo{year}{2002}).
\newblock \eprint{astro-ph/0107510}.

\bibitem{Burtscher:2016ys}
\bibinfo{author}{{Burtscher}, L.} \emph{et~al.}
\newblock {On the relation of optical obscuration and X-ray absorption in
  Seyfert galaxies}.
\newblock \emph{\bibinfo{journal}{\aap}} \textbf{\bibinfo{volume}{586}},
  \bibinfo{pages}{A28} (\bibinfo{year}{2016}).
\newblock \eprint{1511.05566}.

\bibitem{Maiolino:2001ex}
\bibinfo{author}{{Maiolino}, R.} \emph{et~al.}
\newblock {Dust in active nuclei. I. Evidence for ``anomalous'' properties}.
\newblock \emph{\bibinfo{journal}{\aap}} \textbf{\bibinfo{volume}{365}},
  \bibinfo{pages}{28--36} (\bibinfo{year}{2001}).
\newblock \eprint{astro-ph/0010009}.

\bibitem{Barr:1977rr}
\bibinfo{author}{{Barr}, P.}, \bibinfo{author}{{White}, N.~E.},
  \bibinfo{author}{{Sanford}, P.~W.} \& \bibinfo{author}{{Ives}, J.~C.}
\newblock {An increase in the X-ray absorption of NGC 4151.}
\newblock \emph{\bibinfo{journal}{\mnras}} \textbf{\bibinfo{volume}{181}},
  \bibinfo{pages}{43P--46P} (\bibinfo{year}{1977}).

\bibitem{Ives:1976ty}
\bibinfo{author}{{Ives}, J.~C.}, \bibinfo{author}{{Sanford}, P.~W.} \&
  \bibinfo{author}{{Penston}, M.~V.}
\newblock {The variability and absorption of the X-ray emission from NGC 4151.}
\newblock \emph{\bibinfo{journal}{\apjl}} \textbf{\bibinfo{volume}{207}},
  \bibinfo{pages}{L159--L162} (\bibinfo{year}{1976}).

\bibitem{Yaqoob:1989rk}
\bibinfo{author}{{Yaqoob}, T.}, \bibinfo{author}{{Warwick}, R.~S.} \&
  \bibinfo{author}{{Pounds}, K.~A.}
\newblock {Variable X-ray absorption in NGC 4151.}
\newblock \emph{\bibinfo{journal}{\mnras}} \textbf{\bibinfo{volume}{236}},
  \bibinfo{pages}{153--170} (\bibinfo{year}{1989}).

\bibitem{Holt:1980kc}
\bibinfo{author}{{Holt}, S.~S.} \emph{et~al.}
\newblock {X-ray spectral constraints on the broad-line cloud geometry of NGC
  4151.}
\newblock \emph{\bibinfo{journal}{\apjl}} \textbf{\bibinfo{volume}{241}},
  \bibinfo{pages}{L13--L17} (\bibinfo{year}{1980}).

\bibitem{Malizia:1997vy}
\bibinfo{author}{{Malizia}, A.}, \bibinfo{author}{{Bassani}, L.},
  \bibinfo{author}{{Stephen}, J.~B.}, \bibinfo{author}{{Malaguti}, G.} \&
  \bibinfo{author}{{Palumbo}, G.~G.~C.}
\newblock {High-Energy Spectra of Active Galactic Nuclei. II. Absorption in
  Seyfert Galaxies}.
\newblock \emph{\bibinfo{journal}{\apjs}} \textbf{\bibinfo{volume}{113}},
  \bibinfo{pages}{311--331} (\bibinfo{year}{1997}).

\bibitem{Warwick:1988si}
\bibinfo{author}{{Warwick}, R.~S.}, \bibinfo{author}{{Pounds}, K.~A.} \&
  \bibinfo{author}{{Turner}, T.~J.}
\newblock {Variable low-energy absorption in the X-ray spectrum of ESO
  103-G35.}
\newblock \emph{\bibinfo{journal}{\mnras}} \textbf{\bibinfo{volume}{231}},
  \bibinfo{pages}{1145--1152} (\bibinfo{year}{1988}).

\bibitem{Iyomoto:1997af}
\bibinfo{author}{{Iyomoto}, N.}, \bibinfo{author}{{Makishima}, K.},
  \bibinfo{author}{{Fukazawa}, Y.}, \bibinfo{author}{{Tashiro}, M.} \&
  \bibinfo{author}{{Ishisaki}, Y.}
\newblock {Detection of Strong Fe-K Lines from the Spiral Galaxies NGC 1365 and
  NGC 1386}.
\newblock \emph{\bibinfo{journal}{\pasj}} \textbf{\bibinfo{volume}{49}},
  \bibinfo{pages}{425--434} (\bibinfo{year}{1997}).

\bibitem{Risaliti:2000rg}
\bibinfo{author}{{Risaliti}, G.}, \bibinfo{author}{{Maiolino}, R.} \&
  \bibinfo{author}{{Bassani}, L.}
\newblock {The hard X-ray properties of the Seyfert nucleus in NGC 1365}.
\newblock \emph{\bibinfo{journal}{\aap}} \textbf{\bibinfo{volume}{356}},
  \bibinfo{pages}{33--40} (\bibinfo{year}{2000}).
\newblock \eprint{astro-ph/0002169}.

\bibitem{Risaliti:2005cd}
\bibinfo{author}{{Risaliti}, G.}, \bibinfo{author}{{Elvis}, M.},
  \bibinfo{author}{{Fabbiano}, G.}, \bibinfo{author}{{Baldi}, A.} \&
  \bibinfo{author}{{Zezas}, A.}
\newblock {Rapid Compton-thick/Compton-thin Transitions in the Seyfert 2 Galaxy
  NGC 1365}.
\newblock \emph{\bibinfo{journal}{\apjl}} \textbf{\bibinfo{volume}{623}},
  \bibinfo{pages}{L93--L96} (\bibinfo{year}{2005}).
\newblock \eprint{astro-ph/0503351}.

\bibitem{Guainazzi:2002mz}
\bibinfo{author}{{Guainazzi}, M.}, \bibinfo{author}{{Matt}, G.},
  \bibinfo{author}{{Fiore}, F.} \& \bibinfo{author}{{Perola}, G.~C.}
\newblock {The Phoenix galaxy: UGC 4203 re-birth from its ashes?}
\newblock \emph{\bibinfo{journal}{\aap}} \textbf{\bibinfo{volume}{388}},
  \bibinfo{pages}{787--792} (\bibinfo{year}{2002}).
\newblock \eprint{astro-ph/0204052}.

\bibitem{Matt:2009uk}
\bibinfo{author}{{Matt}, G.} \emph{et~al.}
\newblock {Suzaku observation of the Phoenix galaxy}.
\newblock \emph{\bibinfo{journal}{\aap}} \textbf{\bibinfo{volume}{496}},
  \bibinfo{pages}{653--658} (\bibinfo{year}{2009}).
\newblock \eprint{0902.0930}.

\bibitem{Leighly:1999td}
\bibinfo{author}{{Leighly}, K.~M.} \emph{et~al.}
\newblock {An RXTE Observation of NGC 6300: A New Bright Compton
  Reflection-dominated Seyfert 2 Galaxy}.
\newblock \emph{\bibinfo{journal}{\apj}} \textbf{\bibinfo{volume}{522}},
  \bibinfo{pages}{209--213} (\bibinfo{year}{1999}).
\newblock \eprint{astro-ph/9904155}.

\bibitem{Guainazzi:2002oq}
\bibinfo{author}{{Guainazzi}, M.}
\newblock {The formerly X-ray reflection-dominated Seyfert 2 galaxy NGC 6300}.
\newblock \emph{\bibinfo{journal}{\mnras}} \textbf{\bibinfo{volume}{329}},
  \bibinfo{pages}{L13--L17} (\bibinfo{year}{2002}).
\newblock \eprint{astro-ph/0111148}.

\bibitem{Jana:2020yk}
\bibinfo{author}{{Jana}, A.} \emph{et~al.}
\newblock {Probing the nuclear and circumnuclear properties of NGC 6300 using
  X-ray observations}.
\newblock \emph{\bibinfo{journal}{\mnras}} \textbf{\bibinfo{volume}{499}},
  \bibinfo{pages}{5396--5409} (\bibinfo{year}{2020}).
\newblock \eprint{2008.08033}.

\bibitem{Elvis:2004jg}
\bibinfo{author}{{Elvis}, M.} \emph{et~al.}
\newblock {An Unveiling Event in the Type 2 Active Galactic Nucleus NGC 4388:A
  Challenge for a Parsec-Scale Absorber}.
\newblock \emph{\bibinfo{journal}{\apjl}} \textbf{\bibinfo{volume}{615}},
  \bibinfo{pages}{L25--L28} (\bibinfo{year}{2004}).
\newblock \eprint{astro-ph/0407291}.

\bibitem{Risaliti:2009yq}
\bibinfo{author}{{Risaliti}, G.} \emph{et~al.}
\newblock {Variable Partial Covering and A Relativistic Iron Line in NGC 1365}.
\newblock \emph{\bibinfo{journal}{\apj}} \textbf{\bibinfo{volume}{696}},
  \bibinfo{pages}{160--171} (\bibinfo{year}{2009}).
\newblock \eprint{0901.4809}.

\bibitem{Sanfrutos:2013xt}
\bibinfo{author}{{Sanfrutos}, M.} \emph{et~al.}
\newblock {The size of the X-ray emitting region in SWIFT J2127.4+5654 via a
  broad line region cloud X-ray eclipse}.
\newblock \emph{\bibinfo{journal}{\mnras}} \textbf{\bibinfo{volume}{436}},
  \bibinfo{pages}{1588--1594} (\bibinfo{year}{2013}).
\newblock \eprint{1309.1092}.

\bibitem{Risaliti:2011jl}
\bibinfo{author}{{Risaliti}, G.} \emph{et~al.}
\newblock {X-ray absorption by broad-line region clouds in Mrk 766}.
\newblock \emph{\bibinfo{journal}{\mnras}} \textbf{\bibinfo{volume}{410}},
  \bibinfo{pages}{1027--1035} (\bibinfo{year}{2011}).
\newblock \eprint{1008.5067}.

\bibitem{Gallo:2021bw}
\bibinfo{author}{{Gallo}, L.~C.}, \bibinfo{author}{{Gonzalez}, A.~G.} \&
  \bibinfo{author}{{Miller}, J.~M.}
\newblock {Eclipsing the X-Ray Emitting Region in the Active Galaxy NGC 6814}.
\newblock \emph{\bibinfo{journal}{\apjl}} \textbf{\bibinfo{volume}{908}},
  \bibinfo{pages}{L33} (\bibinfo{year}{2021}).
\newblock \eprint{2101.05433}.

\bibitem{Ricci:2016lq}
\bibinfo{author}{{Ricci}, C.} \emph{et~al.}
\newblock {IC 751: A New Changing Look AGN Discovered by NuSTAR}.
\newblock \emph{\bibinfo{journal}{\apj}} \textbf{\bibinfo{volume}{820}},
  \bibinfo{pages}{5} (\bibinfo{year}{2016}).
\newblock \eprint{1602.00702}.

\bibitem{Lopez-Gonzaga:2017kh}
\bibinfo{author}{{L{\'o}pez-Gonzaga}, N.} \emph{et~al.}
\newblock {NGC 1068: No change in the mid-infrared torus structure despite
  X-ray variability}.
\newblock \emph{\bibinfo{journal}{\aap}} \textbf{\bibinfo{volume}{602}},
  \bibinfo{pages}{A78} (\bibinfo{year}{2017}).
\newblock \eprint{1610.03076}.

\bibitem{Chartas:2009fx}
\bibinfo{author}{{Chartas}, G.}, \bibinfo{author}{{Kochanek}, C.~S.},
  \bibinfo{author}{{Dai}, X.}, \bibinfo{author}{{Poindexter}, S.} \&
  \bibinfo{author}{{Garmire}, G.}
\newblock {X-Ray Microlensing in RXJ1131-1231 and HE1104-1805}.
\newblock \emph{\bibinfo{journal}{\apj}} \textbf{\bibinfo{volume}{693}},
  \bibinfo{pages}{174--185} (\bibinfo{year}{2009}).
\newblock \eprint{0805.4492}.

\bibitem{Maiolino:2010fu}
\bibinfo{author}{{Maiolino}, R.} \emph{et~al.}
\newblock {``Comets'' orbiting a black hole}.
\newblock \emph{\bibinfo{journal}{\aap}} \textbf{\bibinfo{volume}{517}},
  \bibinfo{pages}{A47} (\bibinfo{year}{2010}).
\newblock \eprint{1005.3365}.

\bibitem{King:2015oc}
\bibinfo{author}{{King}, A.} \& \bibinfo{author}{{Pounds}, K.}
\newblock {Powerful Outflows and Feedback from Active Galactic Nuclei}.
\newblock \emph{\bibinfo{journal}{\araa}} \textbf{\bibinfo{volume}{53}},
  \bibinfo{pages}{115--154} (\bibinfo{year}{2015}).
\newblock \eprint{1503.05206}.

\bibitem{Kaastra:2014wa}
\bibinfo{author}{{Kaastra}, J.~S.} \emph{et~al.}
\newblock {A fast and long-lived outflow from the supermassive black hole in
  NGC 5548}.
\newblock \emph{\bibinfo{journal}{Science}} \textbf{\bibinfo{volume}{345}},
  \bibinfo{pages}{64--68} (\bibinfo{year}{2014}).
\newblock \eprint{1406.5007}.

\bibitem{Mehdipour:2017tv}
\bibinfo{author}{{Mehdipour}, M.} \emph{et~al.}
\newblock {Chasing obscuration in type-I AGN: discovery of an eclipsing clumpy
  wind at the outer broad-line region of NGC 3783}.
\newblock \emph{\bibinfo{journal}{\aap}} \textbf{\bibinfo{volume}{607}},
  \bibinfo{pages}{A28} (\bibinfo{year}{2017}).
\newblock \eprint{1707.04671}.

\bibitem{Kaastra:2018vu}
\bibinfo{author}{{Kaastra}, J.~S.} \emph{et~al.}
\newblock {Recurring obscuration in NGC 3783}.
\newblock \emph{\bibinfo{journal}{\aap}} \textbf{\bibinfo{volume}{619}},
  \bibinfo{pages}{A112} (\bibinfo{year}{2018}).
\newblock \eprint{1805.03538}.

\bibitem{Beuchert:2015fg}
\bibinfo{author}{{Beuchert}, T.} \emph{et~al.}
\newblock {A variable-density absorption event in NGC 3227 mapped with Suzaku
  and Swift}.
\newblock \emph{\bibinfo{journal}{\aap}} \textbf{\bibinfo{volume}{584}},
  \bibinfo{pages}{A82} (\bibinfo{year}{2015}).
\newblock \eprint{1508.04565}.

\bibitem{Kara:2021bk}
\bibinfo{author}{{Kara}, E.} \emph{et~al.}
\newblock {AGN STORM 2. I. First results: A Change in the Weather of Mrk 817}.
\newblock \emph{\bibinfo{journal}{\apj}} \textbf{\bibinfo{volume}{922}},
  \bibinfo{pages}{151} (\bibinfo{year}{2021}).
\newblock \eprint{2105.05840}.

\bibitem{Matt:2003tc}
\bibinfo{author}{{Matt}, G.}, \bibinfo{author}{{Guainazzi}, M.} \&
  \bibinfo{author}{{Maiolino}, R.}
\newblock {Changing look: from Compton-thick to Compton-thin, or the rebirth of
  fossil active galactic nuclei}.
\newblock \emph{\bibinfo{journal}{\mnras}} \textbf{\bibinfo{volume}{342}},
  \bibinfo{pages}{422--426} (\bibinfo{year}{2003}).
\newblock \eprint{astro-ph/0302328}.

\bibitem{Gilli:2000bw}
\bibinfo{author}{{Gilli}, R.} \emph{et~al.}
\newblock {The variability of the Seyfert galaxy NGC 2992: the case for a
  revived AGN}.
\newblock \emph{\bibinfo{journal}{\aap}} \textbf{\bibinfo{volume}{355}},
  \bibinfo{pages}{485--498} (\bibinfo{year}{2000}).
\newblock \eprint{astro-ph/0001107}.

\bibitem{Guainazzi:1998yy}
\bibinfo{author}{{Guainazzi}, M.} \emph{et~al.}
\newblock {A swan song: the disappearance of the nucleus of NGC 4051 and the
  echo of its past glory}.
\newblock \emph{\bibinfo{journal}{\mnras}} \textbf{\bibinfo{volume}{301}},
  \bibinfo{pages}{L1--L4} (\bibinfo{year}{1998}).
\newblock \eprint{astro-ph/9807213}.

\bibitem{Uttley:1999dy}
\bibinfo{author}{{Uttley}, P.}, \bibinfo{author}{{McHardy}, I.~M.},
  \bibinfo{author}{{Papadakis}, I.~E.}, \bibinfo{author}{{Guainazzi}, M.} \&
  \bibinfo{author}{{Fruscione}, A.}
\newblock {The swan song in context: long-time-scale X-ray variability of NGC
  4051}.
\newblock \emph{\bibinfo{journal}{\mnras}} \textbf{\bibinfo{volume}{307}},
  \bibinfo{pages}{L6--L10} (\bibinfo{year}{1999}).
\newblock \eprint{astro-ph/9905104}.

\bibitem{Puccetti:2007zr}
\bibinfo{author}{{Puccetti}, S.} \emph{et~al.}
\newblock {Rapid N$_{H}$ changes in NGC 4151}.
\newblock \emph{\bibinfo{journal}{\mnras}} \textbf{\bibinfo{volume}{377}},
  \bibinfo{pages}{607--616} (\bibinfo{year}{2007}).
\newblock \eprint{astro-ph/0612021}.

\bibitem{Schmid:2003nk}
\bibinfo{author}{{Schmid}, H.~M.}, \bibinfo{author}{{Appenzeller}, I.} \&
  \bibinfo{author}{{Burch}, U.}
\newblock {Spectropolarimetry of the borderline Seyfert 1 galaxy ESO 323-G077}.
\newblock \emph{\bibinfo{journal}{\aap}} \textbf{\bibinfo{volume}{404}},
  \bibinfo{pages}{505--511} (\bibinfo{year}{2003}).
\newblock \eprint{astro-ph/0304439}.

\bibitem{Markowitz:2014oq}
\bibinfo{author}{{Markowitz}, A.~G.}, \bibinfo{author}{{Krumpe}, M.} \&
  \bibinfo{author}{{Nikutta}, R.}
\newblock {First X-ray-based statistical tests for clumpy-torus models: eclipse
  events from 230 years of monitoring of Seyfert AGN}.
\newblock \emph{\bibinfo{journal}{\mnras}} \textbf{\bibinfo{volume}{439}},
  \bibinfo{pages}{1403--1458} (\bibinfo{year}{2014}).
\newblock \eprint{1402.2779}.

\bibitem{Torricelli-Ciamponi:2014do}
\bibinfo{author}{{Torricelli-Ciamponi}, G.}, \bibinfo{author}{{Pietrini}, P.},
  \bibinfo{author}{{Risaliti}, G.} \& \bibinfo{author}{{Salvati}, M.}
\newblock {Search for X-ray occultations in active galactic nuclei}.
\newblock \emph{\bibinfo{journal}{\mnras}} \textbf{\bibinfo{volume}{442}},
  \bibinfo{pages}{2116--2130} (\bibinfo{year}{2014}).
\newblock \eprint{1405.4660}.

\bibitem{Laha:2020sd}
\bibinfo{author}{{Laha}, S.} \emph{et~al.}
\newblock {The Variable and Non-variable X-Ray Absorbers in Compton-thin Type
  II Active Galactic Nuclei}.
\newblock \emph{\bibinfo{journal}{\apj}} \textbf{\bibinfo{volume}{897}},
  \bibinfo{pages}{66} (\bibinfo{year}{2020}).
\newblock \eprint{2005.06079}.

\bibitem{Marinucci:2013iz}
\bibinfo{author}{{Marinucci}, A.} \emph{et~al.}
\newblock {X-ray absorption variability in NGC 4507}.
\newblock \emph{\bibinfo{journal}{\mnras}} \textbf{\bibinfo{volume}{429}},
  \bibinfo{pages}{2581--2586} (\bibinfo{year}{2013}).
\newblock \eprint{1212.4151}.

\bibitem{Piconcelli:2007bh}
\bibinfo{author}{{Piconcelli}, E.}, \bibinfo{author}{{Bianchi}, S.},
  \bibinfo{author}{{Guainazzi}, M.}, \bibinfo{author}{{Fiore}, F.} \&
  \bibinfo{author}{{Chiaberge}, M.}
\newblock {XMM-Newton broad-band observations of NGC 7582: N$\{$H$\}$
  variations and fading out of the active nucleus}.
\newblock \emph{\bibinfo{journal}{\aap}} \textbf{\bibinfo{volume}{466}},
  \bibinfo{pages}{855--863} (\bibinfo{year}{2007}).
\newblock \eprint{astro-ph/0702564}.

\bibitem{Bianchi:2009au}
\bibinfo{author}{{Bianchi}, S.} \emph{et~al.}
\newblock {How Complex is the Obscuration in Active Galactic Nuclei? New Clues
  from the Suzaku Monitoring of the X-Ray Absorbers in NGC 7582}.
\newblock \emph{\bibinfo{journal}{\apj}} \textbf{\bibinfo{volume}{695}},
  \bibinfo{pages}{781--787} (\bibinfo{year}{2009}).
\newblock \eprint{0901.1973}.

\bibitem{Netzer:2013ix}
\bibinfo{author}{{Netzer}, H.}
\newblock \emph{\bibinfo{title}{{The Physics and Evolution of Active Galactic
  Nuclei}}} (\bibinfo{year}{2013}).

\bibitem{Czerny:2011ek}
\bibinfo{author}{{Czerny}, B.} \& \bibinfo{author}{{Hryniewicz}, K.}
\newblock {The origin of the broad line region in active galactic nuclei}.
\newblock \emph{\bibinfo{journal}{\aap}} \textbf{\bibinfo{volume}{525}},
  \bibinfo{pages}{L8} (\bibinfo{year}{2011}).
\newblock \eprint{1010.6201}.

\bibitem{Shen:2007dh}
\bibinfo{author}{{Shen}, Y.} \emph{et~al.}
\newblock {Clustering of High-Redshift (z >= 2.9) Quasars from the Sloan
  Digital Sky Survey}.
\newblock \emph{\bibinfo{journal}{\aj}} \textbf{\bibinfo{volume}{133}},
  \bibinfo{pages}{2222--2241} (\bibinfo{year}{2007}).
\newblock \eprint{astro-ph/0702214}.

\bibitem{Schawinski:2015cs}
\bibinfo{author}{{Schawinski}, K.}, \bibinfo{author}{{Koss}, M.},
  \bibinfo{author}{{Berney}, S.} \& \bibinfo{author}{{Sartori}, L.~F.}
\newblock {Active galactic nuclei flicker: an observational estimate of the
  duration of black hole growth phases of $\sim10^{5}$ yr}.
\newblock \emph{\bibinfo{journal}{\mnras}} \textbf{\bibinfo{volume}{451}},
  \bibinfo{pages}{2517--2523} (\bibinfo{year}{2015}).
\newblock \eprint{1505.06733}.

\bibitem{Lawrence:2016bi}
\bibinfo{author}{{Lawrence}, A.} \emph{et~al.}
\newblock {Slow-blue nuclear hypervariables in PanSTARRS-1}.
\newblock \emph{\bibinfo{journal}{\mnras}} \textbf{\bibinfo{volume}{463}},
  \bibinfo{pages}{296--331} (\bibinfo{year}{2016}).
\newblock \eprint{1605.07842}.

\bibitem{Rumbaugh:2018iv}
\bibinfo{author}{{Rumbaugh}, N.} \emph{et~al.}
\newblock {Extreme Variability Quasars from the Sloan Digital Sky Survey and
  the Dark Energy Survey}.
\newblock \emph{\bibinfo{journal}{\apj}} \textbf{\bibinfo{volume}{854}},
  \bibinfo{pages}{160} (\bibinfo{year}{2018}).
\newblock \eprint{1706.07875}.

\bibitem{Trakhtenbrot:2019wj}
\bibinfo{author}{{Trakhtenbrot}, B.} \emph{et~al.}
\newblock {A new class of flares from accreting supermassive black holes}.
\newblock \emph{\bibinfo{journal}{Nature Astronomy}}
  \textbf{\bibinfo{volume}{3}}, \bibinfo{pages}{242--250}
  (\bibinfo{year}{2019}).
\newblock \eprint{1901.03731}.

\bibitem{Shen:2021se}
\bibinfo{author}{{Shen}, Y.}
\newblock {Extreme Variability and Episodic Lifetime of Quasars}.
\newblock \emph{\bibinfo{journal}{\apj}} \textbf{\bibinfo{volume}{921}},
  \bibinfo{pages}{70} (\bibinfo{year}{2021}).
\newblock \eprint{2108.05381}.

\bibitem{Timlin:2020bv}
\bibinfo{author}{{Timlin}, I., John~D.} \emph{et~al.}
\newblock {The frequency of extreme X-ray variability for radio-quiet quasars}.
\newblock \emph{\bibinfo{journal}{\mnras}} \textbf{\bibinfo{volume}{498}},
  \bibinfo{pages}{4033--4050} (\bibinfo{year}{2020}).
\newblock \eprint{2008.12778}.

\bibitem{Tohline:1976gd}
\bibinfo{author}{{Tohline}, J.~E.} \& \bibinfo{author}{{Osterbrock}, D.~E.}
\newblock {Variation of the spectrum of the Seyfert galaxy NGC 7603.}
\newblock \emph{\bibinfo{journal}{\apjl}} \textbf{\bibinfo{volume}{210}},
  \bibinfo{pages}{L117--L120} (\bibinfo{year}{1976}).

\bibitem{Ward:1980ed}
\bibinfo{author}{{Ward}, M.}, \bibinfo{author}{{Penston}, M.~V.},
  \bibinfo{author}{{Blades}, J.~C.} \& \bibinfo{author}{{Turtle}, A.~J.}
\newblock {New optical and radio observations of the X-ray galaxies NGC 7582
  andNGC 2992.}
\newblock \emph{\bibinfo{journal}{\mnras}} \textbf{\bibinfo{volume}{193}},
  \bibinfo{pages}{563--582} (\bibinfo{year}{1980}).

\bibitem{Allen:1999yl}
\bibinfo{author}{{Allen}, M.~G.}, \bibinfo{author}{{Dopita}, M.~A.},
  \bibinfo{author}{{Tsvetanov}, Z.~I.} \& \bibinfo{author}{{Sutherland}, R.~S.}
\newblock {Physical Conditions in the Seyfert Galaxy NGC 2992}.
\newblock \emph{\bibinfo{journal}{\apj}} \textbf{\bibinfo{volume}{511}},
  \bibinfo{pages}{686--708} (\bibinfo{year}{1999}).
\newblock \eprint{astro-ph/9809123}.

\bibitem{Mushotzky:1982nt}
\bibinfo{author}{{Mushotzky}, R.~F.}
\newblock {The X-ray spectrum and time variability of narrow emission line
  galaxies.}
\newblock \emph{\bibinfo{journal}{\apj}} \textbf{\bibinfo{volume}{256}},
  \bibinfo{pages}{92--102} (\bibinfo{year}{1982}).

\bibitem{Penston:1984tl}
\bibinfo{author}{{Penston}, M.~V.} \& \bibinfo{author}{{Perez}, E.}
\newblock {An evolutionary link between Seyfert I and II galaxies.}
\newblock \emph{\bibinfo{journal}{\mnras}} \textbf{\bibinfo{volume}{211}},
  \bibinfo{pages}{33P--39} (\bibinfo{year}{1984}).

\bibitem{Gregory:1991dq}
\bibinfo{author}{{Gregory}, S.~A.}, \bibinfo{author}{{Tifft}, W.~G.} \&
  \bibinfo{author}{{Cocke}, W.~J.}
\newblock {Variability of the Broad Line Spectrum of Markarian 372}.
\newblock \emph{\bibinfo{journal}{\aj}} \textbf{\bibinfo{volume}{102}},
  \bibinfo{pages}{1977} (\bibinfo{year}{1991}).

\bibitem{Tran:1992hb}
\bibinfo{author}{{Tran}, H.~D.}, \bibinfo{author}{{Osterbrock}, D.~E.} \&
  \bibinfo{author}{{Martel}, A.}
\newblock {Extreme Spectral Variations of the Seyfert Galaxy Markarian 993}.
\newblock \emph{\bibinfo{journal}{\aj}} \textbf{\bibinfo{volume}{104}},
  \bibinfo{pages}{2072} (\bibinfo{year}{1992}).

\bibitem{Aretxaga:1999ro}
\bibinfo{author}{{Aretxaga}, I.}, \bibinfo{author}{{Joguet}, B.},
  \bibinfo{author}{{Kunth}, D.}, \bibinfo{author}{{Melnick}, J.} \&
  \bibinfo{author}{{Terlevich}, R.~J.}
\newblock {Seyfert 1 Mutation of the Classical Seyfert 2 Nucleus NGC 7582}.
\newblock \emph{\bibinfo{journal}{\apjl}} \textbf{\bibinfo{volume}{519}},
  \bibinfo{pages}{L123--L126} (\bibinfo{year}{1999}).
\newblock \eprint{astro-ph/9905147}.

\bibitem{Eracleous:2001pu}
\bibinfo{author}{{Eracleous}, M.} \& \bibinfo{author}{{Halpern}, J.~P.}
\newblock {NGC 3065: A Certified LINER with Broad, Variable Balmer Lines}.
\newblock \emph{\bibinfo{journal}{\apj}} \textbf{\bibinfo{volume}{554}},
  \bibinfo{pages}{240--244} (\bibinfo{year}{2001}).
\newblock \eprint{astro-ph/0101050}.

\bibitem{Cohen:1986sa}
\bibinfo{author}{{Cohen}, R.~D.}, \bibinfo{author}{{Rudy}, R.~J.},
  \bibinfo{author}{{Puetter}, R.~C.}, \bibinfo{author}{{Ake}, T.~B.} \&
  \bibinfo{author}{{Foltz}, C.~B.}
\newblock {Variability of Markarian 1018: Seyfert 1.9 to Seyfert 1}.
\newblock \emph{\bibinfo{journal}{\apj}} \textbf{\bibinfo{volume}{311}},
  \bibinfo{pages}{135} (\bibinfo{year}{1986}).

\bibitem{McElroy:2016hs}
\bibinfo{author}{{McElroy}, R.~E.} \emph{et~al.}
\newblock {The Close AGN Reference Survey (CARS). Mrk 1018 returns to the
  shadows after 30 years as a Seyfert 1}.
\newblock \emph{\bibinfo{journal}{\aap}} \textbf{\bibinfo{volume}{593}},
  \bibinfo{pages}{L8} (\bibinfo{year}{2016}).
\newblock \eprint{1609.04423}.

\bibitem{Oknyansky:2019le}
\bibinfo{author}{{Oknyansky}, V.~L.} \emph{et~al.}
\newblock {New changing look case in NGC 1566}.
\newblock \emph{\bibinfo{journal}{\mnras}} \textbf{\bibinfo{volume}{483}},
  \bibinfo{pages}{558--564} (\bibinfo{year}{2019}).
\newblock \eprint{1811.06926}.

\bibitem{Alloin:1986iu}
\bibinfo{author}{{Alloin}, D.}, \bibinfo{author}{{Pelat}, D.},
  \bibinfo{author}{{Phillips}, M.~M.}, \bibinfo{author}{{Fosbury}, R.~A.~E.} \&
  \bibinfo{author}{{Freeman}, K.}
\newblock {Recurrent Outbursts in the Broad-Line Region of NGC 1566}.
\newblock \emph{\bibinfo{journal}{\apj}} \textbf{\bibinfo{volume}{308}},
  \bibinfo{pages}{23} (\bibinfo{year}{1986}).

\bibitem{Osterbrock:1977pd}
\bibinfo{author}{{Osterbrock}, D.~E.}
\newblock {Spectrophotometry of Seyfert 1 galaxies.}
\newblock \emph{\bibinfo{journal}{\apj}} \textbf{\bibinfo{volume}{215}},
  \bibinfo{pages}{733--745} (\bibinfo{year}{1977}).

\bibitem{Shapovalova:2010an}
\bibinfo{author}{{Shapovalova}, A.~I.} \emph{et~al.}
\newblock {Long-term variability of the optical spectra of NGC 4151. II.
  Evolution of the broad H{\ensuremath{\alpha}} and H{\ensuremath{\beta}}
  emission-line profiles}.
\newblock \emph{\bibinfo{journal}{\aap}} \textbf{\bibinfo{volume}{509}},
  \bibinfo{pages}{A106} (\bibinfo{year}{2010}).
\newblock \eprint{0910.2980}.

\bibitem{Denney:2014do}
\bibinfo{author}{{Denney}, K.~D.} \emph{et~al.}
\newblock {The Typecasting of Active Galactic Nuclei: Mrk 590 no Longer Fits
  the Role}.
\newblock \emph{\bibinfo{journal}{\apj}} \textbf{\bibinfo{volume}{796}},
  \bibinfo{pages}{134} (\bibinfo{year}{2014}).
\newblock \eprint{1404.4879}.

\bibitem{Ilic:2020jj}
\bibinfo{author}{{Ili{\'c}}, D.} \emph{et~al.}
\newblock {A flare in the optical spotted in the changing-look Seyfert NGC
  3516}.
\newblock \emph{\bibinfo{journal}{\aap}} \textbf{\bibinfo{volume}{638}},
  \bibinfo{pages}{A13} (\bibinfo{year}{2020}).
\newblock \eprint{2004.01308}.

\bibitem{LaMassa:2015rt}
\bibinfo{author}{{LaMassa}, S.~M.} \emph{et~al.}
\newblock {The Discovery of the First {\textquotedblleft}Changing
  Look{\textquotedblright} Quasar: New Insights Into the Physics and
  Phenomenology of Active Galactic Nucleus}.
\newblock \emph{\bibinfo{journal}{\apj}} \textbf{\bibinfo{volume}{800}},
  \bibinfo{pages}{144} (\bibinfo{year}{2015}).
\newblock \eprint{1412.2136}.

\bibitem{Ricci:2020fp}
\bibinfo{author}{{Ricci}, C.} \emph{et~al.}
\newblock {The Destruction and Recreation of the X-Ray Corona in a
  Changing-look Active Galactic Nucleus}.
\newblock \emph{\bibinfo{journal}{\apjl}} \textbf{\bibinfo{volume}{898}},
  \bibinfo{pages}{L1} (\bibinfo{year}{2020}).
\newblock \eprint{2007.07275}.

\bibitem{MacLeod:2016pu}
\bibinfo{author}{{MacLeod}, C.~L.} \emph{et~al.}
\newblock {A systematic search for changing-look quasars in SDSS}.
\newblock \emph{\bibinfo{journal}{\mnras}} \textbf{\bibinfo{volume}{457}},
  \bibinfo{pages}{389--404} (\bibinfo{year}{2016}).
\newblock \eprint{1509.08393}.

\bibitem{Shakura:1973ex}
\bibinfo{author}{{Shakura}, N.~I.} \& \bibinfo{author}{{Sunyaev}, R.~A.}
\newblock {Reprint of 1973A\&A....24..337S. Black holes in binary systems.
  Observational appearance.}
\newblock \emph{\bibinfo{journal}{\aap}} \textbf{\bibinfo{volume}{500}},
  \bibinfo{pages}{33--51} (\bibinfo{year}{1973}).

\bibitem{Trakhtenbrot:2019ay}
\bibinfo{author}{{Trakhtenbrot}, B.} \emph{et~al.}
\newblock {1ES 1927+654: An AGN Caught Changing Look on a Timescale of Months}.
\newblock \emph{\bibinfo{journal}{\apj}} \textbf{\bibinfo{volume}{883}},
  \bibinfo{pages}{94} (\bibinfo{year}{2019}).
\newblock \eprint{1903.11084}.

\bibitem{York:2000kn}
\bibinfo{author}{{York}, D.~G.} \emph{et~al.}
\newblock {The Sloan Digital Sky Survey: Technical Summary}.
\newblock \emph{\bibinfo{journal}{\aj}} \textbf{\bibinfo{volume}{120}},
  \bibinfo{pages}{1579--1587} (\bibinfo{year}{2000}).
\newblock \eprint{astro-ph/0006396}.

\bibitem{Abazajian:2003iw}
\bibinfo{author}{{Abazajian}, K.} \emph{et~al.}
\newblock {The First Data Release of the Sloan Digital Sky Survey}.
\newblock \emph{\bibinfo{journal}{\aj}} \textbf{\bibinfo{volume}{126}},
  \bibinfo{pages}{2081--2086} (\bibinfo{year}{2003}).
\newblock \eprint{astro-ph/0305492}.

\bibitem{Runnoe:2016hb}
\bibinfo{author}{{Runnoe}, J.~C.} \emph{et~al.}
\newblock {Now you see it, now you don't: the disappearing central engine of
  the quasar J1011+5442}.
\newblock \emph{\bibinfo{journal}{\mnras}} \textbf{\bibinfo{volume}{455}},
  \bibinfo{pages}{1691--1701} (\bibinfo{year}{2016}).
\newblock \eprint{1509.03640}.

\bibitem{Yang:2018mn}
\bibinfo{author}{{Yang}, Q.} \emph{et~al.}
\newblock {Discovery of 21 New Changing-look AGNs in the Northern Sky}.
\newblock \emph{\bibinfo{journal}{\apj}} \textbf{\bibinfo{volume}{862}},
  \bibinfo{pages}{109} (\bibinfo{year}{2018}).
\newblock \eprint{1711.08122}.

\bibitem{MacLeod:2019rk}
\bibinfo{author}{{MacLeod}, C.~L.} \emph{et~al.}
\newblock {Changing-look Quasar Candidates: First Results from Follow-up
  Spectroscopy of Highly Optically Variable Quasars}.
\newblock \emph{\bibinfo{journal}{\apj}} \textbf{\bibinfo{volume}{874}},
  \bibinfo{pages}{8} (\bibinfo{year}{2019}).
\newblock \eprint{1810.00087}.

\bibitem{Green:2022cv}
\bibinfo{author}{{Green}, P.~J.} \emph{et~al.}
\newblock {The Time Domain Spectroscopic Survey: Changing-look Quasar
  Candidates from Multi-epoch Spectroscopy in SDSS-IV}.
\newblock \emph{\bibinfo{journal}{\apj}} \textbf{\bibinfo{volume}{933}},
  \bibinfo{pages}{180} (\bibinfo{year}{2022}).
\newblock \eprint{2201.09123}.

\bibitem{Guo2020_hiz_CLQs}
\bibinfo{author}{{Guo}, H.} \emph{et~al.}
\newblock {High-redshift Extreme Variability Quasars from Sloan Digital Sky
  Survey Multiepoch Spectroscopy}.
\newblock \emph{\bibinfo{journal}{\apj}} \textbf{\bibinfo{volume}{905}},
  \bibinfo{pages}{52} (\bibinfo{year}{2020}).
\newblock \eprint{2006.08645}.

\bibitem{Ross:2020gn}
\bibinfo{author}{{Ross}, N.~P.} \emph{et~al.}
\newblock {The first high-redshift changing-look quasars}.
\newblock \emph{\bibinfo{journal}{\mnras}} \textbf{\bibinfo{volume}{498}},
  \bibinfo{pages}{2339--2353} (\bibinfo{year}{2020}).
\newblock \eprint{1912.05310}.

\bibitem{Sun2015}
\bibinfo{author}{{Sun}, M.} \emph{et~al.}
\newblock {The Sloan Digital Sky Survey Reverberation Mapping Project: Ensemble
  Spectroscopic Variability of Quasar Broad Emission Lines}.
\newblock \emph{\bibinfo{journal}{\apj}} \textbf{\bibinfo{volume}{811}},
  \bibinfo{pages}{42} (\bibinfo{year}{2015}).
\newblock \eprint{1506.07886}.

\bibitem{Roig2014}
\bibinfo{author}{{Roig}, B.}, \bibinfo{author}{{Blanton}, M.~R.} \&
  \bibinfo{author}{{Ross}, N.~P.}
\newblock {Unusual Broad-line Mg II Emitters among Luminous Galaxies in the
  Baryon Oscillation Spectroscopic Survey}.
\newblock \emph{\bibinfo{journal}{\apj}} \textbf{\bibinfo{volume}{781}},
  \bibinfo{pages}{72} (\bibinfo{year}{2014}).

\bibitem{Guo:2020ll}
\bibinfo{author}{{Guo}, H.} \emph{et~al.}
\newblock {Understanding Broad Mg II Variability in Quasars with
  Photoionization: Implications for Reverberation Mapping and Changing-look
  Quasars}.
\newblock \emph{\bibinfo{journal}{\apj}} \textbf{\bibinfo{volume}{888}},
  \bibinfo{pages}{58} (\bibinfo{year}{2020}).
\newblock \eprint{1907.06669}.

\bibitem{KoristaGoad2004}
\bibinfo{author}{{Korista}, K.~T.} \& \bibinfo{author}{{Goad}, M.~R.}
\newblock {What the Optical Recombination Lines Can Tell Us about the
  Broad-Line Regions of Active Galactic Nuclei}.
\newblock \emph{\bibinfo{journal}{\apj}} \textbf{\bibinfo{volume}{606}},
  \bibinfo{pages}{749--762} (\bibinfo{year}{2004}).
\newblock \eprint{astro-ph/0402506}.

\bibitem{Yang2020_MgII}
\bibinfo{author}{{Yang}, Q.} \emph{et~al.}
\newblock {Spectral variability of a sample of extreme variability quasars and
  implications for the Mg II broad-line region}.
\newblock \emph{\bibinfo{journal}{\mnras}} \textbf{\bibinfo{volume}{493}},
  \bibinfo{pages}{5773--5787} (\bibinfo{year}{2020}).
\newblock \eprint{1904.10912}.

\bibitem{Clavel1991}
\bibinfo{author}{{Clavel}, J.} \emph{et~al.}
\newblock {Steps toward Determination of the Size and Structure of the
  Broad-Line Region in Active Galactic Nuclei. I. an 8 Month Campaign of
  Monitoring NGC 5548 with IUE}.
\newblock \emph{\bibinfo{journal}{\apj}} \textbf{\bibinfo{volume}{366}},
  \bibinfo{pages}{64} (\bibinfo{year}{1991}).

\bibitem{Cackett2015}
\bibinfo{author}{{Cackett}, E.~M.} \emph{et~al.}
\newblock {Swift/UVOT Grism Monitoring of NGC 5548 in 2013: An Attempt at MgII
  Reverberation Mapping}.
\newblock \emph{\bibinfo{journal}{\apj}} \textbf{\bibinfo{volume}{810}},
  \bibinfo{pages}{86} (\bibinfo{year}{2015}).
\newblock \eprint{1503.02029}.

\bibitem{Runco:2016po}
\bibinfo{author}{{Runco}, J.~N.} \emph{et~al.}
\newblock {Broad H{\ensuremath{\beta}} Emission-line Variability in a Sample of
  102 Local Active Galaxies}.
\newblock \emph{\bibinfo{journal}{\apj}} \textbf{\bibinfo{volume}{821}},
  \bibinfo{pages}{33} (\bibinfo{year}{2016}).
\newblock \eprint{1603.00035}.

\bibitem{Caplar2017}
\bibinfo{author}{{Caplar}, N.}, \bibinfo{author}{{Lilly}, S.~J.} \&
  \bibinfo{author}{{Trakhtenbrot}, B.}
\newblock {Optical Variability of AGNs in the PTF/iPTF Survey}.
\newblock \emph{\bibinfo{journal}{\apj}} \textbf{\bibinfo{volume}{834}},
  \bibinfo{pages}{111} (\bibinfo{year}{2017}).
\newblock \eprint{1611.03082}.

\bibitem{Drake:2009ts}
\bibinfo{author}{{Drake}, A.~J.} \emph{et~al.}
\newblock {First Results from the Catalina Real-Time Transient Survey}.
\newblock \emph{\bibinfo{journal}{\apj}} \textbf{\bibinfo{volume}{696}},
  \bibinfo{pages}{870--884} (\bibinfo{year}{2009}).
\newblock \eprint{0809.1394}.

\bibitem{Law:2009it}
\bibinfo{author}{{Law}, N.~M.} \emph{et~al.}
\newblock {The Palomar Transient Factory: System Overview, Performance, and
  First Results}.
\newblock \emph{\bibinfo{journal}{\pasp}} \textbf{\bibinfo{volume}{121}},
  \bibinfo{pages}{1395} (\bibinfo{year}{2009}).
\newblock \eprint{0906.5350}.

\bibitem{Kaiser:2010wj}
\bibinfo{author}{{Kaiser}, N.} \emph{et~al.}
\newblock {The Pan-STARRS wide-field optical/NIR imaging survey}.
\newblock In \bibinfo{editor}{{Stepp}, L.~M.}, \bibinfo{editor}{{Gilmozzi}, R.}
  \& \bibinfo{editor}{{Hall}, H.~J.} (eds.)
  \emph{\bibinfo{booktitle}{Ground-based and Airborne Telescopes III}}, vol.
  \bibinfo{volume}{7733} of \emph{\bibinfo{series}{Society of Photo-Optical
  Instrumentation Engineers (SPIE) Conference Series}}, \bibinfo{pages}{77330E}
  (\bibinfo{year}{2010}).

\bibitem{Shappee:2014yp}
\bibinfo{author}{{Shappee}, B.~J.} \emph{et~al.}
\newblock {The Man behind the Curtain: X-Rays Drive the UV through NIR
  Variability in the 2013 Active Galactic Nucleus Outburst in NGC 2617}.
\newblock \emph{\bibinfo{journal}{\apj}} \textbf{\bibinfo{volume}{788}},
  \bibinfo{pages}{48} (\bibinfo{year}{2014}).
\newblock \eprint{1310.2241}.

\bibitem{Bellm:2019il}
\bibinfo{author}{{Bellm}, E.~C.} \emph{et~al.}
\newblock {The Zwicky Transient Facility: System Overview, Performance, and
  First Results}.
\newblock \emph{\bibinfo{journal}{\pasp}} \textbf{\bibinfo{volume}{131}},
  \bibinfo{pages}{018002} (\bibinfo{year}{2019}).
\newblock \eprint{1902.01932}.

\bibitem{Gallo:2013vw}
\bibinfo{author}{{Gallo}, L.~C.} \emph{et~al.}
\newblock {1ES 1927+654: a bare Seyfert 2}.
\newblock \emph{\bibinfo{journal}{\mnras}} \textbf{\bibinfo{volume}{433}},
  \bibinfo{pages}{421--433} (\bibinfo{year}{2013}).
\newblock \eprint{1304.7155}.

\bibitem{Gezari:2017md}
\bibinfo{author}{{Gezari}, S.} \emph{et~al.}
\newblock {iPTF Discovery of the Rapid
  {\textquotedblleft}Turn-on{\textquotedblright} of a Luminous Quasar}.
\newblock \emph{\bibinfo{journal}{\apj}} \textbf{\bibinfo{volume}{835}},
  \bibinfo{pages}{144} (\bibinfo{year}{2017}).
\newblock \eprint{1612.04830}.

\bibitem{Frederick:2019vf}
\bibinfo{author}{{Frederick}, S.} \emph{et~al.}
\newblock {A New Class of Changing-look LINERs}.
\newblock \emph{\bibinfo{journal}{\apj}} \textbf{\bibinfo{volume}{883}},
  \bibinfo{pages}{31} (\bibinfo{year}{2019}).
\newblock \eprint{1904.10973}.

\bibitem{Assef:2018vj}
\bibinfo{author}{{Assef}, R.~J.} \emph{et~al.}
\newblock {The WISE AGN Catalog}.
\newblock \emph{\bibinfo{journal}{\apjs}} \textbf{\bibinfo{volume}{234}},
  \bibinfo{pages}{23} (\bibinfo{year}{2018}).
\newblock \eprint{1706.09901}.

\bibitem{Stern:2018su}
\bibinfo{author}{{Stern}, D.} \emph{et~al.}
\newblock {A Mid-IR Selected Changing-look Quasar and Physical Scenarios for
  Abrupt AGN Fading}.
\newblock \emph{\bibinfo{journal}{\apj}} \textbf{\bibinfo{volume}{864}},
  \bibinfo{pages}{27} (\bibinfo{year}{2018}).
\newblock \eprint{1805.06920}.

\bibitem{Ross:2018ah}
\bibinfo{author}{{Ross}, N.~P.} \emph{et~al.}
\newblock {A new physical interpretation of optical and infrared variability in
  quasars}.
\newblock \emph{\bibinfo{journal}{\mnras}} \textbf{\bibinfo{volume}{480}},
  \bibinfo{pages}{4468--4479} (\bibinfo{year}{2018}).
\newblock \eprint{1805.06921}.

\bibitem{Shen2021_bias}
\bibinfo{author}{{Shen}, Y.} \& \bibinfo{author}{{Burke}, C.~J.}
\newblock {A Sample Bias in Quasar Variability Studies}.
\newblock \emph{\bibinfo{journal}{\apjl}} \textbf{\bibinfo{volume}{918}},
  \bibinfo{pages}{L19} (\bibinfo{year}{2021}).
\newblock \eprint{2108.05391}.

\bibitem{Noda:2018jz}
\bibinfo{author}{{Noda}, H.} \& \bibinfo{author}{{Done}, C.}
\newblock {Explaining changing-look AGN with state transition triggered by
  rapid mass accretion rate drop}.
\newblock \emph{\bibinfo{journal}{\mnras}} \textbf{\bibinfo{volume}{480}},
  \bibinfo{pages}{3898--3906} (\bibinfo{year}{2018}).
\newblock \eprint{1805.07873}.

\bibitem{Ricci:2021ro}
\bibinfo{author}{{Ricci}, C.} \emph{et~al.}
\newblock {The 450 Day X-Ray Monitoring of the Changing-look AGN 1ES 1927+654}.
\newblock \emph{\bibinfo{journal}{\apjs}} \textbf{\bibinfo{volume}{255}},
  \bibinfo{pages}{7} (\bibinfo{year}{2021}).
\newblock \eprint{2102.05666}.

\bibitem{Krumpe:2017ss}
\bibinfo{author}{{Krumpe}, M.} \emph{et~al.}
\newblock {The Close AGN Reference Survey (CARS). Mrk 1018 halts dimming and
  experiences strong short-term variability}.
\newblock \emph{\bibinfo{journal}{\aap}} \textbf{\bibinfo{volume}{607}},
  \bibinfo{pages}{L9} (\bibinfo{year}{2017}).
\newblock \eprint{1710.09382}.

\bibitem{Parker:2019sp}
\bibinfo{author}{{Parker}, M.~L.} \emph{et~al.}
\newblock {X-ray spectra reveal the reawakening of the repeat changing-look AGN
  NGC 1566}.
\newblock \emph{\bibinfo{journal}{\mnras}} \textbf{\bibinfo{volume}{483}},
  \bibinfo{pages}{L88--L92} (\bibinfo{year}{2019}).
\newblock \eprint{1811.10289}.

\bibitem{Jana:2021wk}
\bibinfo{author}{{Jana}, A.} \emph{et~al.}
\newblock {Broad-band X-ray observations of the 2018 outburst of the
  changing-look active galactic nucleus NGC 1566}.
\newblock \emph{\bibinfo{journal}{\mnras}} \textbf{\bibinfo{volume}{507}},
  \bibinfo{pages}{687--703} (\bibinfo{year}{2021}).
\newblock \eprint{2107.11127}.

\bibitem{Masterson:2022wy}
\bibinfo{author}{{Masterson}, M.} \emph{et~al.}
\newblock {Evolution of a Relativistic Outflow and X-Ray Corona in the Extreme
  Changing-look AGN 1ES 1927+654}.
\newblock \emph{\bibinfo{journal}{\apj}} \textbf{\bibinfo{volume}{934}},
  \bibinfo{pages}{35} (\bibinfo{year}{2022}).
\newblock \eprint{2206.05140}.

\bibitem{Sheng:2017of}
\bibinfo{author}{{Sheng}, Z.} \emph{et~al.}
\newblock {Mid-infrared Variability of Changing-look AGNs}.
\newblock \emph{\bibinfo{journal}{\apjl}} \textbf{\bibinfo{volume}{846}},
  \bibinfo{pages}{L7} (\bibinfo{year}{2017}).
\newblock \eprint{1707.02686}.

\bibitem{Kokubo:2020kr}
\bibinfo{author}{{Kokubo}, M.} \& \bibinfo{author}{{Minezaki}, T.}
\newblock {Rapid luminosity decline and subsequent reformation of the innermost
  dust distribution in the changing-look AGN Mrk 590}.
\newblock \emph{\bibinfo{journal}{\mnras}} \textbf{\bibinfo{volume}{491}},
  \bibinfo{pages}{4615--4633} (\bibinfo{year}{2020}).
\newblock \eprint{1904.08946}.

\bibitem{Panessa:2019vp}
\bibinfo{author}{{Panessa}, F.} \emph{et~al.}
\newblock {The origin of radio emission from radio-quiet active galactic
  nuclei}.
\newblock \emph{\bibinfo{journal}{Nature Astronomy}}
  \textbf{\bibinfo{volume}{3}}, \bibinfo{pages}{387--396}
  (\bibinfo{year}{2019}).
\newblock \eprint{1902.05917}.

\bibitem{Yang:2021jr}
\bibinfo{author}{{Yang}, J.} \emph{et~al.}
\newblock {A compact core-jet structure in the changing-look Seyfert NGC 2617}.
\newblock \emph{\bibinfo{journal}{\mnras}} \textbf{\bibinfo{volume}{503}},
  \bibinfo{pages}{3886--3895} (\bibinfo{year}{2021}).
\newblock \eprint{2103.04309}.

\bibitem{Koay:2016jy}
\bibinfo{author}{{Koay}, J.~Y.}, \bibinfo{author}{{Vestergaard}, M.},
  \bibinfo{author}{{Bignall}, H.~E.}, \bibinfo{author}{{Reynolds}, C.} \&
  \bibinfo{author}{{Peterson}, B.~M.}
\newblock {Parsec-scale radio morphology and variability of a changing-look
  AGN: the case of Mrk 590}.
\newblock \emph{\bibinfo{journal}{\mnras}} \textbf{\bibinfo{volume}{460}},
  \bibinfo{pages}{304--316} (\bibinfo{year}{2016}).
\newblock \eprint{1602.07289}.

\bibitem{Yang:2021oe}
\bibinfo{author}{{Yang}, J.} \emph{et~al.}
\newblock {A parsec-scale faint jet in the nearby changing-look Seyfert galaxy
  Mrk 590}.
\newblock \emph{\bibinfo{journal}{\mnras}} \textbf{\bibinfo{volume}{502}},
  \bibinfo{pages}{L61--L65} (\bibinfo{year}{2021}).
\newblock \eprint{2101.04629}.

\bibitem{Lyu:2021ag}
\bibinfo{author}{{Lyu}, B.}, \bibinfo{author}{{Yan}, Z.},
  \bibinfo{author}{{Yu}, W.} \& \bibinfo{author}{{Wu}, Q.}
\newblock {Long-term and multiwavelength evolution of a changing-look AGN Mrk
  1018}.
\newblock \emph{\bibinfo{journal}{\mnras}} \textbf{\bibinfo{volume}{506}},
  \bibinfo{pages}{4188--4198} (\bibinfo{year}{2021}).
\newblock \eprint{2106.03059}.

\bibitem{Laha:2022yu}
\bibinfo{author}{{Laha}, S.} \emph{et~al.}
\newblock {A Radio, Optical, UV, and X-Ray View of the Enigmatic Changing-look
  Active Galactic Nucleus 1ES 1927+654 from Its Pre- to Postflare States}.
\newblock \emph{\bibinfo{journal}{\apj}} \textbf{\bibinfo{volume}{931}},
  \bibinfo{pages}{5} (\bibinfo{year}{2022}).
\newblock \eprint{2203.07446}.

\bibitem{Husemann:2016iu}
\bibinfo{author}{{Husemann}, B.} \emph{et~al.}
\newblock {The Close AGN Reference Survey (CARS). What is causing Mrk 1018's
  return to the shadows after 30 years?}
\newblock \emph{\bibinfo{journal}{\aap}} \textbf{\bibinfo{volume}{593}},
  \bibinfo{pages}{L9} (\bibinfo{year}{2016}).
\newblock \eprint{1609.04425}.

\bibitem{Jaffarian:2020qt}
\bibinfo{author}{{Jaffarian}, G.~W.} \& \bibinfo{author}{{Gaskell}, C.~M.}
\newblock {The relationship between X-ray and optical absorbers in active
  galactic nuclei}.
\newblock \emph{\bibinfo{journal}{\mnras}} \textbf{\bibinfo{volume}{493}},
  \bibinfo{pages}{930--939} (\bibinfo{year}{2020}).
\newblock \eprint{2001.08900}.

\bibitem{Sheng:2017fu}
\bibinfo{author}{{Sheng}, Z.} \emph{et~al.}
\newblock {Mid-infrared Variability of Changing-look AGNs}.
\newblock \emph{\bibinfo{journal}{\apjl}} \textbf{\bibinfo{volume}{846}},
  \bibinfo{pages}{L7} (\bibinfo{year}{2017}).
\newblock \eprint{1707.02686}.

\bibitem{Hutsemekers:2019zw}
\bibinfo{author}{{Hutsem{\'e}kers}, D.} \emph{et~al.}
\newblock {Polarization of changing-look quasars}.
\newblock \emph{\bibinfo{journal}{\aap}} \textbf{\bibinfo{volume}{625}},
  \bibinfo{pages}{A54} (\bibinfo{year}{2019}).
\newblock \eprint{1904.03914}.

\bibitem{Bentz:2013cn}
\bibinfo{author}{{Bentz}, M.~C.} \emph{et~al.}
\newblock {The Low-luminosity End of the Radius-Luminosity Relationship for
  Active Galactic Nuclei}.
\newblock \emph{\bibinfo{journal}{\apj}} \textbf{\bibinfo{volume}{767}},
  \bibinfo{pages}{149} (\bibinfo{year}{2013}).
\newblock \eprint{1303.1742}.

\bibitem{Goodrich90}
\bibinfo{author}{{Goodrich}, R.~W.}
\newblock {PA beta Measurements and Reddening in Seyfert 1.8 and 1.9 Galaxies}.
\newblock \emph{\bibinfo{journal}{\apj}} \textbf{\bibinfo{volume}{355}},
  \bibinfo{pages}{88} (\bibinfo{year}{1990}).

\bibitem{Zeltyn:2022gr}
\bibinfo{author}{{Zeltyn}, G.} \emph{et~al.}
\newblock {A Transient `Changing-Look' AGN Resolved on Month Timescales From
  First-Year SDSS-V Data}.
\newblock \emph{\bibinfo{journal}{arXiv e-prints}}
  \bibinfo{pages}{arXiv:2210.07258} (\bibinfo{year}{2022}).
\newblock \eprint{2210.07258}.

\bibitem{Emmering:1992ut}
\bibinfo{author}{{Emmering}, R.~T.}, \bibinfo{author}{{Blandford}, R.~D.} \&
  \bibinfo{author}{{Shlosman}, I.}
\newblock {Magnetic Acceleration of Broad Emission-Line Clouds in Active
  Galactic Nuclei}.
\newblock \emph{\bibinfo{journal}{\apj}} \textbf{\bibinfo{volume}{385}},
  \bibinfo{pages}{460} (\bibinfo{year}{1992}).

\bibitem{Elitzur:2006ec}
\bibinfo{author}{{Elitzur}, M.} \& \bibinfo{author}{{Shlosman}, I.}
\newblock {The AGN-obscuring Torus: The End of the ``Doughnut'' Paradigm?}
\newblock \emph{\bibinfo{journal}{\apjl}} \textbf{\bibinfo{volume}{648}},
  \bibinfo{pages}{L101--L104} (\bibinfo{year}{2006}).
\newblock \eprint{astro-ph/0605686}.

\bibitem{Elitzur:2009hh}
\bibinfo{author}{{Elitzur}, M.} \& \bibinfo{author}{{Ho}, L.~C.}
\newblock {On the Disappearance of the Broad-Line Region in Low-Luminosity
  Active Galactic Nuclei}.
\newblock \emph{\bibinfo{journal}{\apjl}} \textbf{\bibinfo{volume}{701}},
  \bibinfo{pages}{L91--L94} (\bibinfo{year}{2009}).
\newblock \eprint{0907.3752}.

\bibitem{Elitzur:2014kc}
\bibinfo{author}{{Elitzur}, M.}, \bibinfo{author}{{Ho}, L.~C.} \&
  \bibinfo{author}{{Trump}, J.~R.}
\newblock {Evolution of broad-line emission from active galactic nuclei}.
\newblock \emph{\bibinfo{journal}{\mnras}} \textbf{\bibinfo{volume}{438}},
  \bibinfo{pages}{3340--3351} (\bibinfo{year}{2014}).
\newblock \eprint{1312.4922}.

\bibitem{Balbus:1991qv}
\bibinfo{author}{{Balbus}, S.~A.} \& \bibinfo{author}{{Hawley}, J.~F.}
\newblock {A Powerful Local Shear Instability in Weakly Magnetized Disks. I.
  Linear Analysis}.
\newblock \emph{\bibinfo{journal}{\apj}} \textbf{\bibinfo{volume}{376}},
  \bibinfo{pages}{214} (\bibinfo{year}{1991}).

\bibitem{Alloin:1985it}
\bibinfo{author}{{Alloin}, D.}, \bibinfo{author}{{Pelat}, D.},
  \bibinfo{author}{{Phillips}, M.} \& \bibinfo{author}{{Whittle}, M.}
\newblock {Recent spectral variations in the active nucleus of NGC 1566.}
\newblock \emph{\bibinfo{journal}{\apj}} \textbf{\bibinfo{volume}{288}},
  \bibinfo{pages}{205--220} (\bibinfo{year}{1985}).

\bibitem{Lawrence:2012sd}
\bibinfo{author}{{Lawrence}, A.}
\newblock {The UV peak in active galactic nuclei: a false continuum from
  blurred reflection?}
\newblock \emph{\bibinfo{journal}{\mnras}} \textbf{\bibinfo{volume}{423}},
  \bibinfo{pages}{451--463} (\bibinfo{year}{2012}).
\newblock \eprint{1110.0854}.

\bibitem{Lawrence:2018gc}
\bibinfo{author}{{Lawrence}, A.}
\newblock {Quasar viscosity crisis}.
\newblock \emph{\bibinfo{journal}{Nature Astronomy}}
  \textbf{\bibinfo{volume}{2}}, \bibinfo{pages}{102--103}
  (\bibinfo{year}{2018}).
\newblock \eprint{1802.00408}.

\bibitem{Kelly:2009bo}
\bibinfo{author}{{Kelly}, B.~C.}, \bibinfo{author}{{Bechtold}, J.} \&
  \bibinfo{author}{{Siemiginowska}, A.}
\newblock {Are the Variations in Quasar Optical Flux Driven by Thermal
  Fluctuations?}
\newblock \emph{\bibinfo{journal}{\apj}} \textbf{\bibinfo{volume}{698}},
  \bibinfo{pages}{895--910} (\bibinfo{year}{2009}).
\newblock \eprint{0903.5315}.

\bibitem{Jiang:2016ql}
\bibinfo{author}{{Jiang}, Y.-F.}, \bibinfo{author}{{Davis}, S.~W.} \&
  \bibinfo{author}{{Stone}, J.~M.}
\newblock {Iron Opacity Bump Changes the Stability and Structure of Accretion
  Disks in Active Galactic Nuclei}.
\newblock \emph{\bibinfo{journal}{\apj}} \textbf{\bibinfo{volume}{827}},
  \bibinfo{pages}{10} (\bibinfo{year}{2016}).
\newblock \eprint{1601.06836}.

\bibitem{Agol:2000pz}
\bibinfo{author}{{Agol}, E.} \& \bibinfo{author}{{Krolik}, J.~H.}
\newblock {Magnetic Stress at the Marginally Stable Orbit: Altered Disk
  Structure, Radiation, and Black Hole Spin Evolution}.
\newblock \emph{\bibinfo{journal}{\apj}} \textbf{\bibinfo{volume}{528}},
  \bibinfo{pages}{161--170} (\bibinfo{year}{2000}).
\newblock \eprint{astro-ph/9908049}.

\bibitem{Dexter:2019nr}
\bibinfo{author}{{Dexter}, J.} \& \bibinfo{author}{{Begelman}, M.~C.}
\newblock {Extreme AGN variability: evidence of magnetically elevated
  accretion?}
\newblock \emph{\bibinfo{journal}{\mnras}} \textbf{\bibinfo{volume}{483}},
  \bibinfo{pages}{L17--L21} (\bibinfo{year}{2019}).
\newblock \eprint{1807.03314}.

\bibitem{Feng:2021iy}
\bibinfo{author}{{Feng}, J.}, \bibinfo{author}{{Cao}, X.},
  \bibinfo{author}{{Li}, J.-w.} \& \bibinfo{author}{{Gu}, W.-M.}
\newblock {A Magnetic Disk-outflow Model for Changing Look Active Galactic
  Nuclei}.
\newblock \emph{\bibinfo{journal}{\apj}} \textbf{\bibinfo{volume}{916}},
  \bibinfo{pages}{61} (\bibinfo{year}{2021}).
\newblock \eprint{2106.00650}.

\bibitem{Jiang2019}
\bibinfo{author}{{Jiang}, Y.-F.}, \bibinfo{author}{{Blaes}, O.},
  \bibinfo{author}{{Stone}, J.~M.} \& \bibinfo{author}{{Davis}, S.~W.}
\newblock {Global Radiation Magnetohydrodynamic Simulations of sub-Eddington
  Accretion Disks around Supermassive Black Holes}.
\newblock \emph{\bibinfo{journal}{\apj}} \textbf{\bibinfo{volume}{885}},
  \bibinfo{pages}{144} (\bibinfo{year}{2019}).
\newblock \eprint{1904.01674}.

\bibitem{Sniegowska:2020an}
\bibinfo{author}{{Sniegowska}, M.}, \bibinfo{author}{{Czerny}, B.},
  \bibinfo{author}{{Bon}, E.} \& \bibinfo{author}{{Bon}, N.}
\newblock {Possible mechanism for multiple changing-look phenomena in active
  galactic nuclei}.
\newblock \emph{\bibinfo{journal}{\aap}} \textbf{\bibinfo{volume}{641}},
  \bibinfo{pages}{A167} (\bibinfo{year}{2020}).
\newblock \eprint{2007.06441}.

\bibitem{Lin:1986yt}
\bibinfo{author}{{Lin}, D.~N.~C.} \& \bibinfo{author}{{Shields}, G.~A.}
\newblock {Accretion Disks and Periodic Outbursts of Active Galaxies Nuclei}.
\newblock \emph{\bibinfo{journal}{\apj}} \textbf{\bibinfo{volume}{305}},
  \bibinfo{pages}{28} (\bibinfo{year}{1986}).

\bibitem{Ai:2020wv}
\bibinfo{author}{{Ai}, Y.} \emph{et~al.}
\newblock {X-Ray Spectral Shape Variation in Changing-look Seyfert Galaxy SDSS
  J155258+273728}.
\newblock \emph{\bibinfo{journal}{\apjl}} \textbf{\bibinfo{volume}{890}},
  \bibinfo{pages}{L29} (\bibinfo{year}{2020}).

\bibitem{Ruan:2019pa}
\bibinfo{author}{{Ruan}, J.~J.} \emph{et~al.}
\newblock {The Analogous Structure of Accretion Flows in Supermassive and
  Stellar Mass Black Holes: New Insights from Faded Changing-look Quasars}.
\newblock \emph{\bibinfo{journal}{\apj}} \textbf{\bibinfo{volume}{883}},
  \bibinfo{pages}{76} (\bibinfo{year}{2019}).
\newblock \eprint{1903.02553}.

\bibitem{Scepi:2021yc}
\bibinfo{author}{{Scepi}, N.}, \bibinfo{author}{{Begelman}, M.~C.} \&
  \bibinfo{author}{{Dexter}, J.}
\newblock {Magnetic flux inversion in a peculiar changing look AGN}.
\newblock \emph{\bibinfo{journal}{\mnras}} \textbf{\bibinfo{volume}{502}},
  \bibinfo{pages}{L50--L54} (\bibinfo{year}{2021}).
\newblock \eprint{2011.01954}.

\bibitem{Merloni:2015ew}
\bibinfo{author}{{Merloni}, A.} \emph{et~al.}
\newblock {A tidal disruption flare in a massive galaxy? Implications for the
  fuelling mechanisms of nuclear black holes}.
\newblock \emph{\bibinfo{journal}{\mnras}} \textbf{\bibinfo{volume}{452}},
  \bibinfo{pages}{69--87} (\bibinfo{year}{2015}).
\newblock \eprint{1503.04870}.

\bibitem{vanVelzen2020}
\bibinfo{author}{{van Velzen}, S.}, \bibinfo{author}{{Holoien}, T. W.~S.},
  \bibinfo{author}{{Onori}, F.}, \bibinfo{author}{{Hung}, T.} \&
  \bibinfo{author}{{Arcavi}, I.}
\newblock {Optical-Ultraviolet Tidal Disruption Events}.
\newblock \emph{\bibinfo{journal}{\ssr}} \textbf{\bibinfo{volume}{216}},
  \bibinfo{pages}{124} (\bibinfo{year}{2020}).
\newblock \eprint{2008.05461}.

\bibitem{Zhang:2021fc}
\bibinfo{author}{{Zhang}, X.-G.}
\newblock {Further evidence to support a tidal disruption event in the
  changing-look AGN SDSS J0159}.
\newblock \emph{\bibinfo{journal}{\mnras}} \textbf{\bibinfo{volume}{500}},
  \bibinfo{pages}{L57--L61} (\bibinfo{year}{2021}).
\newblock \eprint{2011.06213}.

\bibitem{Kesden2012}
\bibinfo{author}{{Kesden}, M.}
\newblock {Tidal-disruption rate of stars by spinning supermassive black
  holes}.
\newblock \emph{\bibinfo{journal}{\prd}} \textbf{\bibinfo{volume}{85}},
  \bibinfo{pages}{024037} (\bibinfo{year}{2012}).
\newblock \eprint{1109.6329}.

\bibitem{Li:2022pp}
\bibinfo{author}{{Li}, R.} \emph{et~al.}
\newblock {The Host Galaxy and Rapidly Evolving Broad-line Region in the
  Changing-look Active Galactic Nucleus 1ES 1927+654}.
\newblock \emph{\bibinfo{journal}{\apj}} \textbf{\bibinfo{volume}{933}},
  \bibinfo{pages}{70} (\bibinfo{year}{2022}).
\newblock \eprint{2208.01797}.

\bibitem{Karas2007}
\bibinfo{author}{{Karas}, V.} \& \bibinfo{author}{{{\v{S}}ubr}, L.}
\newblock {Enhanced activity of massive black holes by stellar capture assisted
  by a self-gravitating accretion disc}.
\newblock \emph{\bibinfo{journal}{\aap}} \textbf{\bibinfo{volume}{470}},
  \bibinfo{pages}{11--19} (\bibinfo{year}{2007}).
\newblock \eprint{0704.2781}.

\bibitem{McKernan:2022yd}
\bibinfo{author}{{McKernan}, B.} \emph{et~al.}
\newblock {Starfall: a heavy rain of stars in 'turning on' AGN}.
\newblock \emph{\bibinfo{journal}{\mnras}} \textbf{\bibinfo{volume}{514}},
  \bibinfo{pages}{4102--4110} (\bibinfo{year}{2022}).
\newblock \eprint{2110.03741}.

\bibitem{Chan:2019aa}
\bibinfo{author}{{Chan}, C.-H.}, \bibinfo{author}{{Piran}, T.},
  \bibinfo{author}{{Krolik}, J.~H.} \& \bibinfo{author}{{Saban}, D.}
\newblock {Tidal Disruption Events in Active Galactic Nuclei}.
\newblock \emph{\bibinfo{journal}{\apj}} \textbf{\bibinfo{volume}{881}},
  \bibinfo{pages}{113} (\bibinfo{year}{2019}).
\newblock \eprint{1904.12261}.

\bibitem{Chan2020}
\bibinfo{author}{{Chan}, C.-H.}, \bibinfo{author}{{Piran}, T.} \&
  \bibinfo{author}{{Krolik}, J.~H.}
\newblock {Light Curves of Tidal Disruption Events in Active Galactic Nuclei}.
\newblock \emph{\bibinfo{journal}{\apj}} \textbf{\bibinfo{volume}{903}},
  \bibinfo{pages}{17} (\bibinfo{year}{2020}).
\newblock \eprint{2004.06234}.

\bibitem{Wang:2020si}
\bibinfo{author}{{Wang}, J.-M.} \& \bibinfo{author}{{Bon}, E.}
\newblock {Changing-look active galactic nuclei: close binaries of supermassive
  black holes in action}.
\newblock \emph{\bibinfo{journal}{\aap}} \textbf{\bibinfo{volume}{643}},
  \bibinfo{pages}{L9} (\bibinfo{year}{2020}).
\newblock \eprint{2010.04417}.

\bibitem{Kim:2018sa}
\bibinfo{author}{{Kim}, D.~C.}, \bibinfo{author}{{Yoon}, I.} \&
  \bibinfo{author}{{Evans}, A.~S.}
\newblock {Recoiling Supermassive Black Hole in Changing-look AGN Mrk 1018}.
\newblock \emph{\bibinfo{journal}{\apj}} \textbf{\bibinfo{volume}{861}},
  \bibinfo{pages}{51} (\bibinfo{year}{2018}).
\newblock \eprint{1805.05251}.

\bibitem{Ricci:2017aa}
\bibinfo{author}{{Ricci}, C.} \emph{et~al.}
\newblock {Growing supermassive black holes in the late stages of galaxy
  mergers are heavily obscured}.
\newblock \emph{\bibinfo{journal}{\mnras}} \textbf{\bibinfo{volume}{468}},
  \bibinfo{pages}{1273--1299} (\bibinfo{year}{2017}).
\newblock \eprint{1701.04825}.

\bibitem{Yamada:2021to}
\bibinfo{author}{{Yamada}, S.} \emph{et~al.}
\newblock {Comprehensive Broadband X-Ray and Multiwavelength Study of Active
  Galactic Nuclei in 57 Local Luminous and Ultraluminous Infrared Galaxies
  Observed with NuSTAR and/or Swift/BAT}.
\newblock \emph{\bibinfo{journal}{\apjs}} \textbf{\bibinfo{volume}{257}},
  \bibinfo{pages}{61} (\bibinfo{year}{2021}).
\newblock \eprint{2107.10855}.

\bibitem{Merloni:2020ec}
\bibinfo{author}{{Merloni}, A.}, \bibinfo{author}{{Nandra}, K.} \&
  \bibinfo{author}{{Predehl}, P.}
\newblock {eROSITA's X-ray eyes on the Universe}.
\newblock \emph{\bibinfo{journal}{Nature Astronomy}}
  \textbf{\bibinfo{volume}{4}}, \bibinfo{pages}{634--636}
  (\bibinfo{year}{2020}).

\bibitem{Kollmeier:2017dx}
\bibinfo{author}{{Kollmeier}, J.~A.} \emph{et~al.}
\newblock {SDSS-V: Pioneering Panoptic Spectroscopy}.
\newblock \emph{\bibinfo{journal}{arXiv e-prints}}
  \bibinfo{pages}{arXiv:1711.03234} (\bibinfo{year}{2017}).
\newblock \eprint{1711.03234}.

\bibitem{Sanchez-Saez:2021bq}
\bibinfo{author}{{S{\'a}nchez-S{\'a}ez}, P.} \emph{et~al.}
\newblock {Searching for Changing-state AGNs in Massive Data Sets. I. Applying
  Deep Learning and Anomaly-detection Techniques to Find AGNs with Anomalous
  Variability Behaviors}.
\newblock \emph{\bibinfo{journal}{\aj}} \textbf{\bibinfo{volume}{162}},
  \bibinfo{pages}{206} (\bibinfo{year}{2021}).
\newblock \eprint{2106.07660}.

\bibitem{de-Jong:2019fp}
\bibinfo{author}{{de Jong}, R.~S.} \emph{et~al.}
\newblock {4MOST: Project overview and information for the First Call for
  Proposals}.
\newblock \emph{\bibinfo{journal}{The Messenger}}
  \textbf{\bibinfo{volume}{175}}, \bibinfo{pages}{3--11}
  (\bibinfo{year}{2019}).
\newblock \eprint{1903.02464}.

\end{thebibliography}
\end{document}